\documentclass{aa} 
\usepackage{psfig,epsfig,graphicx,times}
\usepackage{aalongtable}
\newcommand{\kms}{{{km\,s}$^{-1}$}}
\newcommand{\teff}{{$T_\mathrm{eff}$}}
\newcommand{\logg}{{log~$g$}}
\newcommand{\logL}{{log~$L$}}
\newcommand{\deltaL}{{$\Delta$log~$L$}}

\newcommand{\vsini}{$v$\,sin\,$i$}
\newcommand{\hei}{He\,{\sc i}}
\newcommand{\msun}{M$_\odot$}

\def\5{\footnotesize V\normalsize}
\def\4{\footnotesize IV\normalsize}
\def\3{\footnotesize III\normalsize}
\def\2{\footnotesize II\normalsize}
\def\1{\footnotesize I\normalsize}

\def\kms{$\mbox{km s}^{-1}$}
\def\p{$\phantom{:}$}

\def\pp{$\phantom{-}$}
\def\o{$\phantom{1}$}

\begin{document}

\title{The VLT-FLAMES Survey of Massive Stars: Stellar parameters 
and rotational velocities in NGC\,3293, NGC\,4755 and NGC\,6611}
\subtitle{}

\author{P.L. Dufton\inst{1}
   \and S.J. Smartt\inst{1}
   \and J.K. Lee\inst{1} 
   \and R.S.I. Ryans\inst{1}
   \and I. Hunter\inst{1}
   \and C.J. Evans\inst{2}
   \and A. Herrero\inst{3}
   \and C. Trundle\inst{1,4}
   \and D.J. Lennon\inst{4,3}
   \and M.J. Irwin\inst{5}
   \and A. Kaufer\inst{6}} 

\institute
{Department of Physics and Astronomy, Queen's University Belfast, BT7 1NN, UK       
\and UK Astronomy Technology Centre, Royal Observatory, Blackford Hill,
Edinburgh EH9~3HJ, Scotland, UK
\and Instituto de Astrof\'isica de Canarias, Calle V\'ia L\`actea, E-38200 
La Laguna, Tenerife, Canary Islands, Spain
\and The Isaac Newton Group of Telescopes, Apartado de Correos 321, E-38700
Santa Cruz de La Palma, Canary Islands, Spain
\and Institute of Astronomy, University of Cambridge,
Madingley Road, Cambridge, CB3 0HA, UK
\and European Southern Observatory, Alonso de Cordova 3107, Santiago 19, Chile
}

\offprints{p.dufton@qub.ac.uk}

\date{Received: 2005 / Accepted}

\abstract{An analysis is presented of VLT-FLAMES spectroscopy for three 
Galactic clusters, NGC\,3293, NGC\,4755 and NGC\,6611. Non-LTE model 
atmosphere calculations have been used to estimate effective temperatures 
(from either the helium spectrum or the silicon ionization equilibrium) 
and gravities (from the hydrogen spectrum). Projected rotational velocities 
have been deduced from the helium spectrum (for fast and moderate rotators) 
or the metal line spectrum (for slow rotators). The origin of the low gravity 
estimates for apparently near main sequence objects is discussed and is 
related to the stellar rotational velocity. The atmospheric parameters 
have been used to estimate cluster distances (which are generally in 
good agreement with previous determinations) 
and these have been used to estimate stellar luminosities and evolutionary 
masses. The observed Hertzsprung-Russell diagrams are compared with 
theoretical predictions and some discrepancies including differences in the
main sequence luminosities are discussed. Cluster ages have been deduced 
and evidence for non-coeval star formation is found for all three of the 
clusters.  Projected rotational velocities for targets in the 
older clusters, NGC\,3293 and NGC\,4755, have been found to be systematically 
larger than those for the field, confirming recent results in other 
similar age clusters. The distribution of projected
rotational velocities are consistent with a Gaussian distribution of
intrinsic rotational velocities. For the relatively unevolved targets
in the older clusters, NGC\,3293 and NGC\,4755, the peak of the velocity
distribution would be 250 \kms with a full-width-half-maximum of 
approximately 180 \kms. For NGC\,6611, the sample size is relatively 
small but implies a lower mean rotational velocity. This may be
evidence for the spin-down effect due to angular momentum loss through 
stellar winds, although our results are consistent with those found
for very young high mass stars. For all three 
clusters we deduce present day mass functions with $\Gamma$-values 
in the range of -1.5 to -1.8, which are similar to other young 
stellar clusters in the Milky Way. 
}

\titlerunning{Stellar parameters in NGC\,3293, NGC\,4755 and NGC\,6611}
\authorrunning{Dufton et al.}
\maketitle

\section{Introduction}

As part of a European Southern Observatory Large Programme,
we are using the Fibre Large Array Multi-Element Spectrograph  
(FLAMES) at the Very Large Telescope to survey the hot stellar populations
in our Galaxy and in the Magellanic Clouds. We have observed in excess of 
50 O-type stars and 500 B-type stars, in a total of seven clusters. These
data have been supplemented with observations of our brighter targets
using the Fibre-Fed Extended Range Optical Spectrograph (FEROS) at La Silla.
Evans et al. (\cite{Eva05}; hereafter Paper I) have given an overview 
of the scientific goals of the project, discussed the target selection
and data reduction techniques, and presented some preliminary analysis 
for three Galactic clusters. The scientific aims of the survey as discussed 
in Paper I, included understanding how rotation and metallicity 
influence stellar evolution and stellar winds, calibration of the 
wind momentum-luminosity relationship and 
the nature of supernova progenitors. In the case of stellar evolution,
theoretical studies (see, for example, Heger \& Langer \cite{Heg00};
Maeder \& Meynet \cite{Mae00}) provide predictions of the surface 
enhancements expected for helium and nitrogen. Of particular relevance to 
our survey are the larger logarithmic enhancements predicted for lower
metallicity regimes (Maeder \& Meynet \cite{Mae01}). Although there are
many previous detailed quantitative 
studies of O- and B-type stars (Crowther at al. 
\cite{Cro02, Cro05}; Bouret et al. \cite{Bou03}; Hillier at al. 
\cite{Hil03}; Evans et al. \cite{Eva04}; Trundle et al. 
\cite{Tru04}; Walborn et al. \cite{Wal04}; Dufton et al. 
\cite{Duf05}; Trundle \& Lennon \cite{Tru05}), these have tended 
to concentrate on narrow lined (and hence possibly slowly rotating) 
stars, whilst the sample sizes have been relatively small. The current 
survey will sample both the main sequence and supergiant regimes and 
include rapidly rotating stars, thereby extending the scope of such 
studies.

In order to have a sample of stars in the Milky Way to compare with the 
FLAMES datasets for the SMC and LMC we have obtained spectra of likely 
cluster members in the three Galactic Clusters, NGC\,3293, NGC\,4755 and 
NGC\,6611. These clusters were chosen so as to have a similar age range 
to their lower metallicity Magellanic Cloud counterparts. Although the 
Milky Way has an abundance gradient (see, for example,
Afflerbach et al. \cite{Aff97}, Rolleston et al. 2000, 
Gummersbach et al. 1998, Daflon \& Cunha 2004) these clusters are 
relatively close to the solar position and appear to have a similar 
metallicity to that of the Sun from a preliminary model atmosphere analysis 
of selected narrow line targets (see Hunter et al. 2006 for details). 
Previous studies have discussed the HR-diagrams, star 
formation timescales, initial mass funtions and rotational velocity 
distributions in young Galactic clusters and have highlighted the value 
of spectroscopic analysis of cluster members rather than spectral type 
assignments from photometric colours (e.g. see Massey et al. 
\cite{1995ApJ...454..151M}, Hillenbrand et al. \cite{Hil93}). 
Additionally there has been an revival of interest in the rotational 
velocity distribution of large samples of massive stars for comparison 
with the predictions of theoretical evolutionary models and also to 
probe initial conditions during star-formation. Strom et al. 
(\cite{Str05}) have measured projected rotational velocities for approximately 
200 B-type stars in h\&$\chi$\,Per and suggest that 
differences between the mean values for the cluster and field stars are due 
to the star-formation process rather than being a consequence of angular 
momentum evolution as they have moved from the ZAMS.  Huang \& Gies 
(\cite{Hua05a}) have presented rotational velocities of stars in 
nine Galactic open clusters with age ranges of approximately 6 to 73 Myr, and 
have provided an interpretation on the basis of evolutionary arguments. 
Additionally the extensive field star sample of  Abt et al. 
(\cite{2002ApJ...573..359A}) remains a benchmark study with which to 
compare cluster rotational velocities. All of these studies illustrate 
the importance of large stellar samples in order to ameliorate statistical 
significance problems arising from the random inclination angle of the 
rotational axes. The availability of the FLAMES spectrograph on an 
8m telescope has allowed studies of large samples to be extended
to the Magellanic Clouds. For example, 
Martayan et al. (\cite{2006A&A...445..931M})
have studied the effects of metallicity on the rotational velocities of 
B-type and Be stars in the LMC and find that the LMC stars rotate
faster than Be stars at solar metallicity. 

In this paper we analyse the early type stars in the three 
Galactic clusters observed with the FLAMES spectrograph. Stellar atmospheric 
parameters are deduced using non-LTE model atmosphere calculations and 
rotational velocities are estimated by comparing observed and theoretical 
rotationally broadened spectra. The former are discussed with particular 
reference to the surface gravities found for near main sequence objects;
evidence is presented that relates the lower gravity estimates to large 
stellar rotational velocities. Our projected rotational velocities are 
discussed both in the context of possible systematic differences between 
field and cluster populations and to infer the underlying rotational 
velocity distributions. The estimates of the atmospheric parameters are 
used to deduce cluster distances leading to estimates of stellar luminosities 
and evolutionary masses.  The former are then used to generate 
Hertzsprung-Russell diagrams and the agreement with theoretical 
predictions is discussed, while present day mass functions are 
estimated from the latter. 

\section{Observational data}

The observational data have been obtained during a European 
Southern Observatory (ESO) Large Programme to study early-type stars in
our Galaxy and the Magellanic Clouds.  For the former, targets were 
observed in three clusters -- NGC\,3293, NGC\,4755 and 
NGC\,6611 -- with the Fibre Large Array Multi-Element Spectrograph
(FLAMES, R $\approx$ 20\,000, see Pasquini et al. \cite{Pas02}) 
on the Very Large Telescope (VLT). The targets
were selected based on astrometry and photometry from the pre-FLAMES 
observations using the wide-field imager (WFI) on the 2.2-m Max Planck 
Gesellschaft (MPG)/ESO telescope, whilst the brightest stars were 
observed separately with the Fibre-Fed, Extended Range Optical Spectrograph 
(FEROS, R $\approx$ 48\,000, see Kaufer et al. \cite{Kau99}) on 
the MPG. Further details of the target 
selection criteria and data acquisition are presented in Paper I.
Using six standard high-resolution GIRAFFE settings, the FLAMES data  
cover a wavelength range of 3950-4755\AA\, and 6380-6620\AA\/ 
with a typical signal-to-noise (S/N) ratio of 100, whilst the FEROS 
spectra  range from 3600 to 9200~\AA\/ with a S/N ratio $>$100.  
Reductions were performed using both IRAF\footnote{IRAF is 
distributed by the National Optical Astronomy Observatories, which 
are operated by the Association of Universities for Research in Astronomy, 
Inc., under cooperative agreement with the National Science Foundation.} 
software and the FLAMES pipeline (girBLDRS, Blecha et al. \cite{Ble03}), 
which showed excellent agreement. Details of the data reduction
procedure, spectral classification and radial velocity measurements 
of the targets were presented in Paper\,I, which also contains
the coordinates and cross-referencing for the designations of
the targets. As an appendix we include the 
finding charts for all objects we observed in these clusters (available
on-line only). 

In this paper, we have only considered targets classified as earlier 
than B9. This cutoff was adopted as for later spectral types the neutral 
helium lines, which have been used as a temperature diagnostic were not 
normally visible. One exception is the target 4755-001 (note we have 
adopted the nomenclature used in Paper I), which despite a classification 
of B9\,Ia had a measurable helium spectrum.

\section{Method of analysis}\label{analysis}

In this section, we discuss the methods that were used to estimate 
the stellar atmospheric parameters, projected rotational velocities, 
masses and luminosities together with the cluster distances. 
We also discuss some problematic stars (Sect. \ref{sec:stars}) and 
also the criteria to identify cluster membership (Sect. \ref{sec:member}). 

\subsection{Initial estimation of atmospheric parameters}\label{sec:atm}

The estimation of the stellar atmospheric parameters --  effective 
temperature (\teff) and logarithmic surface gravity (\logg) -- 
requires an iterative process and we have adopted different 
approaches depending on the stellar effective temperature.   
For B2--B8 stars, which make up the majority of the sample,
the strengths of the hydrogen and neutral helium lines were utilized, 
while the ionization balance of Si {\sc iii/iv} was used for the 
earlier B spectral types.  In the case of the O-type stars, 
the hydrogen and neutral and ionized helium line profiles provided 
diagnostics for the atmospheric parameters, together with those 
of the wind. The different methodologies are discussed below:

\subsubsection{B2 -- B8 stars: H \& He lines}\label{sec:atm_hhe}

As the FLAMES dataset contains a relatively large number of mid 
and late B-type stars, we have utilized relatively simple methods 
to estimate their atmospheric parameters. The ratio of the strengths 
of the neutral helium and hydrogen lines is sensitive to effective 
temperature for spectral types later than B1, whilst the strength 
of the hydrogen lines can be used to estimate the surface gravity 
(see for example Dufton et al.\ \cite{Duf00}).  For hydrogen, 
the equivalent widths of the H\,$\gamma$ and H\,$\delta$ lines 
were measured with the continuum defined at $\pm$16\AA\/ from 
the line centre. For any given star these normally agreed 
to within 3\% (which is consistent with our theoretical calculations) 
and hence their average  was adopted.  For helium, while a 
number of different features were available, the best observed line 
which was seen in effectively all our spectra was that at 4026\AA. Moreover,
this line has been found to yield reliable effective temperature estimates 
in a study of hot stars towards the Galactic centre by Dufton et al.\ 
(\cite{Duf00}). The equivalent widths (EWs) of this line were estimated
with the continuum normally defined at $\pm$6\AA\/ from the line centre, 
although for some of the sharper and weaker \hei\/ lines a narrower range 
was adopted where this was believed to increase the reliability of 
the measurements. 

\begin{figure*}\caption{Diagnostic plots used for the initial estimation of
atmospheric parameters assuming a microturbulence of 5 \kms: ({\it left}) 
the ratio of the equivalent width of the He\,{\sc i} line at 4026\AA\ to 
the mean of the Hydrogen H\,$\gamma$ and H\,$\delta$ equivalent widths; the 
ratio is plotted as a function of effective temperature (\teff) for three 
logarithmic surface gravities:({\it right}) the mean equivalent widths 
of the H\,$\gamma$ and H$\delta$ lines on a logarithmic scale plotted as 
a function of effective temperature (\teff) for five logarithmic surface 
gravities.}
\label{fig:diagnostics}

\begin{center}
\begin{tabular}{ll}
      &     \\
  (a) & (b) \\
 \psfig{file=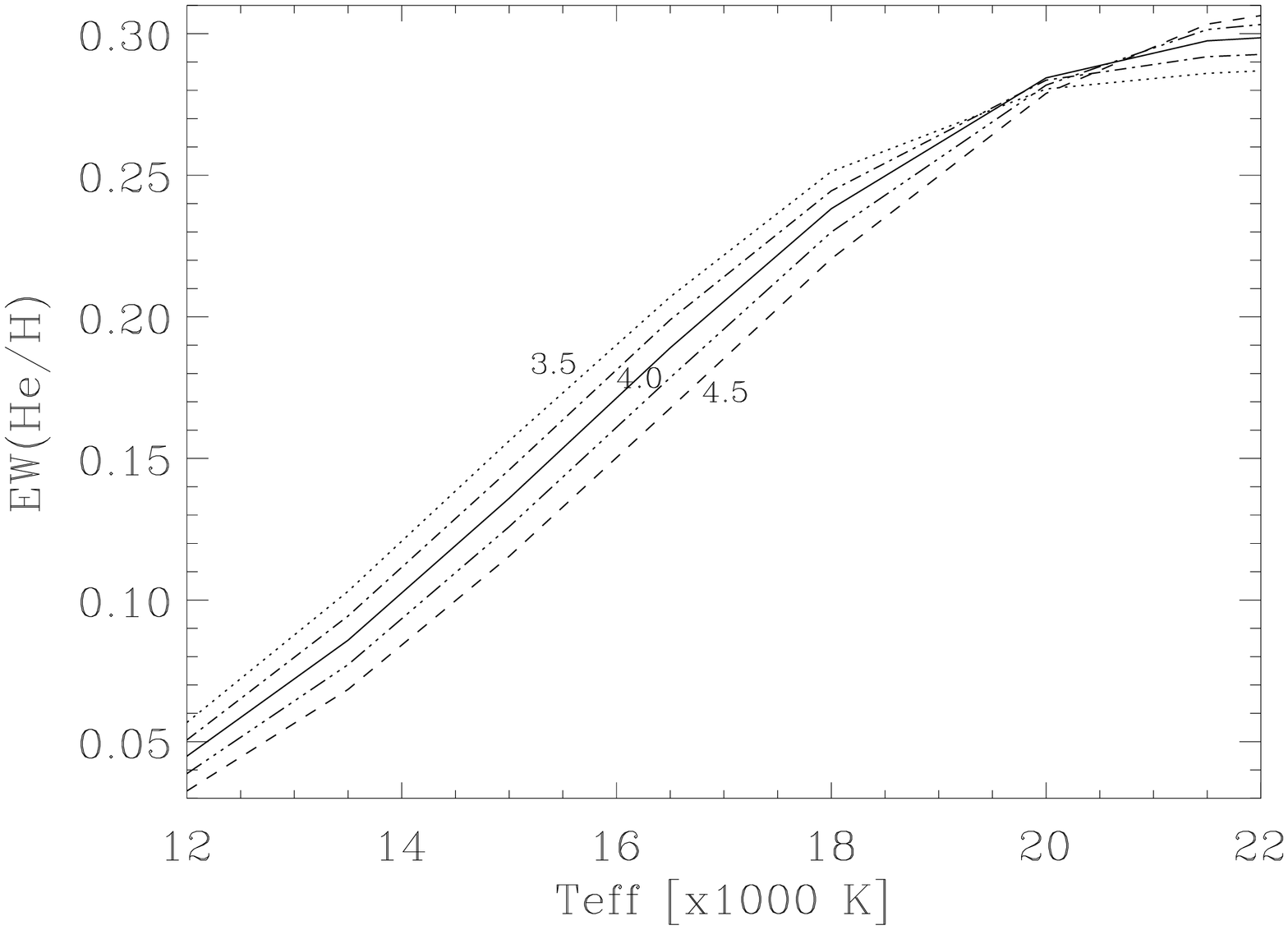,angle=90,width=8cm}&
 \psfig{file=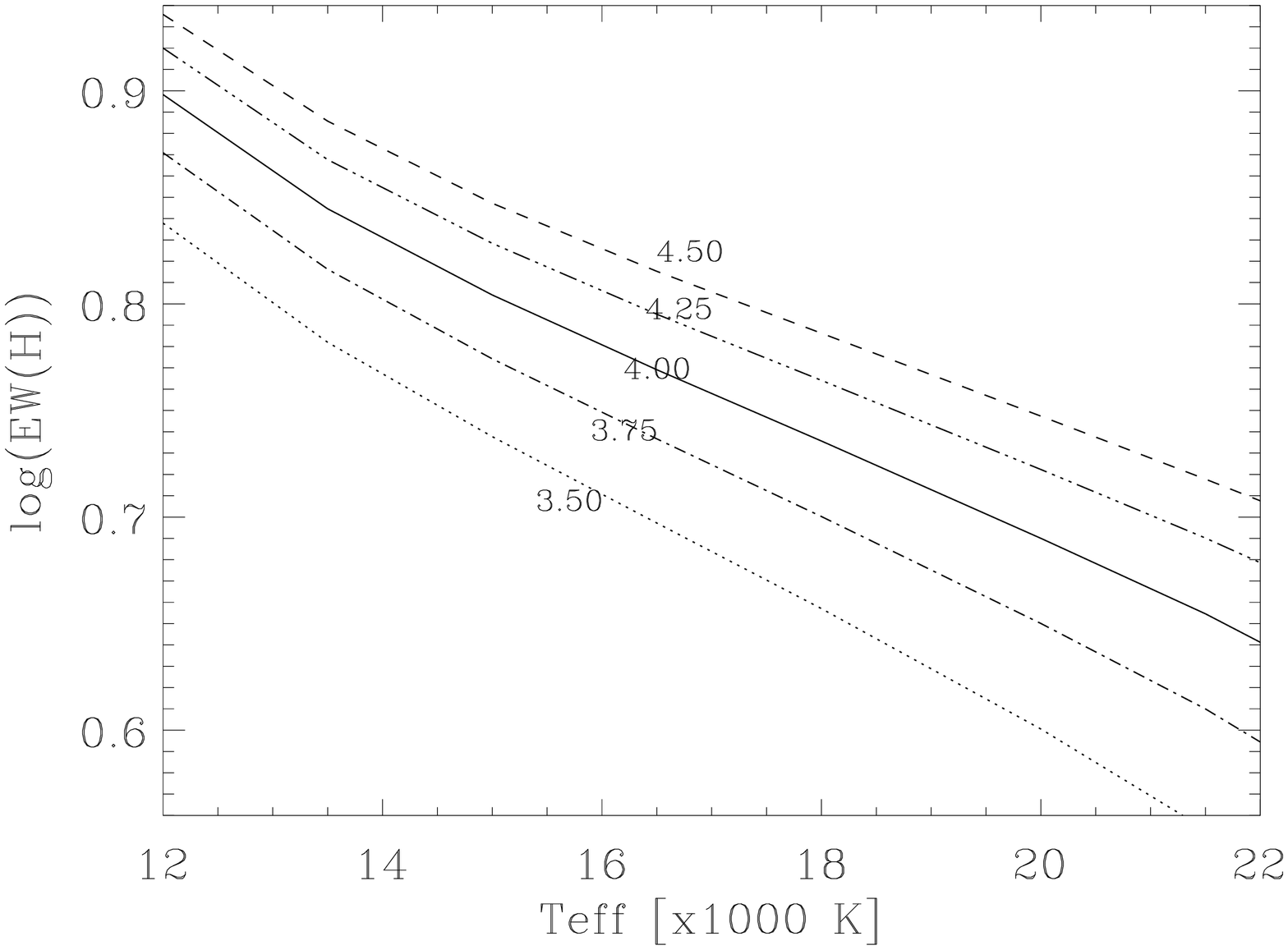,angle=90,width=8cm}\\
\end{tabular}
\end{center}
\end{figure*}

Diagnostic plots (see Fig.~\ref{fig:diagnostics}) were deduced from 
synthetic spectra that had been calculated using the non-LTE model 
atmosphere codes {\sc tlusty/synspec} (Hubeny \cite{Hub88}; Hubeny 
\& Lanz  \cite{Hub95}; Hubeny et al. \cite{Hub98}; see also 
Ryans et al.\ \cite{Rya03}). A microturbulent velocity of 
5~\kms\/ was adopted, consistent with most of our targets having 
spectral types that imply that they are on and/or near the
main sequence (see, for example, Gies \& Lambert \cite{Gie92}).  
However, as discussed by Dufton et al.\ (\cite{Duf00}), this choice 
is not crucial for these broad hydrogen and diffuse helium lines.  
Fig.~\ref{fig:diagnostics} shows the variation of the relative strength of 
the neutral helium and hydrogen lines (EW(He/H)), and of the mean of the 
equivalent width of the hydrogen lines (EW(H)), for the range of 
effective temperatures that were available from our 
grid of model atmospheres. Note that all measurements of the theoretical 
equivalent widths have used the same continuum definitions as for the 
observations. For estimating the gravity, the hydrogen lines 
provide reliable diagnostics across the entire temperature range.  
By contrast the helium line diagnostic flattens out at temperatures 
$>$20\,000~K, which sets the effective upper limit for which this 
method can be used. 

For each star, estimates of the atmospheric parameters were 
obtained by an iterative procedure, using the diagnostic plots illustrated 
in Fig.~\ref{fig:diagnostics}.  Firstly the observed value EW(He/H) was
compared to theoretical predictions for a surface gravity of \logg\/ = 
4.0 dex ({\it g} in cm\,s$^{-2}$) to give an initial estimate of 
the effective temperature (Fig.~\ref{fig:diagnostics}a).  The 
observed value of log(EW(H)) was then compared with theoretical 
predictions to give an estimate of surface gravity 
(Fig.~\ref{fig:diagnostics}b).  The above procedure  was then repeated
until convergence was found for both effective temperature and surface 
gravity.

\begin{table*}
\caption[]{Methods used to estimate the effective temperature scale for the
B0.5--1.5 spectral-types}
\label{tab:si4temp}
\begin{tabular}{lcl}
\hline\hline\noalign{\smallskip}
Spec.\ Type& \teff     & Method \\
\noalign{\smallskip}\hline\noalign{\smallskip}
B0.5 V     & 27\,000 K & Si {\sc iii/iv} ionization balance (6611-012)\\
B1.0 V     & 25\,000 K & Si {\sc iii/iv} ionization balance 
(3293-018; 4755-015; 6611-021 \& -033)\\
B1.5 V     & 22\,500 K & Interpolation between B1 V and B2 V \\
B2.0 V	   & 20\,000 K & Average of the effective temperature estimate for 
11 stars based on their helium spectra  \\
\noalign{\medskip}
B1.0 III   & 21\,500 K & Si {\sc iii/iv} ionization balance (3293-003, 
-007 \& -010) \\
B1.5 III   & 20\,500 K & Interpolation between B1 III and B2 III \\
B2.0 III   & 19\,500 K & Average of the effective temperature estimate for 
2 stars based on their helium spectra  \\
\noalign{\smallskip}\hline
\end{tabular}
\end{table*}

\subsubsection{B0.5 -- B1.5: Si ionization balance}

The neutral helium spectrum is not a useful effective temperature diagnosis 
for stars earlier than approximately B2 (\teff\/ $\ge$20\,000~K, 
see Fig.~\ref{fig:diagnostics}) and hence for these objects, 
the silicon ionization balance was utilized instead.  However, atmospheric 
parameters could not be estimated for stars with large rotational velocities 
where the silicon spectra was poorly observed. Therefore we
only considered narrow-lined stars and adopted their temperature estimates as 
representative for other stars with the same spectral type.

By adopting an appropriate surface gravity (4.0 dex for dwarfs and 3.0 dex  
for giants) and a microturbulence (5 \kms but see Sect. \ref{sec:stars}
for a discussion of the microturbulence of the supergiant targets) 
an initial effective temperature could be estimated using the
relative strengths of the Si {\sc iii/iv} lines and the {\sc tlusty}
(Hubeny \cite{Hub88}; Hubeny \& Lanz  \cite{Hub95}) non-LTE model 
atmosphere grids discussed in Sect. \ref{sec:atm_hhe} (see also
Dufton et al.\ \cite{Duf05} for details on the calculation of the
silicon spectrum). These temperature estimates 
were in turn used to estimate surface gravity by the comparison of 
observed and theoretical hydrogen Balmer line profiles.  
The new gravity estimate was then used as the starting point in the 
next iteration to estimate the effective temperature and the process 
was repeated until convergence was obtained. 

Narrow-lined stars in the clusters were selected with spectral types 
between B0.5 and B1.5 for both dwarfs and giants. For B0.5 V, the target
6611-012 was analyzed, whilst for B1 V, four narrow-lined stars were 
available (3293-018, 4755-015, 6611-021 and 6611-033).  For B1.5 V, the
spectra of all three targets (4755-011, 4755-017 and 6611-032) only show
Si {\sc iii} lines and the effective temperature for this spectral type 
was adopted as the average of those of B1 and B2 type stars.  
For B1 III, the effective temperature estimates of three stars 
(3293-003, 3293-007 and 3293-010) were averaged.  No narrow lined star 
was present in our sample with a spectral type B1.5 III.  As in 
the case of dwarfs, for this spectral type, the effective temperature 
was taken to be the average of those for our B1 III 
and B2 III stars. In the case of the latter spectral type, the targets 
3293-034 and 4755-003 were excluded from the estimation of mean effective
temperature. This was because the former has a peculiar spectrum 
(see paper I), whilst the estimate for the latter (\teff=24000~K)
was beyond the high temperature limit for using the helium spectra
as a reliable diagnostic.
 
These effective temperature estimates are summarized in 
Table~\ref{tab:si4temp} where they have been rounded to the 
nearest 500K. They were adopted for all other stars with the same 
spectral classification but with individual surface gravities 
being deduced from the hydrogen Balmer line profiles.

\subsubsection{O and B0} \label{sec:Ostars}

The B0 and O-type stars have temperatures at the edge of, or outside, the 
range of our {\sc tlusty} grid. In addition many of these objects 
show wind features in their optical spectra. For these reasons 
the hottest objects in our sample have been analyzed with the unified model 
atmosphere code {\sc fastwind} as described in Puls et al. (\cite{Pul05}). 
In order to reduce convergence times, the theoretical calculations included 
only the H and He or the  H, He and Si model atoms (depending on the 
strength of the Si lines in the spectra and their usefulness as a \teff\/ 
diagnostic). However in order to incorporate the line blanketing from 
important metals not included in the rate equations an approximate treatment 
was invoked (see Puls et al.\ for details and comparisons with a more 
detailed technique used by the model atmosphere code {\sc cmfgen}). 
The analysis method  described below follows closely the iterative 
processes outlined in Herrero, Puls \& Najarro (\cite{Her02}) and 
references therein.

For the majority of the O-type stars the effective temperature 
was determined by profile fitting of the neutral and singly ionized 
helium spectra, with more emphasis given to the latter. For the 
cooler objects in this subset (i.e.\ O9 - B0), estimates were also 
obtained using the silicon ionization equilibrium (Si {\sc iii}/{\sc iv}), 
with good agreement being obtained between the two approaches.
Profile fitting to the wings of the hydrogen lines was used to determine the
surface gravity. Greater weight was given to the higher order 
Balmer lines where the wind has less influence on the core of the lines. The 
wind properties were described by the mass-loss rate, the $\beta$-parameter 
and the terminal velocity, the latter was adopted from the spectral-type - 
terminal velocity scale introduced by Kudritzki \& Puls (\cite{Kud00}). The 
$\beta$-parameter was initially assumed to be 0.8 and iterated along with 
the mass-loss rate to obtain a good fit to the H$_\alpha$ profile. 

\begin{figure} \caption{An example of the chi-square minimization procedure
used to estimate the projected rotational velocity for the star 4755-030.}
\label{fig:chi}

\centering
\begin{tabular}{c}
 \\ \psfig{file=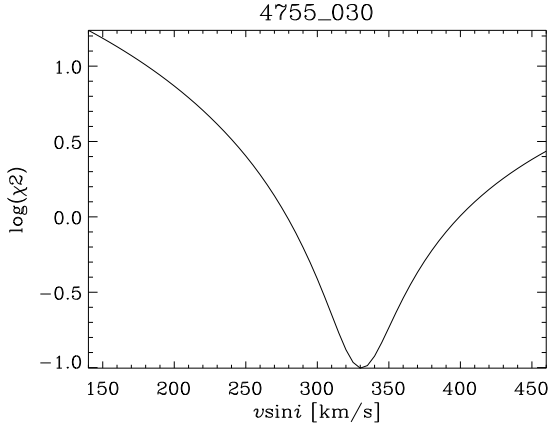,angle=90,width=8cm}\\
\end{tabular}
\end{figure}

\begin{figure}[b] \caption{Observed and rotationally broadened theoretical
helium line profiles. 4755-033 (upper panel) and 
4755-030 (lower panel) have estimated values of \vsini\/  of 75 and 330 \kms, 
respectively.}\label{fig:vrot_eg}

\begin{center}
\begin{tabular}{c}
  \psfig{file=5392_fig3a,angle=270,width=7cm}\\
  \psfig{file=5392_fig3b,angle=270,width=7cm}\\
\end{tabular}
\end{center}
\end{figure}

\subsection{Projected rotational velocities} \label{sec:vsini}

Stellar rotational velocities for the B-type stars were determined 
by comparing observed line profiles to theoretical counterparts that 
had been convolved with a rotational broadening function (see, 
for example, Gray \cite{Gra92}). Although the hydrogen lines are well 
observed throughout our sample, they are intrinsically broad, 
making it difficult to disentangle effects due to rotation from those due 
to the atmospheric parameters, whilst there are no metal lines that are 
usable across all the observed B spectral sub-types.
The helium spectrum was a suitable compromise between being well
observed  whilst not being too intrinsically broad.  In particular the 
\hei\/ 4026\AA\/ line was relatively strong and thus measurable in 
most of the program stars. Its theoretical line shape also 
appeared to be well calculated when its forbidden components 
were taken into account.  In addition, the line having been already 
used  for the temperature estimation provided an additional consistency 
in the analysis.

For each program star, a theoretical spectrum was selected at a 
matching, or the nearest, point in our grid of non-LTE model 
atmospheres to the estimated atmospheric parameters with a 
microturbulence of 5 \kms\/ again being adopted.  To allow for 
any difference in line strength between the model and the observed 
spectra, the ratio of their equivalent widths (which was generally  
near to unity) was used to scale the model spectrum.  We note 
that a broader continuum range of $\pm$10\AA\/ from the line 
centre was adopted than that used for the estimation of the
atmospheric parameters.  This was to ensure the inclusion of the wings 
of the profile which were important when estimating the rotational 
velocity.  The scaled model spectrum was then convolved with various 
projected rotational velocity values, and the best fit to
the observation, was obtained through a chi-square minimization test
(see Fig.~\ref{fig:chi} for an example).  
This procedure is a simplified version of that used by
Ryans et al.\ (\cite{Rya02}) to estimate the contribution of rotation and
macroturbulence in the broadening of the spectra of early-type supergiants. 
Tests reported by Ryans et al.\ indicate that the scaling of the equivalent
widths and the values adopted for the atmospheric parameters (including the
microturbulence) are not likely to lead to significant errors for stars with
significant rotational broadening. Fig.~\ref{fig:vrot_eg} shows 
examples of the optimal fits obtained when estimating the 
projected rotational velocity for two stars with relatively
low and high projected rotational velocities, with all the estimates
listed in Tables \ref{tab:3293vrot} to \ref{tab:6611vrot} having been 
rounded to the nearest 5 \kms.

We note that the estimates for the very narrow-lined objects may be 
unreliable due to the intrinsic width of the \ion{He}{i} diffuse lines. 
Hence for all stars where
we deduced a \vsini\, of less than 50 \kms, we have undertaken a re-analysis
of the stellar spectra. In these cases the \ion{Si}{iii} line at
4552\AA\,in the hotter stars and the \ion{Mg}{ii} line at 4481\AA\, in 
the cooler objects were considered. A similar methodology as for the
helium lines was adopted with the theoretical profiles again being taken 
from our non-LTE grid with a microturbulent velocity of 5 \kms\ 
being adopted and the doublet splitting for the \ion{Mg}{ii} line being
explicitly included. In most cases the values obtained from the
helium and metal lines were similar but in some cases they differed by up
to 20 \kms. For these narrow lined stars the values listed in Tables
\ref{tab:3293vrot} to \ref{tab:6611vrot} are those estimated from the
metal lines.

Instrumental broadening has not been included in our analysis and
we have investigated how this might affect our estimates as follows. 
Representative theoretical profiles have been convoluted with either just a
rotational broadening function or with a rotational broadening function
and a Gaussian profiles to represent instrumental broadening (with a FWHM
equivalent to the inverse of the spectral resolution). For the FLAMES data, 
the profiles only become significantly different for projected rotation 
velocities less than 20 \kms, with a correspondingly smaller value for the
higher resolution FEROS data. Such small estimates of the projected
rotational velocity have only been found 
for two targets in our sample and we have not attempted  
to correct these two estimates as they will be subject to additional
uncertainties due to the choice of the intrinsic profiles and in 
particular the adopted microturbulence.

For the O-type stars in NGC\,6611, the projected rotational velocities were
determined simultaneously with the atmospheric parameters (see Sect.
\ref{sec:Ostars}) using a method similar to the B-type stars. For these
stars the \ion{N}{iii} lines at 4510-4518\AA\, and the \ion{Mg}{ii} 
at 4481\AA\, line were considered. 

\begin{figure}\caption{Initial estimates of atmospheric parameters for targets
in NGC\,3293, deduced from the hydrogen and helium 
equivalent widths.}\label{fig:hr}

\centering \psfig{file=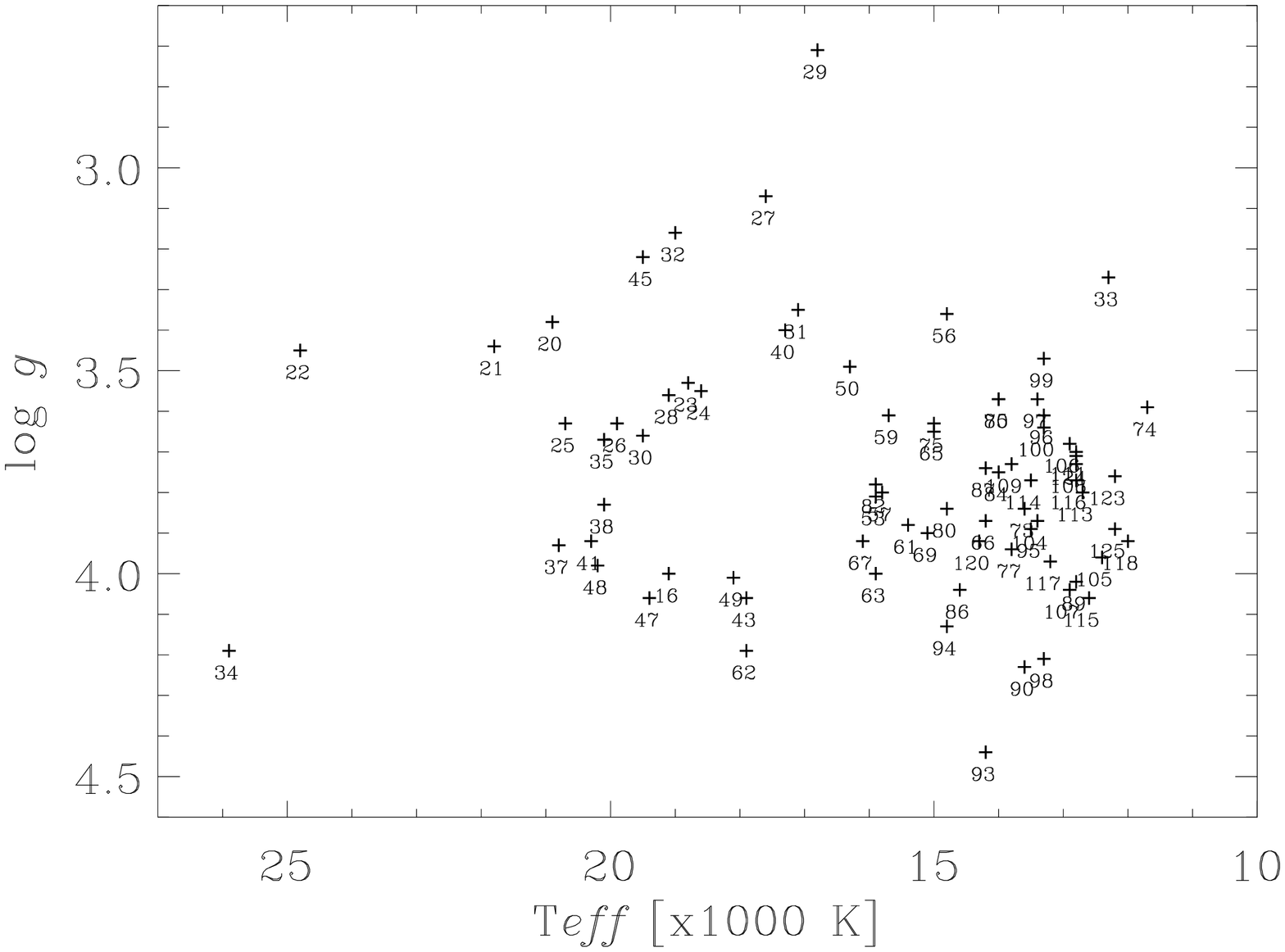,angle=90,width=9cm}
\end{figure}

\begin{figure}\caption{The procedure to correct the atmospheric parameter
estimates for stellar rotation is illustrated for a value of  
\vsini\/ of 300 km/s. Plus sign
(+) mark the atmospheric parameters deduced when the effects of rotation
are included.  Diamonds symbols
 ($\diamond$) connected by a line, represent the original 
atmospheric parameters, which are designated by the adjacent pairs of numbers 
(note \teff\/ is in $\times$1000~K). The correction factors are given 
in parenthesis.}\label{fig:vscor}

\vspace{5mm} \centering \psfig{file=5392_fig5.ps,angle=90,width=9cm}
\end{figure}

\subsection{Correction of Atmospheric Parameters}
\label{sec:vscor}

Fig.~\ref{fig:hr} shows the distribution of our initial estimates of the
atmospheric parameters of stars observed in NGC\,3293, deduced from the 
hydrogen and helium line strengths as discussed in Sec.~\ref{sec:atm_hhe}.  
The distribution of surface gravities is surprising in that there appears
to be a significant number of low gravity objects. Given the age of this
cluster, it would be expected that main sequence stars would make up the most 
of the sample for this range of spectral types. In particular forty percent 
of the stars appear to have surface gravities less than 3.70 dex, which is
closer to that expected for giants or supergiants than to main 
sequence stars.   We suspect this may have arisen from the use of 
a restricted wavelength region when measuring the equivalent widths of 
the hydrogen and helium lines. Stellar rotation broadens line profiles,
and when a line is normalized at  a fixed continuum position, 
the measured EW could be  different from its intrinsic value, with the
discrepancy increasing with the projected rotational velocity.  As we
have compared such  observational equivalent widths with values measured 
from theoretical spectra with no rotational broadening, the estimated 
atmospheric parameters  might be incorrect.  

To investigate this effect, we have constructed a grid of correction 
factors for different projected rotational velocities. Theoretical 
spectra have been convolved with rotational broadening functions  
and then the equivalent widths of hydrogen and helium lines were 
measured. These have been used to deduce atmospheric parameters from 
our calibration to estimate how much they deviate from the actual model
parameters. This deviation is indicative of the effects of rotation
on our adopted methodology, and has been used to correct our initial 
estimates of the atmospheric parameters. For example, 
Fig.~\ref{fig:vscor} illustrates the corrections for a projected 
rotational velocity of \vsini\/ = 300 \kms\/ -- theoretical spectra 
taken from grid points are marked with a diamond ($\diamond$) and after 
convolution, they imply the set of atmospheric parameters marked with a 
plus sign (+).  The pairs of numbers next to the latter are the derived 
atmospheric parameters, with the amount of deviation presented in 
parenthesis. Correction factors were also calculated for projected 
rotational velocities from 0 to 360 \kms, covering all the \vsini\/ 
values of our program stars. The atmospheric parameters of the 
program stars were then corrected through a series of linear 
interpolations between the nearest grid points in \teff, \logg\/ 
and \vsini\ and these corrected values are summarized in 
Tables~\ref{tab:3293vrot} -- \ref{tab:6611vrot}. Tests show that
use of these corrected atmospheric parameters did not lead to any 
significant changes in the values previously deduced for the 
projected rotational velocities.

\begin{table}\caption[]{Stars that have been excluded from the analysis. 
Objects later than B8 are not listed and can be identified from Paper I.}
\label{tab:excl}

\begin{tabular}{lp{1.5cm}p{1.5cm}p{2.5cm}}
\hline\hline\noalign{\smallskip}
             & {\bf 3293} & {\bf 4755} & {\bf 6611} \\
\noalign{\smallskip}\hline\noalign{\smallskip}
Binaries     &  --    & 24   & 7, 13, 14, 30, 68\\
\noalign{\smallskip}
Emission     & 11   & 14, 18& 10, 22, 28\\
\noalign{\smallskip}
Others       &  72, 121 &&\\
\noalign{\smallskip}\hline\noalign{\smallskip}
\end{tabular}
\end{table}

\subsection{Comments on individual stars}\label{sec:stars}

Some stars have been excluded from the analysis as the measurements of their
hydrogen and helium equivalent widths were either not possible or unreliable. 
They largely fall into three categories, viz. targets later than B8, binaries 
or stars with very strong emission feature in their Balmer line profiles 
and are summarized in Table~\ref{tab:excl}. This leaves 92 out of 
126 stars observed in  NGC\,3293, with the numbers for the other clusters 
being  87/108 (NGC\,4755) and 44/85 (NGC\,6611).
Some comments on individual stars are given below:

{\bf 3293-001, -002 and 4755-004:} the effective temperature of these
supergiants were estimated utilizing the silicon ionization balance.  
Their microturbulence has been determined to be 20, 14 and 15 km$^{-1}$, 
respectively, by minimizing the slope in the abundance estimates from 
the Si {\sc iii} lines versus their equivalent widths. 

{\bf 3293-027, -029 and -032:} the spectral types of these stars remain
uncertain due to their very broad line profiles, and they were classified as
B0.5-1.5 in Paper I.  We assume that they are B1.0 and adopt a `calibrated'
effective temperature of 23\,000~K (Table~\ref{tab:si4temp}). 

{\bf 3293-034:}  The helium spectrum implies that this B2 IIIh star 
has an effective temperature of 26\,000~K, much higher than the limit 
of validity for this method and incompatible with its spectral type.  
We suspect that these inconsistencies arise from its spectral peculiarities. 

{\bf 6611-010 \& -022}  These Be stars have no useful effective temperature
indicators, making it 
difficult to estimate their atmospheric parameters.  Theoretical profiles 
were fitted to H$\gamma$ line to produces loci of possible atmospheric 
parameters.   For representative parameter pairs, \vsini\/ values
were estimated and their average is presented in Table~\ref{tab:6611vrot}


\subsection{Stellar masses and luminosities}
\label{sec:distance}
In order to determine stellar luminosities, it is important that we
have reliable cluster distances. Several studies 
have estimated these distances and some of the more recent
are summarized in Table\,\ref{distances}. We have chosen to list 
results from wide-field CCD and spectroscopic surveys, which supersede 
the older photoelectric methods based on less reliable samples. 

We have also independently determined a distance to each cluster 
from a spectroscopic parallax method, applied to the main-sequence 
targets in our clusters. Firstly intrinsic colours ($B-V$)$_o$ 
were calculated using standard Johnson $BV$ filter curves and 
flux calibrated {\sc tlusty} spectra for a range of effective 
temperatures and two gravities. As can be seen from 
Table~\ref{tab:b-vo} there are relatively small differences 
between the values for $\log g=3.0$ and 4.0 dex . Hence we
adopted the former for all stars with $\log g < 3.5 $dex, and the latter
for the higher gravity objects. A standard Galactic reddening law
of A$_v$ = 3.1\,$E$($B-V$) was adopted for NGC\,3293 and NGC\,4755 
as this has been found previously to be a suitable value 
(Baume et al \cite{Bau03}). However there is evidence that the 
reddening towards NGC\,6611 is anomalous and a value of 
$A_V$ = 3.75\,$E$($B-V$) has been found to best match the 
optical and infra-red magnitudes of the cluster members when combined 
with the colour excesses from stars with known spectral types 
(Hillenbrand \cite{Hil93}). 

\begin{table*}\caption{Our distance and age estimates together with 
selected values from the literature.} 
\label{distances}
\begin{center}
\begin{tabular}{llll}
\hline\hline\noalign{\smallskip}
Cluster & Method & Distance & Age \\
        &        & kpc      & Myr \\
NGC\,3293  && \\
Freyhammer et al. (\cite{Fre05}) & Eclipsing Binary     
&  2.8$\pm$0.3  & 10-13  \\
Baume et al. (\cite{Bau03})      & $UBVRI$ MS fitting           
&  2.8$\pm$0.3  & 6.5-10 \\
Shobbrook (\cite{Sho83})         & $uvby\beta$ sequence fitting 
&  2.5$\pm$0.1  & -      \\
This paper (33 stars)    & Spectroscopic parallax      
&  2.9$\pm$0.1  & 10-20  \\
Mean of all results      &                              
&  2.8$\pm0.2$  &   10     \\ 
Mean and SD of $A_{V}$   &  &  0.80$\pm$0.29\\
\\
\noalign{\smallskip}\hline
\\
NGC\,4755  && \\
Sanner et al. (\cite{San01})     & $UBVRI$ MS fitting           
&  2.1$\pm$0.2 & 8-12 \\
Balona \& Koen (\cite{Bal94b})    & $uvby\beta$ MS fitting       
&  2.0$\pm$0.2 & - \\
Sagar \& Cannon (\cite{Sag95})   & $UBVRI$ MS fitting           
&  2.1$\pm$0.2 & 10\\
This paper (39 stars)    & Spectroscopic parallax       
&  2.3$\pm$0.1 & 10-15 \\
Mean of all results      &                              
&  2.1$\pm$0.2 & 10 \\ 
Mean and SD of $A_{V}$   &  &  1.15$\pm$0.15\\
\\
\noalign{\smallskip}\hline
\\
NGC\,6611  && \\
Hillenbrand et al. (\cite{Hil93}) & Spectroscopic parallax      
& 2.0$\pm$0.2 & 1-3 \\
Belikov et al. (\cite{Bel99})     & $UBV$ MS fitting            
& 2.1$\pm$0.2 & - \\ 
This paper (24 stars)     & Spectroscopic parallax      
& 1.8$\pm$0.1 & 2-4 \\
Mean of all results       &                             
& 2.0$\pm$0.2 & 2 \\
Mean and SD of $A_{V}$   &  & 2.81$\pm$0.88\\
\\
\hline\hline
\end{tabular}
\end{center}
\end{table*}

From the non-rotating evolutionary tracks of 
Schaller et al. (\cite{Sch92}) and Meynet et al. (\cite{Mey94}) 
we have inferred surface gravities and then by interpolating 
in the \logg-\teff\ plane, estimated stellar luminosities. 
These have been combined with bolometric corrections and the 
$E(B-V$) values to determine spectroscopic distances for each target. 
For $\log$~\teff\/ $>$4.45, the bolometric corrections  
of Vacca et al.\ (\cite{Vac96}) were used, while for lower effective 
temperatures, those of Balona (\cite{Bal94a}) were adopted.
As discussed above, the determination of the surface gravity 
in our fastest rotating targets is subject to some uncertainty. 
Hence we restricted our sample of stars to those, having  
$4.5 \leq {\rm log} g \leq 3.95$. Effectively we are estimating 
cluster distances by applying a hybrid spectroscopic and evolutionary 
parallax method for stars which we are confident are lying close 
to the main-sequence and have well determined surface gravities. 
The distance estimates are listed in Table\,\ref{distances}, 
and are in good agreement with those found previously from 
other methods. We have adopted a mean value  of these estimates 
for use in this paper. 

Using these distances, we have re-determined the luminosity of each 
star from their $V$-band magnitude, a bolometric correction (as 
discussed above) and their $E(B-V)$ value. This allowed stars 
to be placed on a HR-diagram, and by interpolation between the 
Geneva evolutionary tracks led to estimates of
their evolutionary masses, which are summarized in 
Tables~\ref{tab:3293vrot}--\ref{tab:6611vrot}. 

\subsection{Cluster membership}\label{sec:member}

In order to investigate whether an individual star is a cluster member we 
have used two criteria. One is radial velocity (listed in 
Tables\,\ref{tab:3293vrot}--\ref{tab:6611vrot} and taken directly from 
Paper I) and the other is distance from the cluster centre. 
In Table \ref{tab:member}, we summarize the mean radial
velocity of the clusters (again as reported in Paper\,I), together with 
the adopted cluster centres and radii (NGC\,3293: Baume et al. 
\cite{Bau03}; NGC\,4755: Lynga \cite{Lyn87}; NGC\,6611: Belikov et al. 
\cite{Bel99}). The use of radial velocities alone is not a reliable 
way to determine cluster membership as stars with discrepant 
(but apparently non-variable) radial velocities could be
binaries with periods significantly longer than our sampling 
frequency (which for NGC\,3293 and NGC\,4755 is less than one night).
Hence we have only identified a star as a non-member if its
radial velocity is more than 2$\sigma$ away from that of the
mean value {\em and}
its distance from the adopted cluster centre (designated $r\arcmin$\,
in Tables \ref{tab:3293vrot}--\ref{tab:6611vrot}) is more than twice 
the cluster radius. Such objects have been identified by a $\dagger$\,
and we note that they are only a small fraction of our total sample.
In turn this may imply that our rejection criteria may not be sufficiently
stringent but we have found that changing our criteria (e.g. to
radial distances of less than one cluster radius) had little
effect on our sample.

\begin{table}\caption{The mean cluster radial velocities (v$_r$), cluster
centres and radii (see text for details)}
\label{tab:member}
\begin{tabular}{lrccl}
\hline\hline\noalign{\smallskip}
Cluster & v$_r$ & \multicolumn{2}{c}{Centre} & Radius \\
        & \kms  & RA & Dec                   &\\
\hline\noalign{\smallskip}
NGC\,3293  & -12$\pm$5  & $10^{h}\,35^{m}\,49.3^{s}$  
                    & $-58\degr\,13\arcmin\,28\arcsec$ & 4.1\arcmin\\
NGC\,4755  & -20$\pm$5 & $10^{h}\,53^{m}\,39.0^{s}$ 
                    & $-60\degr\,21\arcmin\,39\arcsec$ & 5.0\arcmin\\
NGC\,6611  & +10$\pm$8 & $18^{h}\,18^{m}\,40.0^{s}$ 
                    & $-13\degr\,47\arcmin\,06\arcsec$ & 4.8\arcmin\\
\hline\hline
\end{tabular}
\end{table}

\section{Discussion}
\subsection{Atmospheric parameters} \label{disc:atm_par}
We have used a variety of methods based on non-LTE
model atmosphere techniques to constrain the
stellar atmospheric parameters. As discussed in Sect.\ 
\ref {sec:Ostars}, the use of different effective temperature
criteria for the O and B0-type stars led to excellent agreement.
Additionally previous analyzes of the spectra of B-type supergiants
using the {\sc tlusty} grid adopted here (Dufton et al. \cite{Duf05})
and the code {\sc fastwind} (Trundle et al. \cite{Tru05}) have also 
yielded encouraging agreement. Hence we believe that in general 
our estimates of effective temperature should have an accuracy of 
typically 5\% and should be generally more reliable than those 
based on the calibrations of broad band colours. Exceptions may
occur for the small number of stars that are peculiar (see Sect.
\ref{sec:stars}) or where we have adopted effective temperatures 
based on spectral types. 

\begin{table}[b]\caption{Johnson UBV colours deduced from the {\sc tlusty} 
spectra}\label{tab:b-vo}
\begin{tabular}{lllllll}
\noalign{\smallskip}\hline\hline
\label{tlusty_ubv}
    && \multicolumn{2}{c}{log~$g=4.0$}  && \multicolumn{2}{c}{log~$g=3.0$}\\
\teff &&   $B-V$  &   $U-V$  &&   $B-V$ &  $U-V$ \\
\noalign{\smallskip}\hline\noalign{\smallskip}
13500 &&  -0.102  &  -0.402  &&  -0.119 & -0.478 \\
15000 &&  -0.125  &  -0.500  &&  -0.140 & -0.571 \\
18000 &&  -0.161  &  -0.649  &&  -0.174 & -0.724 \\
20000 &&  -0.181  &  -0.732  &&  -0.188 & -0.813 \\
23000 &&  -0.203  &  -0.840  &&  -0.198 & -0.917 \\
25000 &&  -0.212  &  -0.899  &&  -0.209 & -0.958 \\
27500 &&  -0.228  &  -0.947  &&  -0.209 & -0.989 \\
30000 &&  -0.244  &  -0.984  && 	&        \\ 
35000 &&  -0.258  &  -1.035  && 	&        \\
\noalign{\smallskip}\hline\hline
\end{tabular}
\end{table}

For our gravity estimates, the situation was complicated by the 
corrections required for targets with large projected rotational 
velocities (see Sect. \ref{sec:vscor}). Whilst the corrections 
for the effective temperature estimate are relative small 
(typically less than 1\,000 K), those for the surface 
gravity can be significant (see Fig. \ref{fig:vscor}). To 
investigate the reliability of these corrections we have considered 
how the gravity estimates vary with rotational broadening. Our
procedure has been to divide our targets into two bins --- those that 
appear to be near to the main sequence (log g $\geq$ 3.9 dex) and 
those that may have evolved. For the latter, we have excluded 
five supergiants (classified as Ia or Ib in Paper I), as these 
targets are clearly evolved. Additionally we have excluded the O-type 
stars for NGC\,6611 as these were analyzed separately from the B-type 
stellar sample and employed detailed fitting of the hydrogen line profiles.
In Table \ref{tab:vsini_g}, we summarize the mean and standard deviations for
the projected rotational velocities of our two samples for each individual
cluster and for all our targets.

\begin{table}\caption{The mean projected rotational velocities for our higher
gravity targets (log g $\geq$ 3.9; designated `High gravity') and our apparently
evolved targets excluding known supergiants (log g $<$ 3.9; designated 
`Low gravity'). The number of stars in each sample is given by n.}
\label{tab:vsini_g}
\begin{tabular}{lrcrc}
\hline\hline\noalign{\smallskip}
Cluster & High gravity & n & Low gravity & n \\
        & \kms  & \kms                  &    \\
\hline\noalign{\smallskip}
NGC\,3293     & $184\pm99$ & 47  & $196\pm95$  & 41 \\
NGC\,4755     & $173\pm71$ & 47  & $211\pm96$  & 38 \\
NGC\,6611     & $132\pm75$ & 22  & $187\pm73$  &  6 \\
\\
All targets   & $170\pm86$ & 116 & $202\pm94$  & 85 \\
\hline\hline
\end{tabular}
\end{table}

For all three clusters (although the sample sizes for NGC\,6611 are 
small), there is a systematic trend in the sense that the low gravity objects 
have the higher mean projected rotational velocity and this is confirmed
when all the targets are considered. This result appears anomalous
as if the lower gravity targets had evolved from the main sequence, their
rotational velocities might have been expected to decrease due both 
to the increase in their moment of inertia and to the loss of 
angular momentum from mass loss. Rather this trend may arise
from the estimation of surface gravities for rapidly rotating stars
and we have identified two possible causes. Firstly, the correction 
procedure for the atmospheric parameters discussed in Sect. 
\ref{sec:vscor} may not have been successful. To test this hypothesis, 
we have used the atmospheric parameters and projected rotational 
velocities given in Tables \ref{tab:3293vrot}--\ref{tab:6611vrot} 
to generate theoretical hydrogen line profiles. In effectively 
all cases these were in good agreement with our observed line profiles 
and hence we believe that our correction procedure is not the
main cause of this discrepancy.

A second explanation is that for stars with large rotational velocities, 
the effective surface gravity will vary over the stellar surface due 
to the centrifugal force increasing as one moves from the pole to the 
equator. To a first order of approximation, our methodology will 
measure the mean effective surface gravity, which will be smaller than 
that of an analagous slowly rotating object. Hence the anti-correlation 
between projected rotational velocity and estimated surface gravity
could be a manifestation of such an effect. For example, Howarth \& Smith 
(\cite{How02}) found a variation of 0.4 dex between the effective surface
gravity at the pole and the equator for three rapidly rotating O9-O9.5 type
stars. 

To test this hypothesis we have computed models and synthetic spectra 
for a star with an B spectral type (corresponding to \teff=20000; 
\logg=4.0 at the stellar poles) using the methodology discussed by 
Townsend (\cite{Tow97}) which  
includes the effects of rotation on the pole to equator temperature 
structure and on the stellar radius. The synthetic input spectra were 
based on LTE atmospheres, calculated with ATLAS9 (Kurucz \cite{Kur05}) 
and LTE line formation, which should be adequate for assessing
the effects of stellar rotation. Further details can be found in 
Kaufer et al. (\cite{Kau06}).

For models with different rotational velocities, we have measured 
the equivalent widths of the features used to estimate atmospheric
parameters such as the \ion{He}{i} line at 4026\AA\ and the hydrogen
Balmer lines. The results are difficult to interpret but
the effects of rotation on these spectral features appear negligible 
for equatorial rotational velocities less than 250 \kms. For faster 
rotators, the equivalent width of the \ion{He}{i} 
remains effectively unchanged but the strength of the hydrogen
lines decreases. However the changes are relatively small with
those for the H$\delta$\ equivalent width being approximately 10-15\% 
for equatorial rotational velocities upto 400 \kms, which corresponds
to 80\% of the critical equatorial velocity. In turn this would 
lead to a change in the estimated surface gravity of 0.2 to 0.3 dex. 

In Sect.\ \ref{disc:vrot}, we estimate the mean equatorial rotational
velocity for the cluster NGC\,3293 and NGC\,4755 as approximately 
225-250 \kms. Assuming a Gaussian distribution of velocities, the 
$\frac{1}{e}$\ width would be approximately 110 \kms. Hence the number
of targets with equatorial velocities greater than 250 \kms would be
40-50\%, whilst those with a rotational velocity of 400 \kms 
or more would be of the order of 5-10\%. From the above, although
high rotational velocities may contribute to the spread of gravity 
estimates found in our two older clusters, it is unclear whether
it can fully explain these variations.

In summary, we believe that our effective temperature estimates are 
secure and find no evidence that our methodology has introduced 
systematic errors in our gravity determinations. Rather the significant 
number of relatively low gravity B-type objects may reflect in part 
a decrease in the effective surface gravity due to centrifugal 
acceleration.

\subsection{Hertzsprung-Russell diagrams}\label{disc:hr_diag}

\begin{figure*}
\begin{center}
\psfig{file=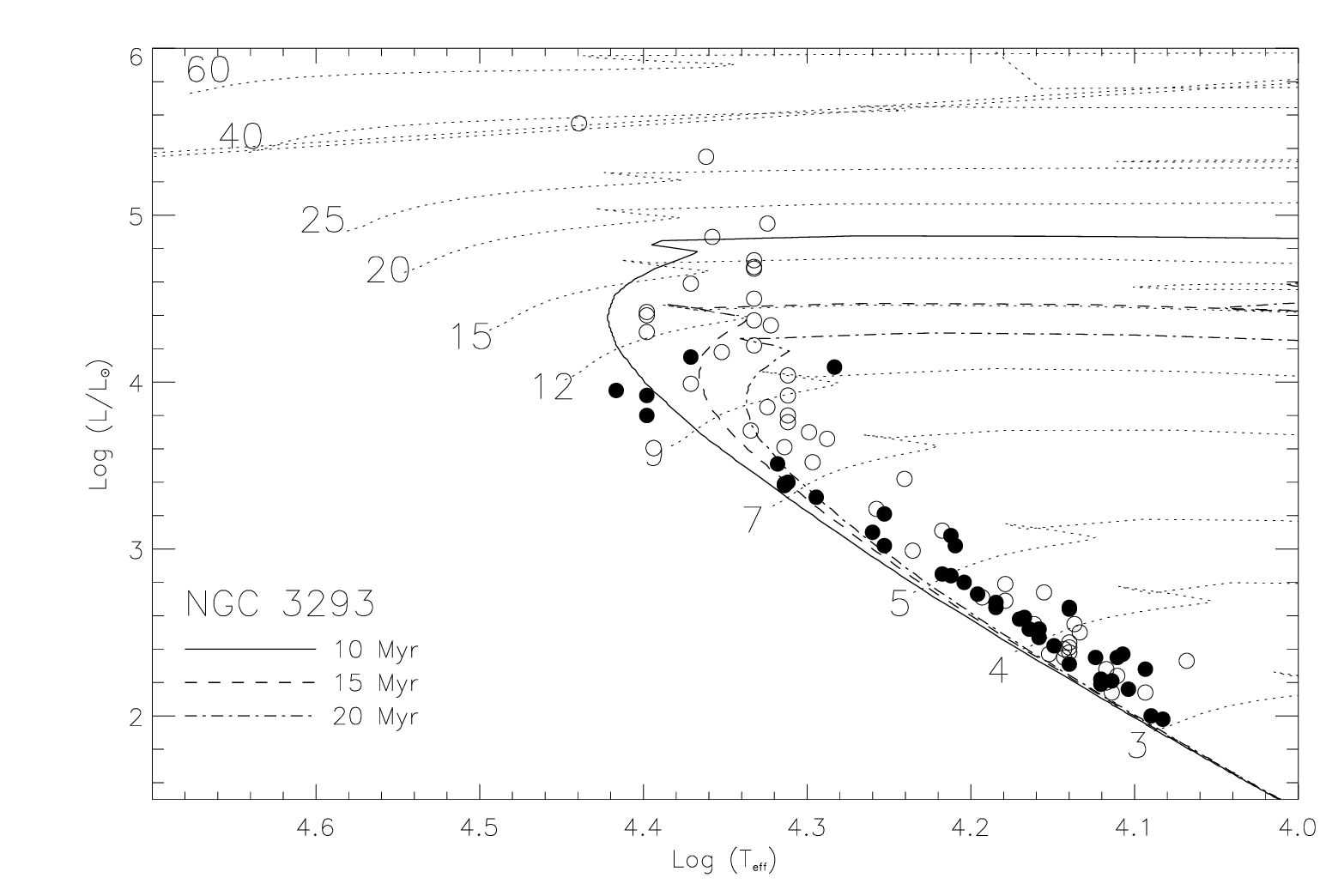,,angle=0,width=12cm}
\psfig{file=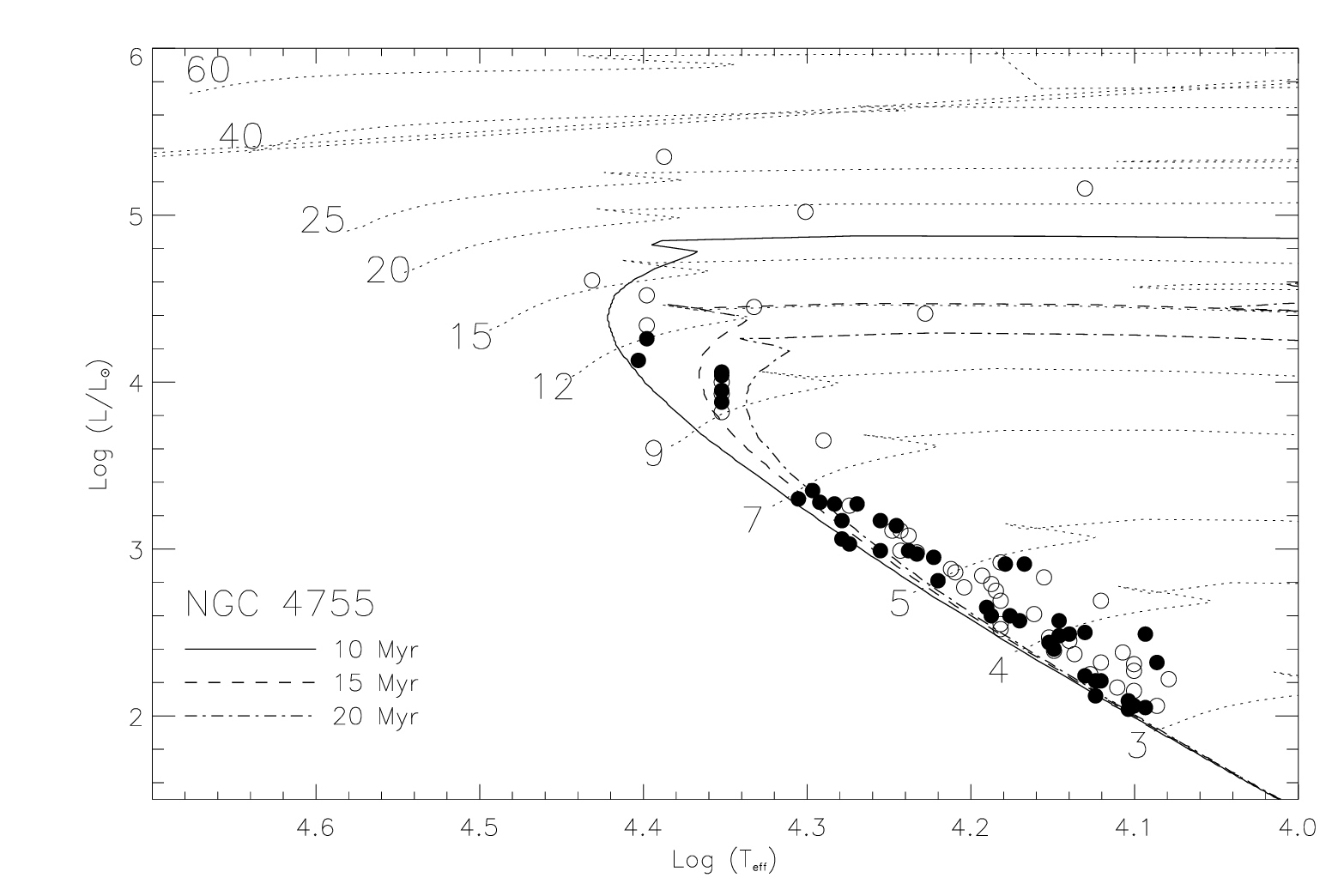,,angle=0,width=12cm}
\psfig{file=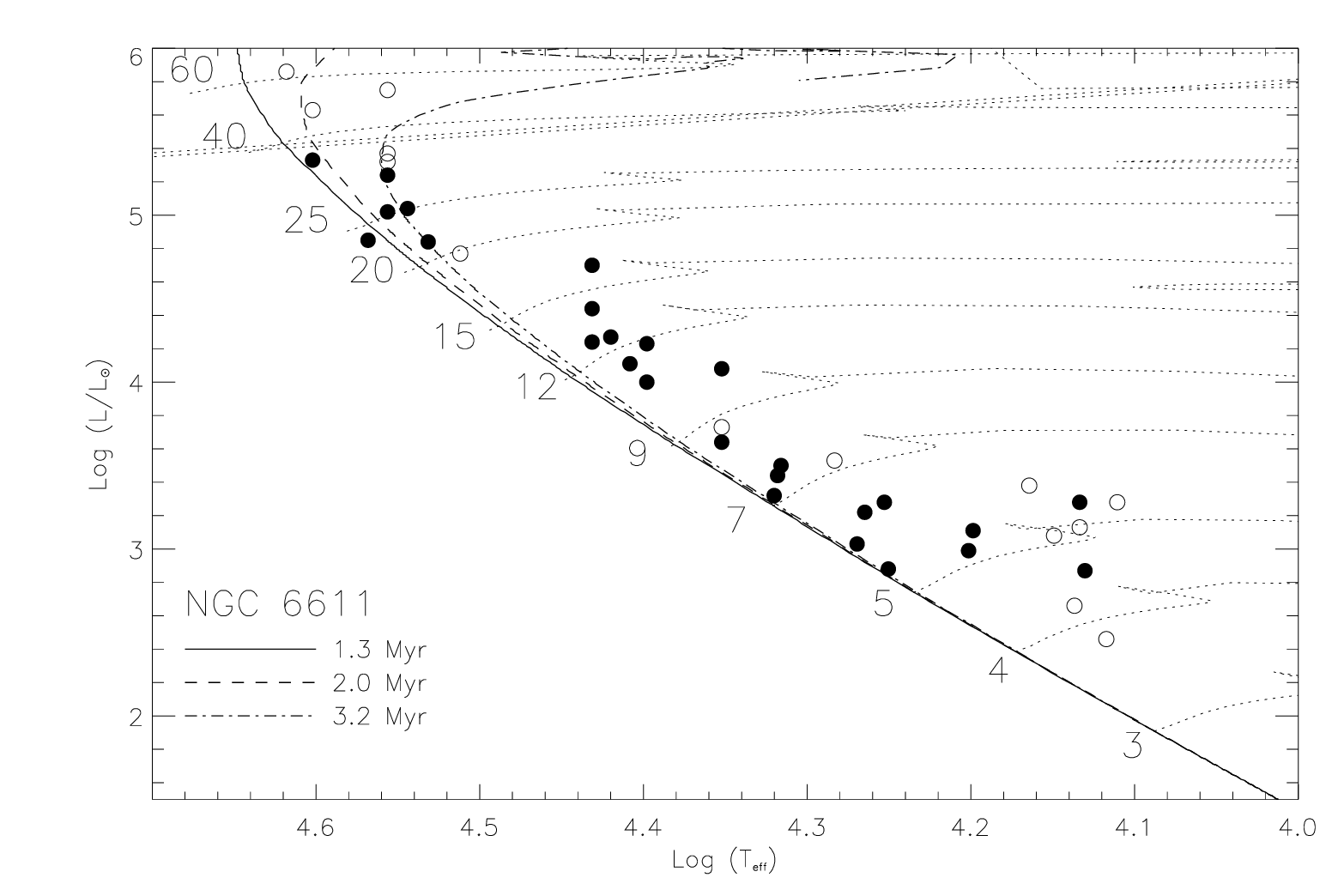,,angle=0,width=12cm}
\caption{Hertzsprung-Russell diagrams for the three clusters, together with
evolutionary tracks and isochrones taken from the Geneva database. Stars with
estimated gravities of less than 3.9 dex are shown as open circles with higher
gravity stars represented by closed circles.}
\label{hr_diag}
\end{center}
\end{figure*}

For all three cluster, we have used the luminosities and effective 
temperatures listed in Tables \ref{tab:3293vrot}--\ref{tab:6611vrot} to 
construct Hertzsprung-Russell (HR) diagrams and these are presented 
in Fig. \ref{hr_diag}. We have used different symbols to distinguish 
between those targets with estimated logarithmic surface gravities 
less than and greater than 3.9 dex. These are effectively the same 
as the samples discussed in Sect. \ref{disc:atm_par} but with the 
supergiants now being included in the former. Also shown are the 
evolutionary tracks and isochrones of Schaller et al. (\cite{Sch92}) 
and Meynet et al. (\cite{Mey94}), taken from the database 
of the Geneve group 
(see  http://obswww.unige.ch/$\sim$mowlavi/evol/stev\_database.html;
also available from the Centre de Données Astronomiques de Strasbourg,
http://cdsweb.u-strasbg.fr/CDS.html) for a metallicity, Z = 0.02.

In general there is a good qualitative agreement between the observed and 
theoretical HR diagrams although the observed main sequences appear to be 
systematically more luminous than predicted. For the late-B type stars, 
where the choice of isochrone is not important, this discrepancy would 
appear to be approximately, \deltaL $\simeq 0.4$, for the older 
clusters. The discrepancy may be larger for NGC\,6611 but in this case 
the sample size is relatively small. This systematic difference 
in luminosity would correspond to an error in our adopted distances 
of approximately 60\%, which although possible is significantly larger 
than the range of values summarized in Table \ref{distances}. 
Slesnick et al. (\cite{Sle02}) have undertaken an extensive photometric 
and spectroscopic study of the double cluster h and $\chi$\,Persei, 
which is similar in age to those of NGC\,3293 and NGC\,4755.
Inspection of the HR diagrams (see their Fig. 6 and 7) imply good 
agreement between the observed late-B type stars and the theoretical 
predictions of Schaller et al. (\cite{Sch92}).  However, recently, 
Strom et al (\cite{Str05}) have undertaken a study of projected 
stellar rotational velocities in h and $\chi$\,Persei and used the
results of Slesnick et al. For this smaller sample, there is again
evidence of a discrepancy (see Fig. 1 of Strom et al.) similar to 
that found here for the later B-type spectral types.

\begin{figure}\caption{Present day mass functions for the three clusters.}
\label{imfs}
\begin{center}
\begin{tabular}{ll}
      &     \\  
 \psfig{file=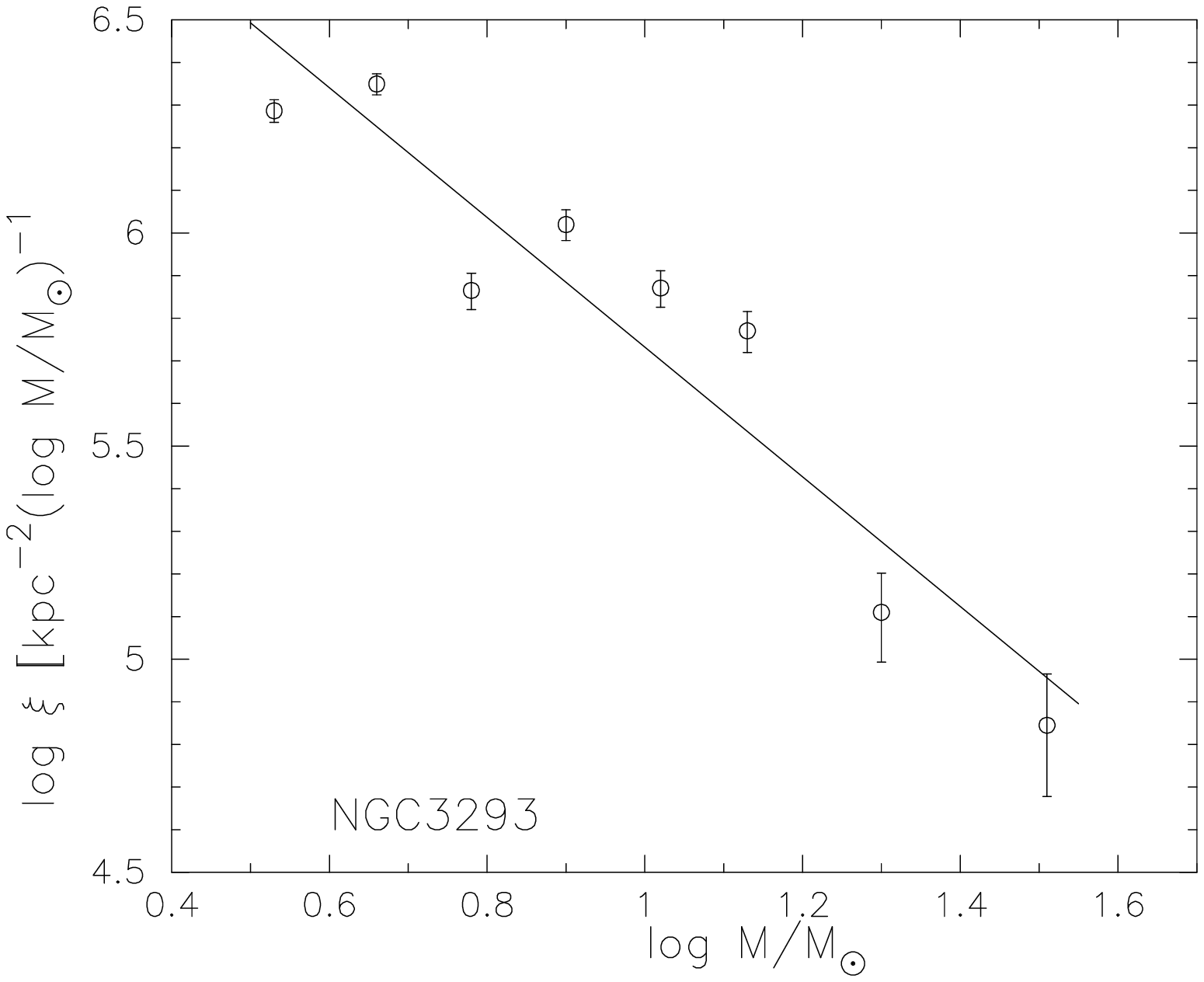,width=8cm} & \\
 \psfig{file=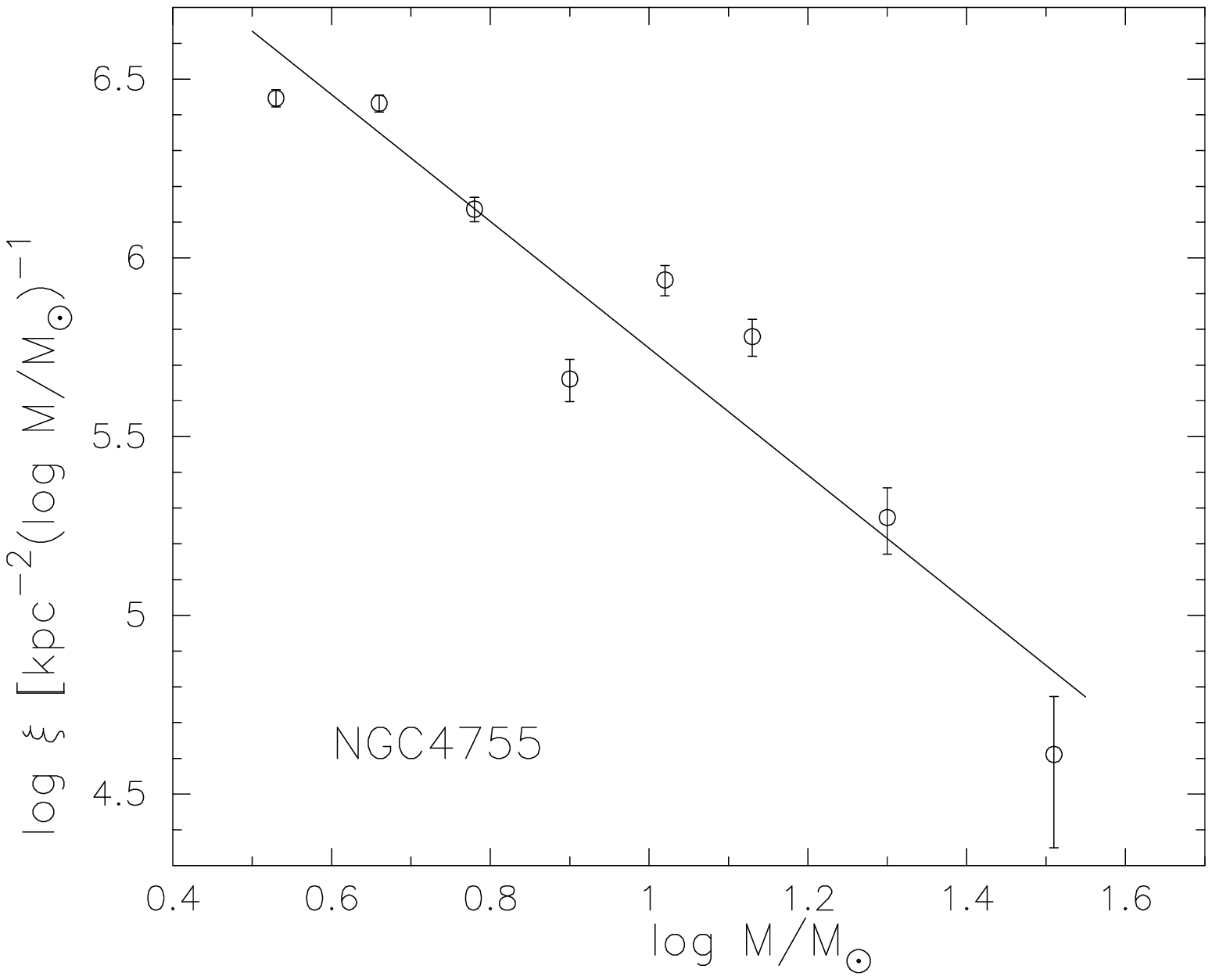,width=8cm} & \\
 \psfig{file=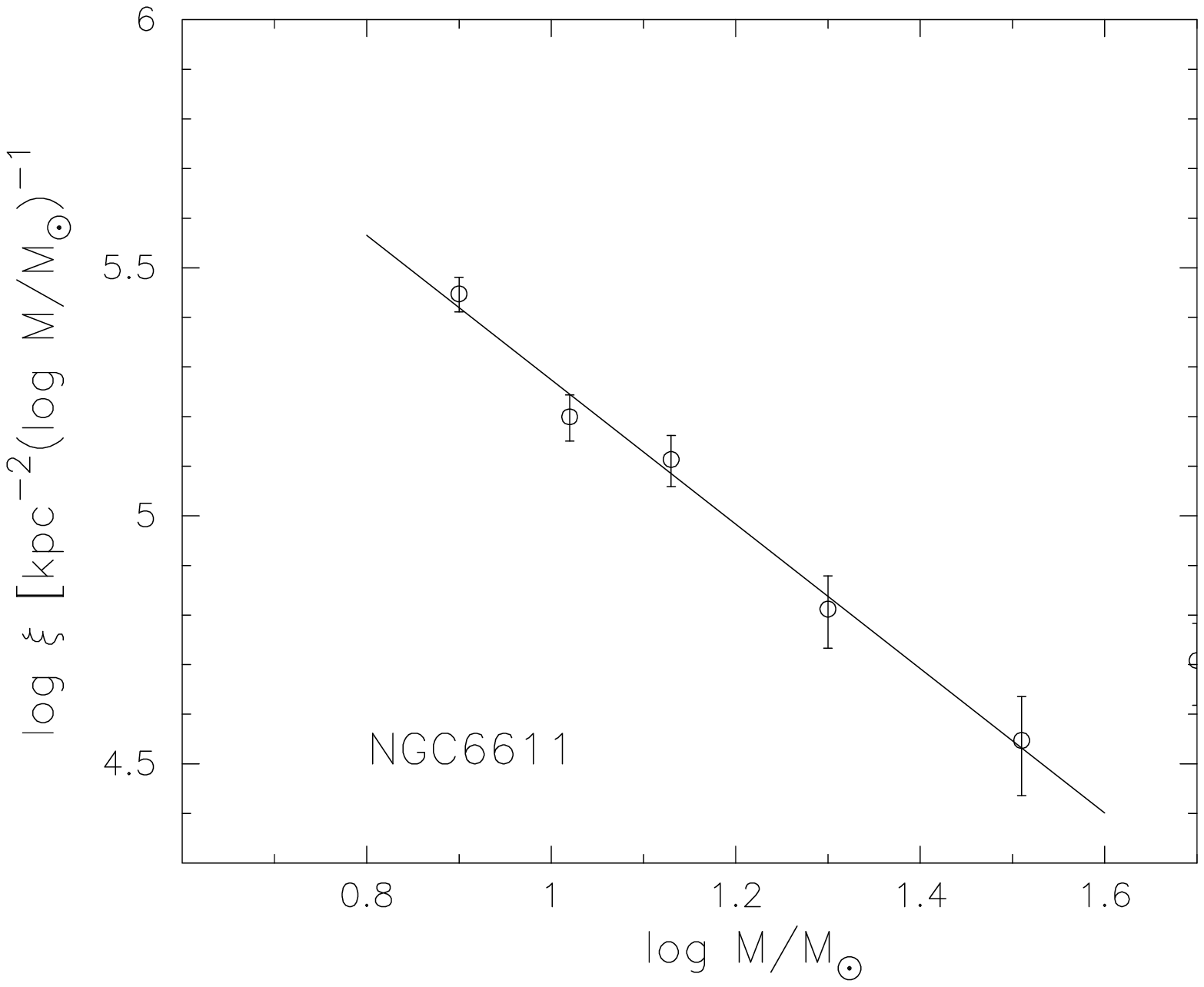,width=8cm} & \\
\end{tabular}
\end{center}
\end{figure}

In our HR diagrams, we have distinguished between apparently low and high 
gravity objects as discussed above. For our two older cluster, the effects of
stars evolving from the main sequence can be seen for our early-B type objects.
However for the later B-spectral types, there is no clear segregation of the
targets as would be expected if the lower gravity objects had evolved
from the main sequence. We have further investigated this by
refining the analysis discussed in Sect.\  \ref{disc:atm_par} to 
only consider targets with effective temperatures less than 20\,000 K, 
where evolutionary effects should be negligible. For all our targets, 
the mean projected rotational velocities are 178$\pm$72 and 
217$\pm$92\kms for the high and low gravity stars respectively. The former 
is similar to that listed in Table \ref{tab:vsini_g} for the corresponding 
sample, whilst the latter is larger due to the exclusion of hotter evolved 
objects. The position of the cool lower gravity objects in the HR diagrams 
and their high mean rotational velocities both support the hypothesis that 
our gravity estimates are reliable (in the sense that they are measuring the
mean surface gravity) and that their range may reflect the different
contributions of the centrifugal acceleration. An important consequence of 
this is that the lower effective gravity has led to many of these objects 
being classified as giants (see Tables \ref{tab:3293vrot}--\ref{tab:6611vrot})
and hence calibrations based on spectral type of, for example, the luminosity 
would led to systematic errors.

From the HR diagrams it was also possible to estimate cluster ages (assuming
coeval star formation). For all three clusters the fitting of the isochrones 
was problematic as discussed below.

{\bf NGC\,6611:} As discussed above, the observed main sequence (which
for this young cluster is still discernible at B0 spectral type) 
appears systematically brighter than predicted. There is evidence that the 
extinction towards this cluster is anomalous (see, for example, Chini \&
Wargau, \cite{Chi90} and references therein) and indeed we adopted a different
value for the ratio of the total extinction to the reddening, R (see Sect.
\ref{sec:distance}) than for the other clusters. Hence some of the discrepancy
may arise from uncertainties in the interstellar extinction. We note that 
the HR diagram of Hillenbrand et al. (\cite{Hil93}; Fig. 8) 
implies a similar discrepancy. These authors also considered isochrones for
pre-main sequence evolutionary tracks but for our targets with effective
temperature greater than 20\,000 K the timescales for reaching the main 
sequence would appear too small to explain this discrepancy -- for example
300\,000 years at an effective temperature of 20\,000K. 
However Hillenbrand et al. also found evidence that some intermediate mass 
stars  (3-8 solar masses) could be as young as 250\,000 years.
If such ages were also present amongst the more massive stars, this 
would help explain the enhanced luminosity of our objects.
We find a best fit isochrone for this cluster of approximately 
2 an , with some evidence for age spread of 1 Myr, consistent 
with the results of Hillenbrand et al. 

{\bf NGC\,3293 and NGC\,4755:} Given the similarity in their HR diagrams, 
these clusters are best considered together. As discussed above, their observed
main sequences (for effective temperatures of less than 20\,000 K) appear to
be more luminous than predicted by evolutionary models. In these cases, this
would not appear to be due to anomalous extinction. Also, given the larger
cluster ages, it would appear unlikely that we are observing stars evolving 
to the main sequence, although Sagar \& Cannon (\cite{Sag95}) identified lower 
mass pre-main sequence objects with ages between 3-10 million years in 
NGC\,4755. Inspection of their Fig. 7 implies that Sagar \& Cannon also 
found the B-type main sequence to be overluminous compared with the isochrones 
of Maeder \& Maynet (\cite{Mae91}). By contrast, Fig. 7 of Baume et al.
(\cite{Bau03}) implies good agreement between the observed HR diagram of
NGC\,3293 and the theoretical models of Girardi et al. (\cite{Gir00}). 
However this comparison is complicated by Baume et al. extending the 
envelope of the predicted B-type main sequence upwards by 0.75 magnitudes 
corresponding to binary systems with two similar stars. 
Additionally the published isochrones of Girardi et al. have a minimum 
age larger than those quoted by Baume et al., who presumably
had access to additional unpublished information further complicating 
any comparisons. A binary population could also
help explain our discrepancy but given our use of high dispersion spectroscopy
to estimate effective temperatures (compared with Baume et al.\ who used
(B-V)$_0$), we would expect any such contamination to be small.
From the position of the main sequence turnoff,
both clusters appear to have ages between 10 and 20 million years. The
theoretical isochrones fit the giant population reasonably well in NGC\,4755 
but in NGC\,3293, there are a group of stars with an implied evolutionary 
age greater than 20 million years. Additionally both clusters contain
supergiants (the highest luminosity stars in Fig. \ref{hr_diag}), which 
if they have evolved as single stars appear to have evolutionary lifetimes 
of less than 10 million years. Indeed the lower age estimate obtained by 
Baume et al. may be at least partly influenced by these targets.

In summary, for all three clusters, the agreement between the observed 
HR diagrams and the theoretical predictions is rather disappointing. 
This is particularly the case as our effective temperature estimates 
should be more reliable than other studies that have used photometric 
colours. Indeed for all three clusters it would appear that there 
is either non-coeval star formation or the evolution of individual 
targets is being influenced by other factors. The former is a plausible 
explanation for NGC\,6611 given its youth and the detection by 
Hillenbrand et al. (1993) of very young pre-main sequence 
intermediate mass objects. Although the bulk of the star formation for  
NGC\,3293 and NGC\,4755 is consistent with ages of 15\,Myrs, the most 
massive stars in the cluters have smaller evolutionary ages of approximately
5 Myrs. This would imply a very extended period of star
formation and hence these discrepancies may be due, at least in part, 
to other factors such as binarity.

\begin{figure}\caption{Histograms of projected rotational velocities for each
of the three cluster.}\label{fig:vsini}
\begin{center}
\begin{tabular}{ll}
      &     \\  
 \psfig{file=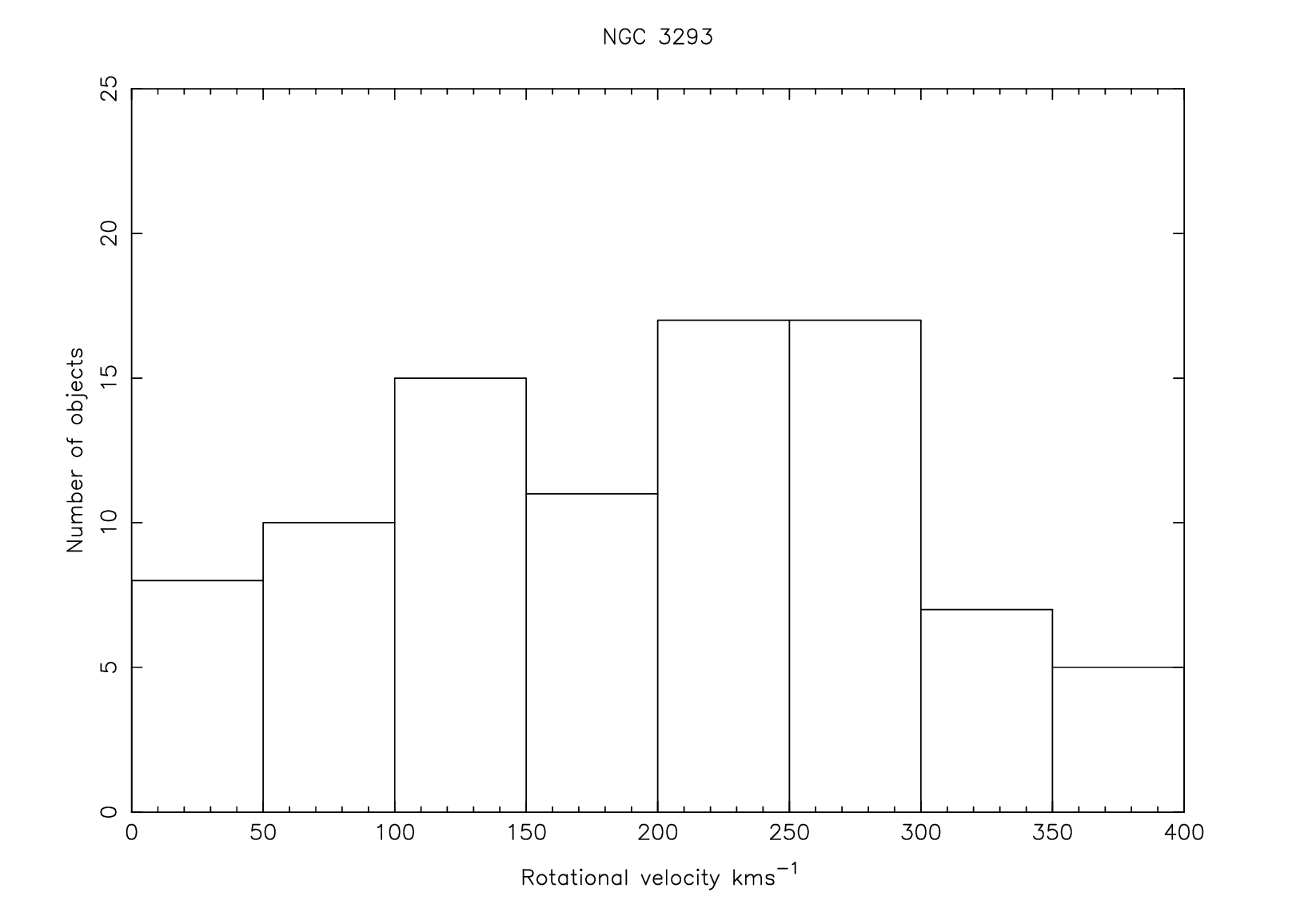,angle=270,width=8cm} & \\
 \psfig{file=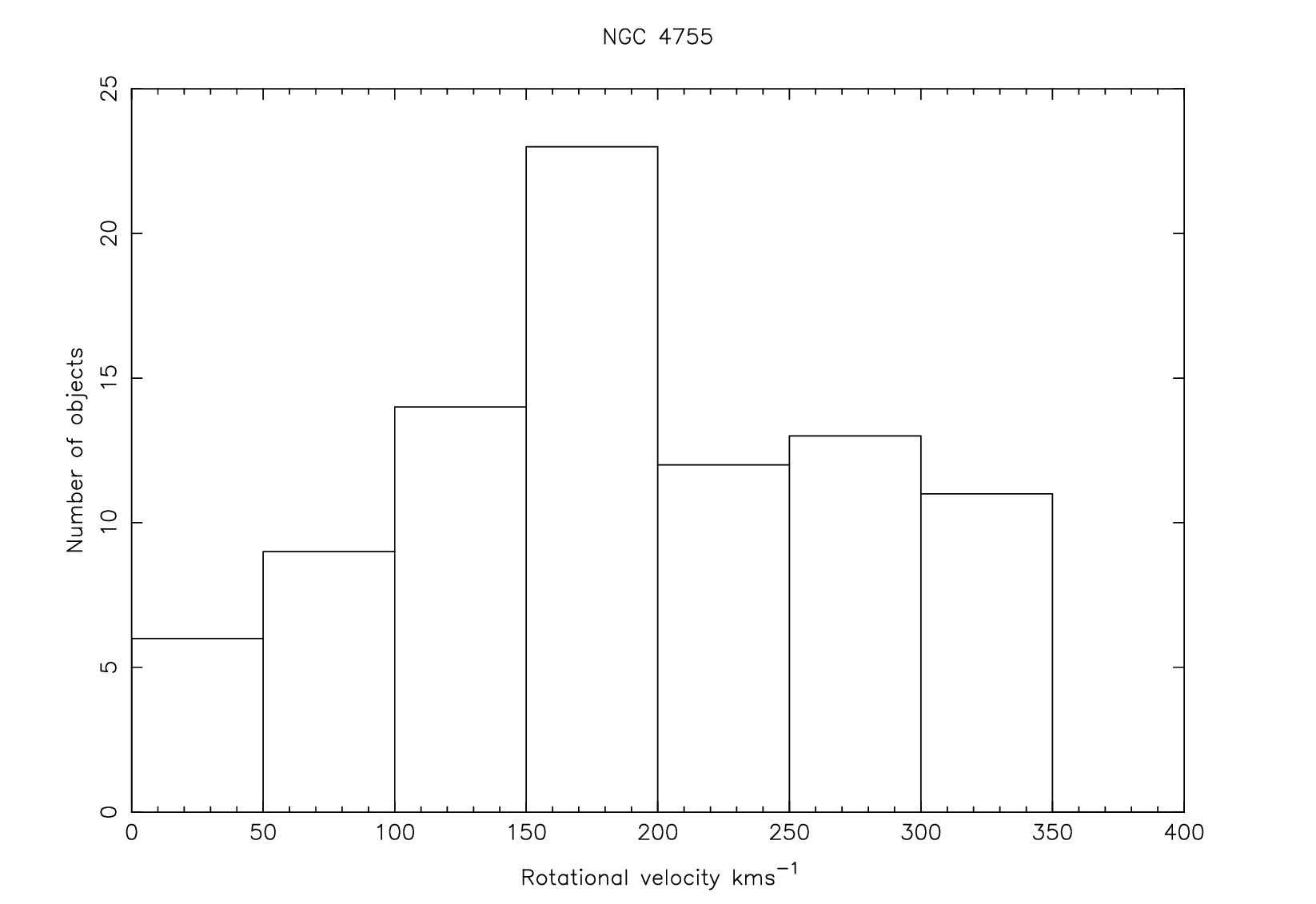,angle=270,width=8cm} & \\
 \psfig{file=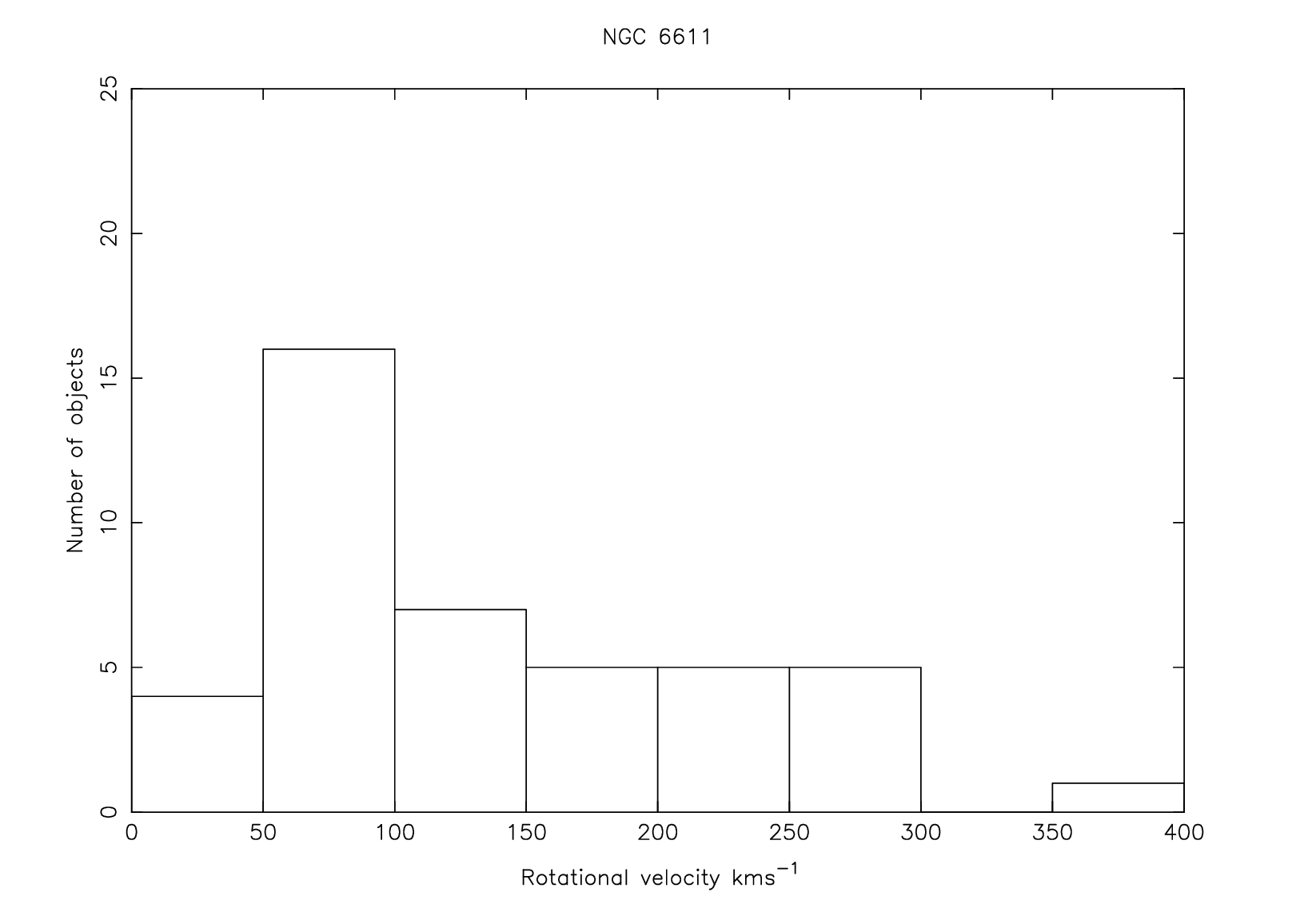,angle=270,width=8cm} \\
\end{tabular}
\end{center}
\end{figure}

\begin{figure*}\caption{Cumulative probability function for all targets in
our two older clusters and in h and $\chi$\ Persei and for those in the 
different classes as specified by Strom et al. (2005).  
Group 1 -  relatively unevolved stars with an effective temperature, 
\teff $<$ 17000; Group 2  - stars with 17000 $\leq$ \teff $<$ 21000; 
Group 3 - stars that have evolved significantly from the main sequence, 
\teff $\geq$ 21000. The solid lines are for field stars, the dotted lines 
are for the h and $\chi$\ Persei sample of Strom et al.\ (\cite{Str05})
and the dot-dash lines are our combined sample for the clusters 
NGC\,3293 and NGC\,4755.}\label{fig:cpf}
\begin{center}
\psfig{file=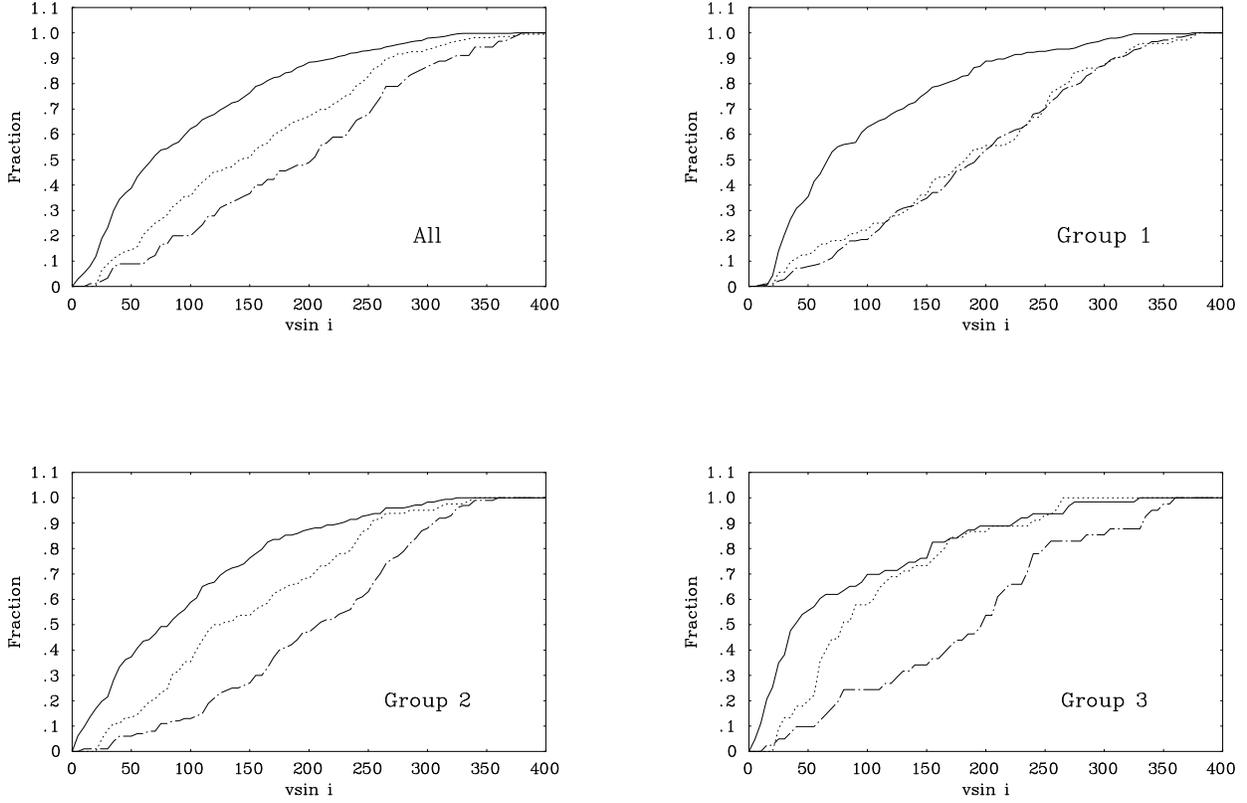,angle=270,width=18cm}
\end{center}
\end{figure*}

\subsection{Initial Mass Functions} 
\label{disc:imf}
The atmospheric parameters and masses determined in the previous sections 
allow the mass functions of the three clusters to be re-evaluated with the
most extensive data and information to date. As discussed in 
Paper I, we are not spectroscopically ``complete'' in any of the clusters, 
in the sense that we were not able to place fibres on all stars within 
a given colour-magnitude range. However in the case of NGC\,3293 and 
NGC\,4755 the reddening towards the cluster appears fairly homogeneous and 
hence we can estimate the effective temperatures and luminosities from 
the stellar $BV$ photometry. In the case of NGC\,6611, which shows a 
scattered colour-magnitude diagram, such a procedure would not be reliable 
and we have chosen to increment the Hillenbrand et al. 
(\cite{Hil93}) study with 
our additional spectroscopy in order to improve the statistical significance 
of the results. 

\subsubsection{NGC\,3293} 
We have selected all the stars from the colour magnitude diagram 
(see Paper I, Fig. 3) for which we do not have spectroscopic data 
and which were in the range $V\leq13.3$ and $(B-V)_{0}\leq0$ 
(where  $(B-V)_{0}$ is the intrinsic colour estimated using the 
data in Table~\ref{distances}). Effective temperatures were 
estimated by fitting a cubic to the \teff $- (B-V)_0$ 
results in Table\,~\ref{tlusty_ubv} (assuming that $\log g=4.0$). 
We then calculated bolmetric luminosities and masses using the 
method employed in Section\,\ref{sec:distance}. Thirty six stars were
selected, all with masses in the range 3-10\msun. There is also one 
confirmed red supergiant cluster member of NGC\,3293 which must be 
included. Feast (\cite{Fea58}) gives an M0\,Iab classification with 
$M_{\rm V}=-5.9$ and using the bolometric corrections of Elias et al. 
(\cite{1985ApJS...57...91E}), we have estimated a mass from the Geneva 
evolutionary tracks of 15\msun. 

To determine the IMF, 
we follow the method of Hillenbrand et al. (\cite{Hil93}) and Massey et al. 
(\cite{1995ApJ...454..151M}), so that our results can be directly
compared to those studies. Stars were divided into mass bins, 
whose size was chosen to give comparable numbers of objects as far 
as possible. The number of stars in each bin was  normalised to unit 
logarithmic mass interval and to unit area (kpc$^{2}$) and 
the Scalo (\cite{1986FCPh...11....1S}) notation was adopted. 
%
%
The area sampled was taken from the cluster radius (we used twice the 
quoted radius, to be consistent with the membership criteria applied above).

In Fig.\,\ref{imfs} we plot the present day mass function (PDMF), which 
given the relatively small age of NGC\,3293 and the fact that the majority
of stars are on the main-sequence, is effectively the IMF. A linear least 
squares fit was performed, with each point being weighted by $\sqrt N$, 
where $N$ is the number of points in the bin.  We find 
a slope of $\Gamma = -1.5 \pm 0.2$ over the range 2.8-40\msun. Baume et al. 
(\cite{Bau03}) estimate $\Gamma = -1.2$ for the mass range 1.4-45\msun, but
find a steeper slope of  $\Gamma = -1.6$ for the higher mass stars in the
range 8-45\msun. Our results imply that this steeper value for the high 
mass objects extends down to $\sim$3\msun\ and are consistent with the 
range of IMFs found by Massey et al.  (\cite{1995ApJ...454..151M}) 
for young clusters in the Northern Milky Way.

\subsubsection{NGC\,4755} 
The PDMF for NGC\,4755, and hence by implication the IMF, was estimated 
in a similar way as that for NGC\,3293 . We used the same magnitude
and colour criteria ($V\leq13.3$ and $(B-V)_{0}\leq0$) and identified 51
stars, with masses in the range 2-4\msun. Again there is one red supergiant 
that is a cluster member, with an M2Ib spectral type (Feast \cite{Fea63}), 
for which we have estimated a mass of 15\msun. In Fig.\,\ref{imfs}, we 
plot the IMF for NGC\,4755 and estimate a slope of $\Gamma = -1.8 \pm 0.2$, 
which again is consistent with the range reported in Massey et al.  
(\cite{1995ApJ...454..151M}) and similar to that for NGC\,3293. 
 
\subsubsection{NGC\,6611} 
For NGC\,6611, because of the highly variable reddening across the field, it
was not possible to use a similar method to that for NGC\,3293 and NGC\,4755.
From Hillenbrand et al. (\cite{Hil93}), we can identify 22 stars 
with spectral types earlier than B3, which are likely cluster members but
for which we do not have spectroscopy. All but one of these are 
classified as B0.5V-B3V and we have adopted effective temperature estimates
based on the data in Table\,\ref{tab:si4temp}. The other target is 
BD$-13^{\circ}$4912, which was classified as B2.5I, and for this we have
adopted $T_{\rm eff}=16\,500$ (Crowther et al. \cite{Cro05}).
Masses were then calculated following the same procedure as discussed 
in Section\,\ref{sec:distance}.

We have also estimated effective temperatures, luminosities and masses 
for likely cluster members in the Hillenbrand et al.\ sample without 
spectral types by calculating $Q$, and using the $Q-T_{\rm eff}$ relation 
of Massey et al. (\cite{Mas89}). None of these targets had masses 
above 7\msun\, and hence it 
is reasonable to assume that the combination of our spectroscopy and 
that of Hillenbrand et al. is virtually complete above 7\msun. 
In Fig.\,\ref{imfs} we plot the IMF for NGC\,6611 and determine a 
slope of $\Gamma = -1.5 \pm 0.1$, which is slightly steeper than the 
value of $\Gamma = -1.1 \pm 0.3$ of  Hillenbrand et al., but is 
consistent with a Salpeter type value and again similar to our other
clusters and those in the Massey et al. (\cite{1995ApJ...454..151M}) 
survey.

\subsection{Projected Rotational Velocities} \label{disc:rotation}

In Figure \ref{fig:vsini}, the distribution of projected rotational 
velocities are plotted as histograms for all three clusters. Given 
the similarity in their ages and HR diagrams, 
it is not surprising that the distributions for 
NGC\,3293 and NGC\,4755 appear similar. There appears to be an excess 
of stars in NGC\,4755 with projected rotational velocities between 
100-150 \kms but assuming Poisson statistics, this is not statistically 
significant and indeed the mean values for the two clusters are almost 
identical (NGC\,3293: 188 \kms; NGC\,4755: 185 \kms). 
We have  calculated cumulative probability functions (CPFs) for our 
combined samples in NGC\,3293 and NGC\,4755 in order to compare 
our results with the study of Strom et al. (\cite{Str05}) for the
double cluster, h and $\chi$~Persei, which appears to have a similar 
age (of approximately 13 million years for their nuclear regions 
- Slesnick et al., \cite{Sle02}). We follow the methodology of
Strom et al. by dividing our sample into three groups (see Fig. 1 
of Strom et al.), viz. (1) relatively unevolved stars with an 
effective temperature, \teff $<$ 17000; (2) stars with 
17000 $\leq$ \teff $<$ 21000; (3) stars that have evolved 
significantly from the main sequence, \teff $\geq$ 21000 K.
In Fig. \ref{fig:cpf}, we plot the CPFs for all our targets and for 
each group, together with the CPFs for stars in h and $\chi$\ Persei 
and Strom et al's sample of field stars. Strom et al. make two major 
conclusions about their CPFs. The first is that the clusters stars 
on the whole tend to rotate faster than the field stars counterparts. 
Secondly they find that by dividing their sample into these different
mass bins they can see a trend in which the lower mass stars 
($4-9$\msun) have $v \sin i$ values clearly higher than the field stars 
but that the higher mass stars ($9-15$\msun) 
show a very similar $v \sin i$ distribution to 
their field star counterparts. 

Inspection of Fig. \ref{fig:cpf} shows that the projected rotational 
velocities are systematically higher in the NGC\,3293 and
NGC\,4755 clusters than in the field populations. This difference has 
also been found previously in other studies 
(see, for example, Bernacca \& Perinotto \cite{Ber74}; Wolff et al.
\cite{Wof82}; Keller \cite{Kel04}) and has been interpreted in several ways
including differences in the ages of the field and cluster targets. 
However our results are significantly different to the detailed
work by Strom et al. Their CPFs 
for the h and $\chi$\ Persei double cluster and the field show a clear 
differentiation for group (1) targets, decreasing for groups (2) and 
(3). Strom et al. discussed this behaviour at length and postulated 
that it reflected the star formation environment rather than evolutionary 
effects during the stellar lifetimes. Their conclusion that
evolutionary effects were unlikely was based on the fact that the
mean $v \sin i$ of the high mass stars (group 3) was nearly a factor of two higher
than that for the lowest mass objects (group 1). 
Meynet \& Maeder (\cite{2000A&A...361..101M})
and Heger \& Langer (2000) calculations of the rotational
velocity evolution of $12-15$\msun\ stars on the main-sequence suggests a
decrease in velocity of between $25-30$\% within $10-15$\,yrs. Hence
Strom et al.  speculated that the differences 
reflected the effects of higher accretion rates that may be characteristic 
of star formation in these dense clusters. In particular they suggested 
that enhanced aggregation will have three effects, viz. (1) higher initial
rotation speeds, (2) higher initial radii along the stellar birth line (when
deuterium shell burning commences), resulting in greater spin up as the star
evolves to the main sequence and (3) that the higher initial birthline 
radii are especially pronounced for mid- to late-B-type stars. This 
latter leads naturally to the larger differences with respect 
to field stars observed for their group (1) objects in h and $\chi$\ Persei.

In our combined sample for NGC\,3293 and NGC\,4755, the rotational velocity
distribution  of our group (1) objects is essentially the same as that found 
by Strom et al. However our group (2) and (3) objects show similar enhanced
projected rotational velocities and in particular do not tend towards those of
the field stars. Hence contrary to the results for h and $\chi$ Persei, any
effect would appear to be operating throughout our stellar sample.
Huang \& Gies (\cite{Hua05a}) have surveyed 496 OB stars in clusters
within the approximate age range $6-73$\,Myrs, and also find fewer
slower rotators in the cluster stars than in the field. They also present 
some evidence for the higher mass stars (late-O to early-B) 
spinning slower than the lower mass late B-types, and
suggest that this can be 
explained by the spin-down of massive stars through angular momentum 
loss via stellar winds e.g. as modelled by 
Meynet \& Maeder (\cite{2000A&A...361..101M}) and Heger \& Langer 
(\cite{Heg00}).
 
\begin{figure*}
\begin{center}
\psfig{file=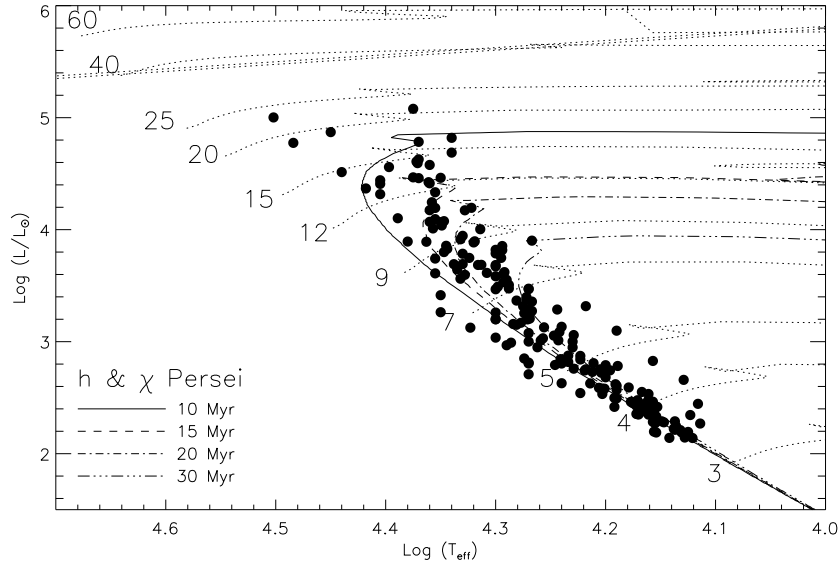,,angle=0,width=12cm}
\caption{Hertzsprung-Russell diagrams for the h and $\chi$ Persei cluster,
together with evolutionary tracks and isochrones taken from the Geneva 
database.}
\label{hr_hk}
\end{center}
\end{figure*}

One possible explanation for the different projected rotational 
velocity distributions for groups (2) and (3) in our combined 
NGC\,3293 and NGC\,4755 sample and that of Strom et al. (\cite{Str05}) 
for h and $\chi$\ Persei is that the clusters have different ages. 
Hence we have used the data presented by Strom et al.\ to construct an 
HR diagram similar to those shown in Fig. \ref{hr_diag}. Our procedure was 
to use their absolute visual magnitudes and the bolometric
corrections of Balona (\cite{Bal94a}) to deduce stellar luminosities
and the HR diagram for h and $\chi$\ Persei is shown in Fig.\ 
\ref{hr_hk}, together with isochrones again taken from the Geneve database.

The observational HR diagram for h and $\chi$\ Persei is qualitatively
similar to those for NGC\,3293 and NGC\,4755. However for the former the
agreement between observed and theoretical main sequences is better for
the later B-type stars. As for our clusters, the evolved star populations
in h and $\chi$\ Persei imply a range of ages with an estimate of 
approximately 15 million years being reasonable, consistent with the
age of 12.8$\pm$1.0 million years deduced by Slesnick et al.
(\cite{Sle02}) from photometry and moderate resolution spectroscopy.
These estimates are similar to
our adopted ages for NGC\,3293 and NGC\,4755. 
Strom et al.\ postulate that their projected
rotational velocity distributions were present when the stars were formed.
Hence given that the ages of NGC\,3293 and NGC\,4755 are comparable with
that of the h and $\chi$\ Persei sample, the differences
in projected rotational velocity distributions cannot be explained by
variations in the stellar ages of the samples.

The mean of the \vsini\ estimates for all the observed stars in our 
younger cluster NGC\,6611 is 144 \kms, which is lower than that for 
NGC\,3293 and NGC\,4755. In NGC\,6611 there are $v \sin i$ measurements 
of 25 stars with masses less than 15\msun\ and 13 stars with masses 
between 15-60\msun, with mean values of 192 and 116\kms 
respectively. That for the lower mass group is in good agreement 
with those for both NGC\,3293 and NGC\,4755, where we should be 
sampling the same stellar mass range albeit at a different age. 
For the massive stars, evolutionary models (Meynet \& Maeder 
\cite{2000A&A...361..101M}; Heger \& Langer \cite{Heg00})
predict that initial rotational velocities of 300\kms are reduced 
to between 60-200\kms during the main-sequence phase due to angular 
momentum loss via their winds, with the most massive luminous objects 
showing the most pronounced effect. The fact that we find a lower mean 
value of $v \sin i $ for the more massive stars supports the existence
of this effect.

We can test this hypothesis in a simple and illustrative way. For the 13 
stars in the 15-60\msun\ range we assume that they had an initial rotational 
velocity of 250\kms (see Sect. \ref{disc:vrot}) and estimate the rotational 
velocity for each star after 2\,Myrs (being our estimated age for
NGC\,6611) using the mass dependent evolution of surface equatorial 
velocities predicted by Meynet \& Maeder (\cite{2000A&A...361..101M}) 
and Heger \& Langer (2000).  This gives a mean equatorial 
velocity of 166 \kms, and hence a mean \vsini\ value of 125\kms\ 
(assuming that the spin axes are randomly distributed).
This is similar to the observed mean \vsini\ of 116\kms, which suggests 
that an initial typical equatorial velocity of about 250\kms and spin 
down according to the evolutionary models is consistent with that observed. 
We caution that this is an illustrative and simple comparison as it assumes
a single value for the initial velocity and ignores any macroturbulent 
broadening in the atmospheres of the luminous objects.
 
The mean \vsini\ of the stars with masses less than 15\msun in each of the 
three clusters is consistently around 190\kms, which would suggest a mean 
equatorial velocity of 240\kms (also see Section\,\ref{disc:vrot}). The 
initial velocity of these objects is then likely to be in the range 
250-300\kms, depending on the evolution of the rotational velocity
over the course of 2-15Myrs. At present this is uncertain for models 
less than 12\msun, as there are no detailed published models for 
the evolution of the rotational velocity. 
For the 12-15\msun\ models, the rotational velocity does
decrease during the first 1-2 Myrs but then remains effectively constant
for the remainder of the main sequence lifetime. Hence it may be reasonable 
to compare the lower mass NGC\,6611 targets with those in NGC\,3293
and NGC\,4755.

Recently Wolff et al. (\cite{Wol06}) have investigated the projected
rotational velocities in a sample of very young stars with masses ranging
from 0.2 to 50 \msun. Their principal aim was to investigate the 
formation mechanism for high mass stars and in particular if it differed
from that for lower mass stars. For the full range of spectral types
they found
that the ratio of the projected rotational velocity to the equatorial
breakup velocity showed no significant variations. Assuming a random
orientation of rotational axes, this led to a median value for the
ratio of the rotational velocity to the equatorial
breakup velocity, R, of 0.14, with values for B-type stars 
($8<M<25$\msun) of 0.13 and for O-type stars ($25<M<90$\msun) of 0.20.

A similar analysis can be undertaken for our samples with the small age 
and presence of O-type stars making that for NGC\,6611 the most 
appropriate. We have again assumed a random orientation of rotational 
axes and restricting our sample to stars that appear to be main sequence 
with \logg$\geq 3.9$ and obtain a ratio, R, of 0.22 for 27 targets.
For our 7 O-type stars, the ratio is 0.16, whilst for the 20 B-type 
stars the ratio is 0.28. Hence our ratio for O-type stars is in 
good agreement with that of Wolff et al. but our B-type value is higher
although given our relatively small sample size this may not be
significant. For our combined B-type sample for older clusters
NGC\,3293 and NGC\,4755, we obtain a far higher median estimate
of 0.61 for 83 targets. The reason
for this higher ratio is unclear, as our targets would not be expected
to have evolved to higher rotational velocities during the clusters'
lifetimes. However it is consistent
with the different mean velocities found for the three clusters and
discussed above.

In summary for our older clusters, we confirm that the cluster stars
show higher projected rotational velocities than their field star 
counterparts. However we do not reproduce the results of Strom et al. 
for h and $\chi$\ Persei who found that the higher mass stars 
tend to rotate significantly more slowly than the lower mass objects. 
If their explanation of enhanced accretion causes
the different behaviour between the cluster and field samples, then at 
least for NGC\,3293 and NGC\,4755 this mechanism is present for all 
B-type objects. The mean equatorial velocity of the most massive stars 
(15-60\msun) in NGC\,6611 is significantly less than the lower mass stars 
in this cluster and those in NGC\,4755 and NGC\,3293. This could 
be due to the more massive stars losing significant angular
momentum through stellar winds during the first 2\,Myrs of their lifetime
but is also consistent with the rotational velocities 
found by Wolff et al. (\cite{Wol06}) for very young O-type stars.

\subsection{The intrinsic rotational velocity distribution} 
\label{disc:vrot}

The sample of stars for which we have reliable \vsini\ values is large
enough that we may be able to model the intrinsic rotational velocity
distribution by assuming that the inclination angle $i$ is randomly
distributed in space. Following the formulation of Chandrasekhar \&
M\"{u}nch (\cite{1950ApJ...111..142C}), which is also discussed in 
Brown \& Verschueren (\cite{1997A&A...319..811B}), we employ the integral 
equation which relates the distribution, $f(x)$, of equatorial velocities,
$x$, for a sample of stars to the corresponding distribution of 
projected rotational velocities $\phi(v\sin i)$. 

\begin{equation}
\phi(v\sin i) =  v\sin i \int_{v\sin i}^{\infty} \frac{f(x)}{x\sqrt{x^{2} 
                 - (v\sin i)^{2}}}dx
\end{equation}

There are two $f(x)$ distributions for which this equation can be 
integrated analytically. The first is a $\delta$-function and the 
other is a uniform distribution in which all values of $x$ occur with equal
probability upto a maximum velocity, $v_{max}$. These analytic solutions
are given in  Chandrasekhar \& M\"{u}nch (\cite{1950ApJ...111..142C}).
We have also considered as an illustrative example a Gaussian function 
for $f(x)$ and integrated the equation numerically in incremental steps 
of $v\sin i$. Mokiem et al. (\cite{Mok06}) have discussed the distribution
of rotational velocities for a sample of O-type stars in the SMC obtained 
during our FLAMES survey. They show that different choices for the distribution
of rotational velocities, $f(x)$, can lead to similar distributions for the
projected rotational velocities, $\phi(v\sin i)$ (see Fig. 10 of their paper). 
Hence we stress that our adoption of a Gaussian distribution should only 
be considered illustrative and does not rule out the possibility that 
the actual distribution may be different.

\begin{figure}\caption{The histogram of the observed projected rotational 
velocities for targets in the clusters NGC\,3293 and NGC\,4755 
is compared with models assuming an intrinsic rotational velocity 
distribution of a delta-function (a) and a uniform distribution (b).  
The histogram data includes only stars in the mass range 3-15\msun\ 
i.e. the evolved objects have been removed. In panel (a) all stars 
have been assumed to rotate at
velocities of 350\kms (solid curve) and 250\kms (dashed curve). 
In panel (b) continuous distributions of velocities between zero 
and a maximum of 350\kms
(solid curve) and 250\kms (dashed curve) are assumed. It is clear 
that neither
rotational velocity distribution provides convincing 
agreement with the observed velocity histograms.}
\label{fig:anal}
\begin{center}
\begin{tabular}{ll}
      &     \\  
 \psfig{file=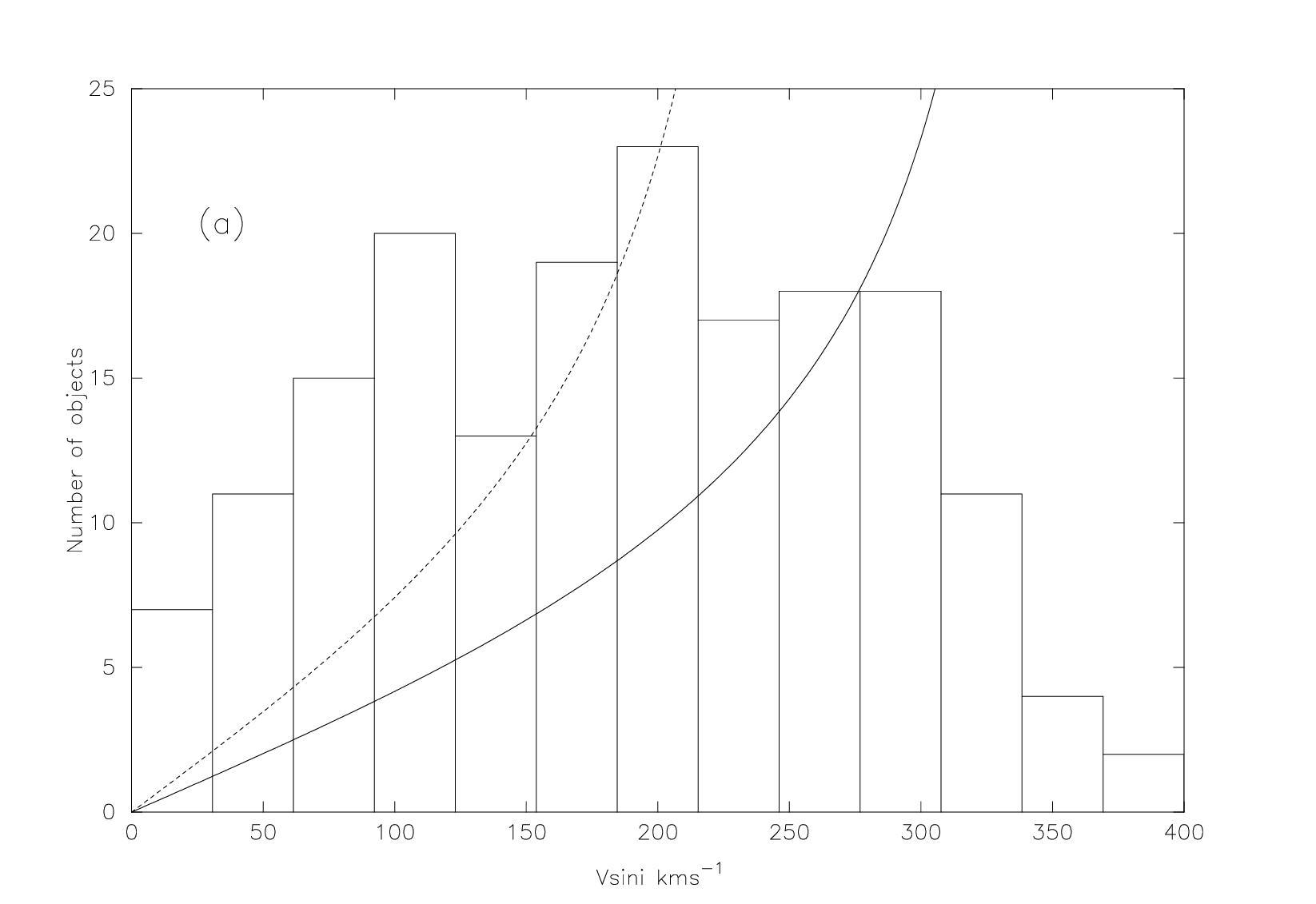,angle=270,width=8cm} & \\
 \psfig{file=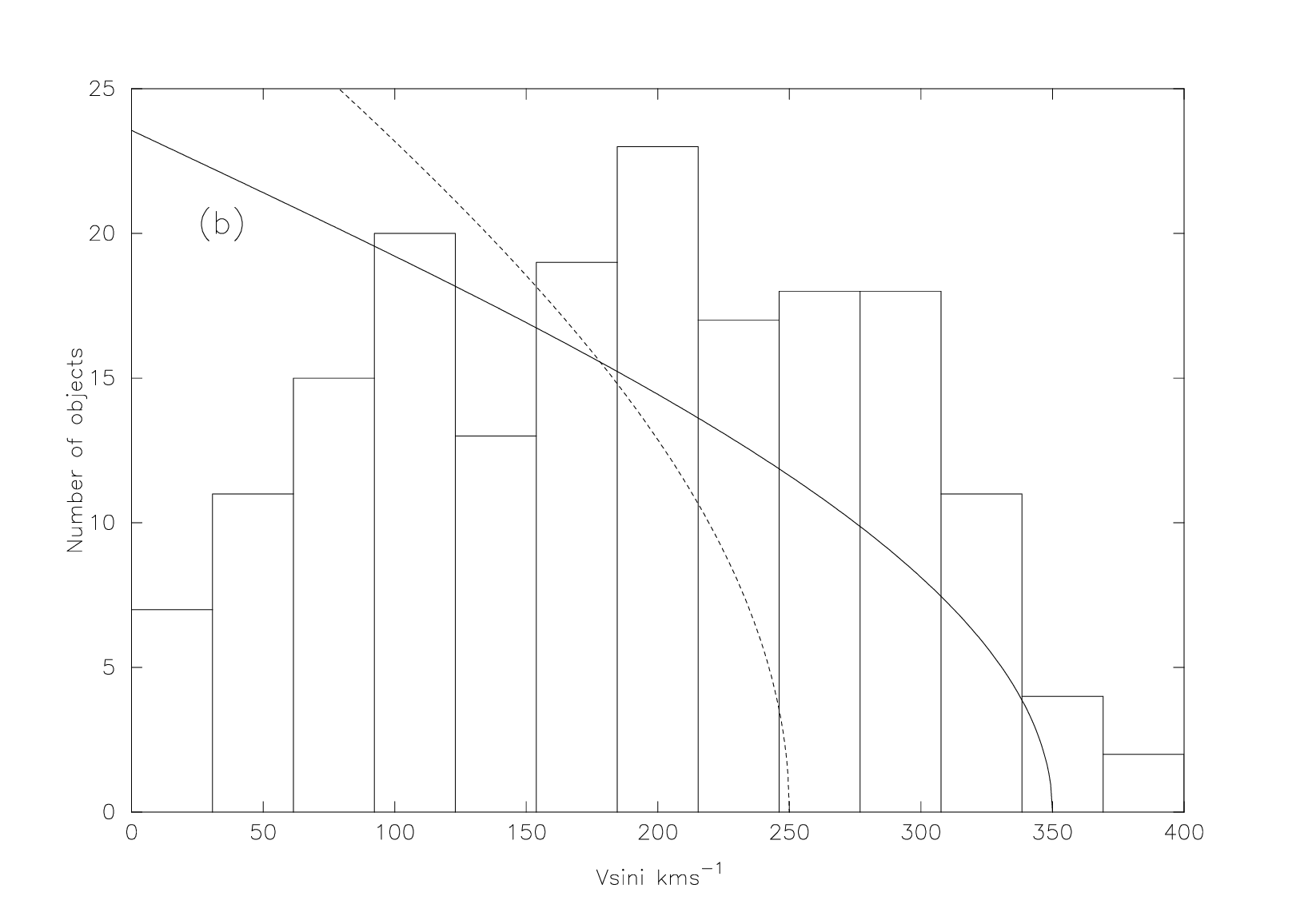,angle=270,width=8cm} & \\
\end{tabular}
\end{center}
\end{figure}

\begin{figure}\caption{Upper Panel:
The histogram of observed projected rotational velocities for targets
in the clusters NGC\,3293 and NGC\,4755 is compared with the model assuming 
a Gaussian distribution of rotational velocities. The Gaussian has a maximum
at a rotational velocity 
250 \kms and a $\sigma$=110\kms. The histogram data includes only stars 
in the mass range 3-15\msun\ i.e. the evolved objects
have been removed. Lower Panel: The NGC\,6611 stars, with a Gaussian
distribution with a
maximum at 175\kms and a $\sigma$=100\kms}
\label{fig:vrot}
\begin{center}
\begin{tabular}{ll}
      &     \\  
 \psfig{file=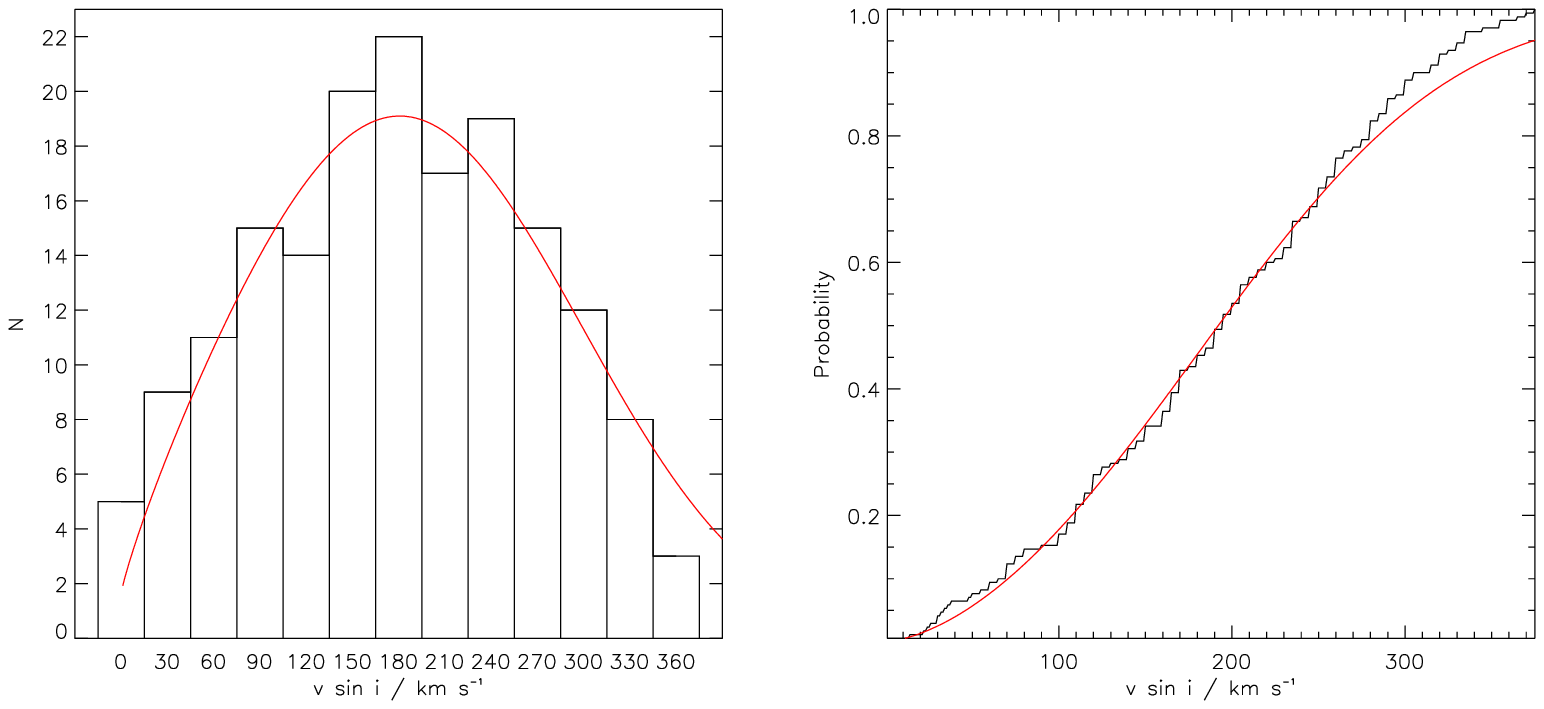,width=8cm} & \\
 \psfig{file=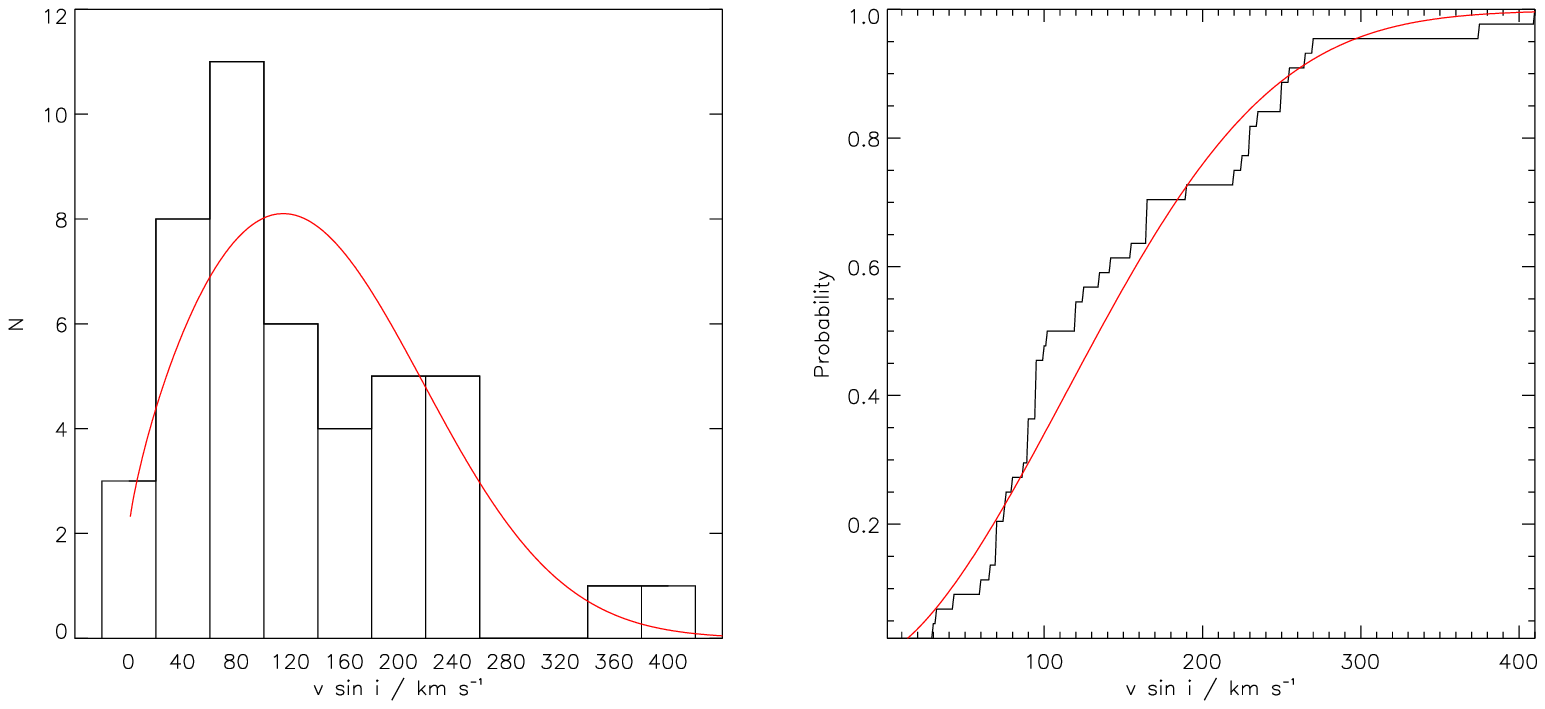,width=8cm} & \\
\end{tabular}
\end{center}
\end{figure}

In order to increase the statistical significance of the model fitting
we have combined the projected rotational velocity distributions for the
samples in NGC\,4755 and NGC\,3293. These two clusters are very similar 
in age, mass function, metallicity and total mass as discussed
above. Hence we should have a sample of stars born in
similar conditions and of a comparable age. 
This leads to a combined sample of 178 stars and the histogram of 
observed $\phi(v\sin i)$ along with the corresponding CPFs can 
be compared with our different predictions for the projected
rotational velocity distribution. We can confidently rule out the 
two analytical solutions of the $\delta$-function, or the continuous
distribution as they provide clearly inconsistent fits with the 
observed histogram and CPF as illustrated in Fig.\,\ref{fig:anal}. 
In Fig.\,\ref{fig:vrot}, we show the 
predictions for an intrinsic rotational velocity distribution, $f(x)$,
which is a Gaussian function with a peak of 225 \kms and a $\frac{1}{e}$
half width, $\sigma$, of 110 \kms (corresponding to a 
full-width-half-maximum, FWHM, of approximately 180 \kms). 
This is the best fit obtainable using a Gaussian profile, both from a 
subjective comparison and a $\chi^{2}$ test for a grid of possible solutions. 
A poorer subjective fit (with significantly higher $\chi^{2}$)  
is found when the peak of the Gaussian profile is changed by 
$\pm$25\kms or the width, $\sigma$, by $\pm$20\kms. As noted 
in Section\,\ref{disc:hr_diag}, there are 7 stars which have 
evolutionary masses of between 7-40\msun, and  which have evolved 
significantly from the main-sequence. When these are removed from 
the sample, we get a slightly higher value of 250\kms 
for the peak of the distribution and a similar width. 
Hence we believe that this is a reasonable solution for the rotational 
velocity distribution of stars in the range 3-15\msun\ at an age of 
approximately 15\,Myrs in these two clusters. However with our methodology 
we cannot rule other distributions of a similar nature. For 
example Mokiem et al. (2006) show that a Gaussian 
function would give a similar model distribution
as a rectangular function of similar width and that a very 
large number of observed stars would be required to definitively distinguish 
between the two input distributions. Although we
cannot distinguish between such model distributions, or indeed other 
distributions of similar form, we can conclude that the underlying
distribution must be a broad function with a centre and width
similar to our best fit Gaussian. 
Additionally
we were not able to exclude possible binaries from this sample. 
For example Brown \& Verschueren (\cite{1997A&A...319..811B}) and 
Huang \&  Gies (\cite{Hua05a}) found that early-type stars 
with radial velocity variations on the whole rotated more slowly than the 
likely single star sample. Our observational time sampling is not extensive
enough to determine likely binary fractions, although we should be able to 
identify binaries in our LMC and SMC clusters (Evans et al. 2006). 

In Fig.\,\ref{fig:vrot} a similar analysis is shown for NGC\,6611. 
Only forty four stars were included, which reduces 
the significance of the fit considerably. A model with a peak of 175\kms 
and $\sigma$=100\kms (FWHM$\simeq$ 165 \kms) appears to be the best 
fit for a Gaussian distribution of rotational velocities. The mean
rotational velocity is somewhat lower than that for the combined data for 
the older clusters, and we note that 
previous studies on the rotational velocities of O-type stars in the Galaxy 
(Howarth et al. \cite{1997MNRAS.284..265H}, Penny \cite{1996ApJ...463..737P}) 
have indicated that rotational velocity distributions are peaked around 
100\kms. Indeed an analogous Gaussian fit to the unevolved sample of 
Penny (1996) gives a peak of 155\kms and 
$\sigma$=100\kms. Hence there is a suggestion that the more massive 
stars (15-60\msun) in the Galaxy have typically smaller rotational 
velocity distributions than their lower mass counterparts (3-15\msun). 
As discussed in Section\,\ref{disc:rotation} using a simple illustrative
example,  this can qualitatively be explained by the models of massive 
rotating stars losing significant angular momentum through strong 
stellar winds (Meynet \& Maeder \cite{2003A&A...404..975M}, 
Heger \& Langer \cite{Heg00}).

\subsection{Comparison with stellar evolutionary models including rotation} 
\label{disc:rotmodels}

\begin{figure*}
\begin{center}
\psfig{file=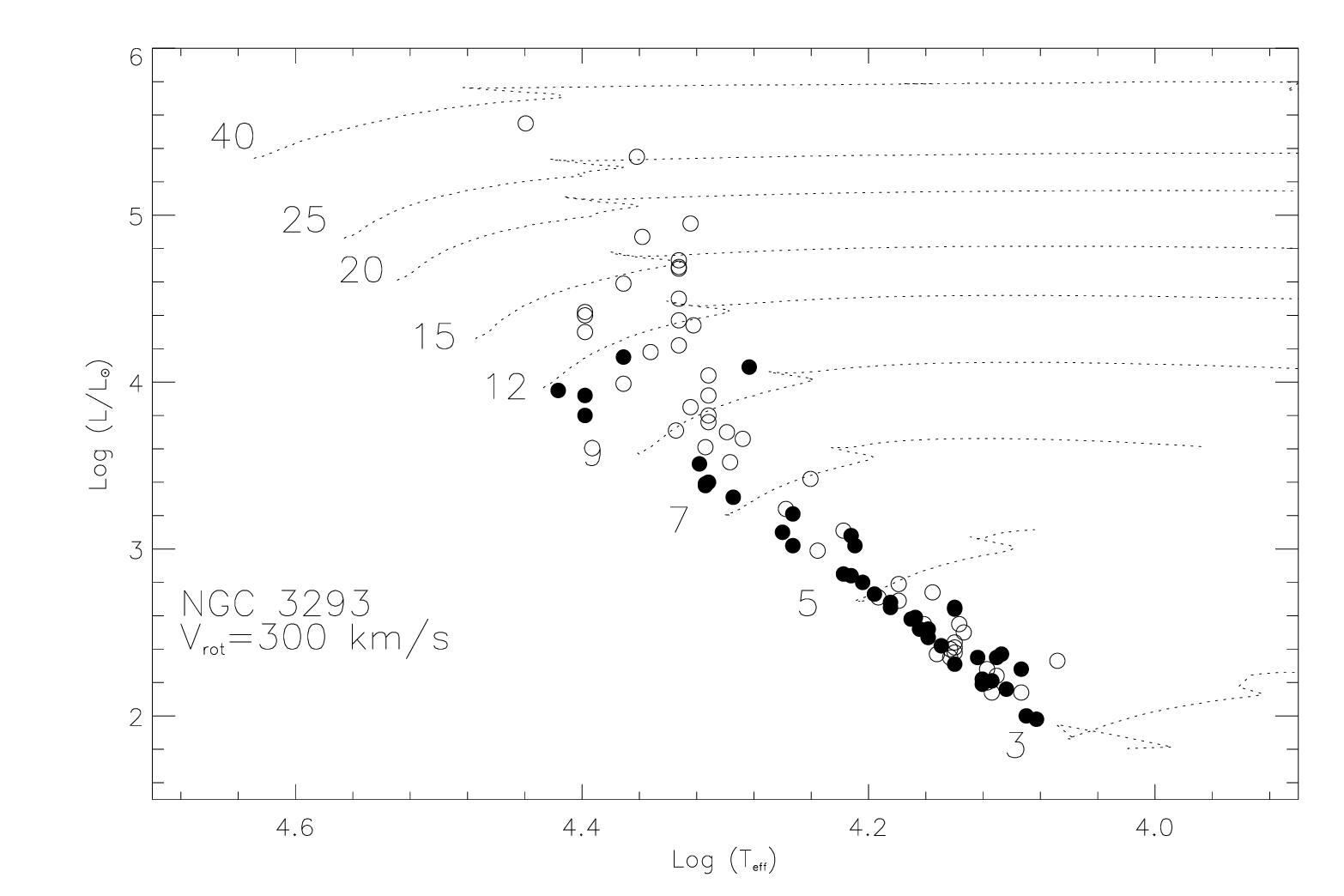,,angle=0,width=12cm}
\psfig{file=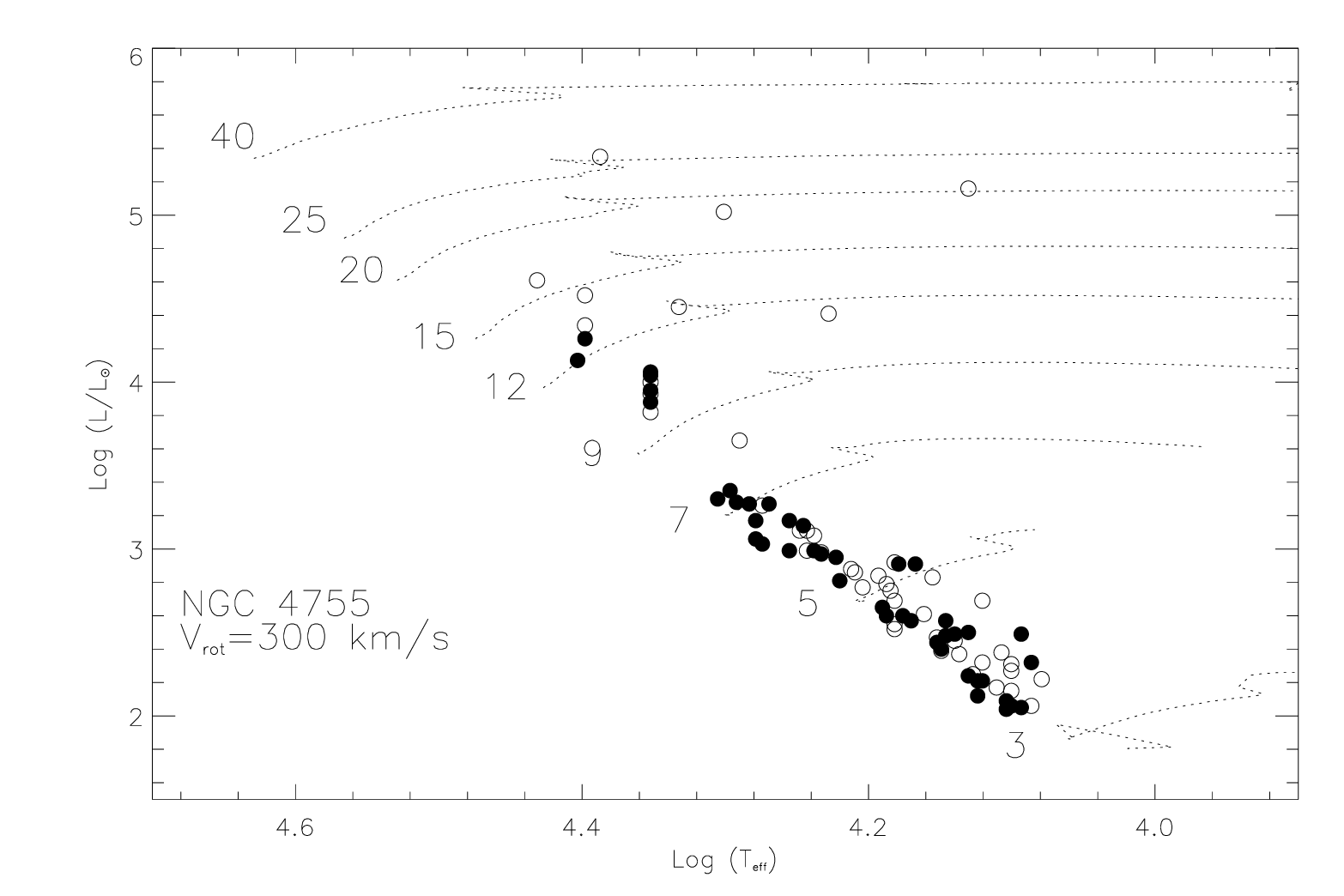,,angle=0,width=12cm}
\psfig{file=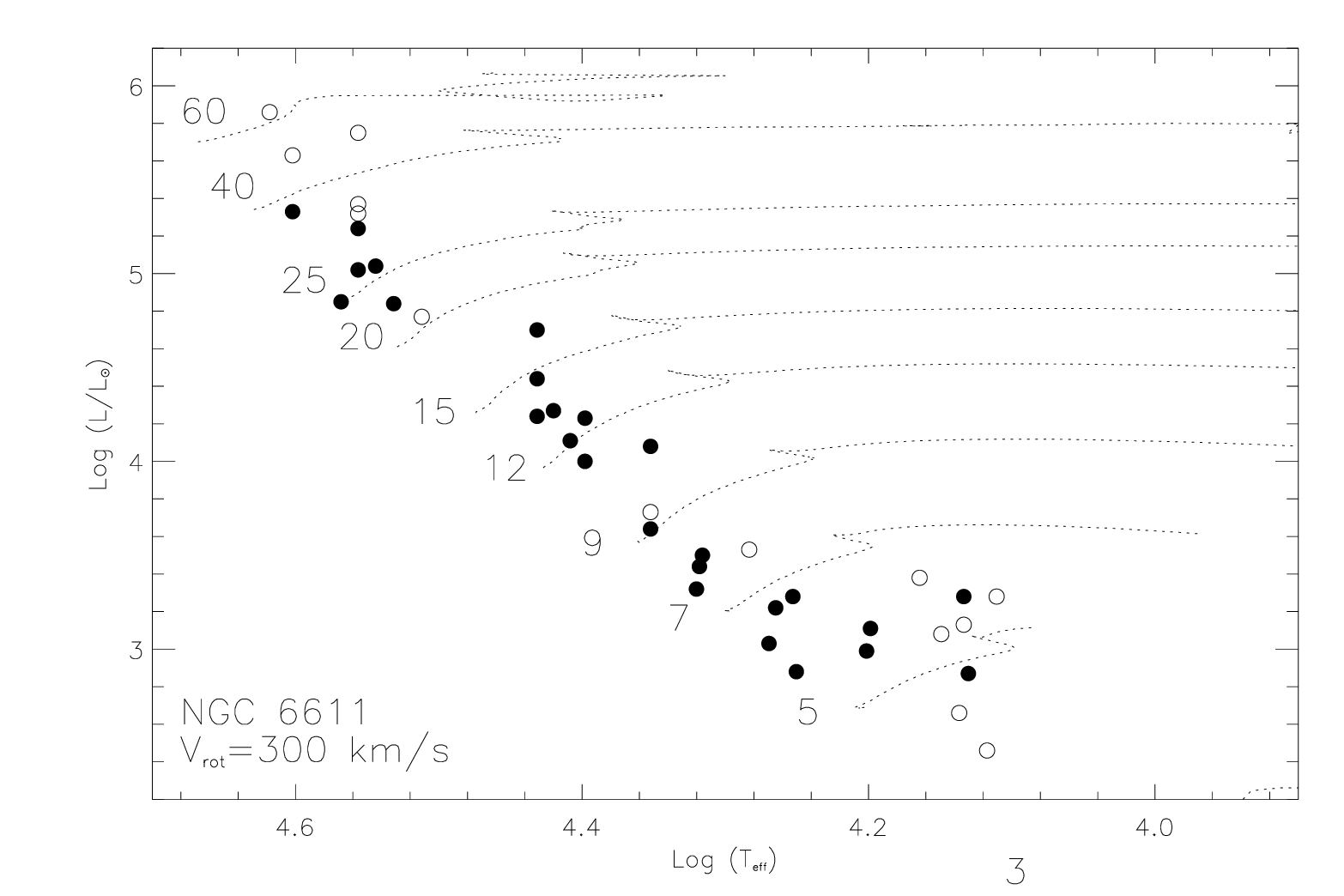,,angle=0,width=12cm}
\caption{Hertzsprung-Russell diagrams for the three clusters together with
rotating stellar evolutionary models from the Geneva group. The symbols are 
as in Fig.\ref{hr_diag}, and all the models have an initial equatorial 
rotational velocity of 300\kms. The tracks for the 25-60\msun models have 
been truncated before the bluewards evolution towards the WR phase, to aid 
clarity.}
\label{hr_rot}
\end{center}
\end{figure*}

The rotational velocity distributions found in the previous section
suggest that we should compare our observational HR diagrams to stellar 
evolutionary calculations which include rotation. In Fig.\ref{hr_rot} 
we reproduce Fig\,\ref{hr_diag}, but with evolutionary tracks taken from
Meynet \& Maeder (\cite{2003A&A...404..975M}) which have an initial
rotational velocity on the ZAMS of 300\kms (we are grateful to Georges 
Meynet for providing the lower mass tracks 3-7\msun; these are unpublished 
at present and are computed without core overshooting).
Meynet \& Maeder point out that these models would correspond to average 
equatorial rotational velocities of between 180 and 240\kms 
during the main sequence phase, which is well matched to
our preferred solutions for the intrinsic rotational velocity distributions. 
For the higher mass stars we have truncated the tracks before they move 
bluewards towards the WR phase in order to aid clarity in the figures. 
The position of the ZAMS in NGC\,3293 and NGC\,4755 was significantly 
below the stellar positions for the non-rotating models as discussed 
in Section\,\ref{disc:hr_diag}. The rotating models do provide 
somewhat better agreement, as the ZAMS moves towards lower
effective temperature. The width 
of the main-sequence phase also increases for a rotating star 
which may help to explain the scattering of points 
above the ZAMS. However one would still expect the ZAMS to fit through 
the middle of the observed distribution rather than the lower envelope
as observed and the cause of this is still unclear.  If we were to 
use the rotating models to estimate the mass of the stars, we 
would obtain different values to those for the non-rotating models, 
but not in any simple scalable fashion. However the differences 
are at most 10\%, and  are not likely to influence the PDMF 
determinations. 

\section{Conclusions}
\label{sec:concl}
We have analysed high resolution spectroscopy of targets in three Galactic
clusters to estimate atmospheric parameters and projected rotational
velocities. These have been used to derive stellar luminosities and
evolutionary masses. Our principle conclusions are as follows:
\begin{enumerate}
\item 
Atmospheric parameters have been deduced 
from non-LTE model atmosphere calculations and should be more 
reliable than those inferred from photometry or spectral types.
Hence the observed HR-diagrams should be most the reliable yet 
determined for these clusters. 
\item 
A significant number of relatively cool low gravity objects
have been identified and are found on average to have large
projected rotational velocity. The low gravities may then reflect 
at least in part 
the significant centrifugal accelerations experienced by these stars.
\item
Observed Hertzsprung-Russell diagrams have been compared with 
theoretical predictions for both non-rotating and rotating models.
The observed main sequence luminosities are in general larger than predicted.
For NGC\,6611, this may reflect non co-eval star formation. Other possible
explanation include stellar rotation and the presence of a significant
number of binary systems.
\item
Cluster ages have been deduced and some evidence for non-coeval star 
formation is considered especially for the young cluster NGC\,6611.
Indeed for none of the clusters does a single isochrone provide a 
good fit to the observed HR-diagram.
\item 
Projected rotational velocities for targets in the 
older clusters, NGC\,3293 and NGC\,4755, have been found to be systematically 
larger than those for the field, confirming the conclusion of Strom et al.\ 
(\cite{Str05}) for h and $\chi$\,Persei.  Contrary to the finding of 
Strom et al., this phenomenon appears to be present at all the B-type 
spectral types present in our sample. 
\item 
The distribution of projected rotational velocities are consistent 
with a Gaussian distribution of current rotational velocities. 
For the relatively unevolved targets
in the older clusters, NGC\,3293 and NGC\,4755, the peak of the velocity
distribution would be 250\kms with a full-width-half-maximum of 
approximately 180 \kms. For NGC\,6611, the corresponding values are
175\kms and 165 \kms.
\item
We find that the mean projected rotational velocity of stars which 
have strong winds
(15-60\msun) is lower than that of the lower mass stars. This is 
consistent with predictions of the rotating models which show 
the spin-down effect due angular momentum loss through stellar winds. 
\item For all three clusters we deduce present day mass functions with 
$\Gamma$-values in the range -1.5 to -1.8.
\end{enumerate}

\begin{acknowledgements}

We thank the referee, Steven Strom, for his valuable and
constructive comments on an earlier version of this paper.
We are grateful the staff at both Paranal and La Silla for
their invaluable assistance with the observational programme.  We also
thank Jonanthan Smoker and Roheid Mokiem for discussions on rotational
velocities and Georges Meynet for access to some of the low
mass rotating models in advance of publication.  CJE and JKL
acknowledge financial support from the UK Particle Physics and
Astronomy Research Council (PPARC) under grants PPA/G/S/2001/00131 and
PPA/G/O/2001/00473 and IH acknowledges financial support from the
Northern Ireland Department of Education and Learning.  AH, CAT and DJL
thank the Spanish Ministerio de Educaci\'on y Ciencia for support under 
project PNAYA 2004-08271-C02-01. SJS acknowledges support from the ESF 
in the form of a EURYI award. This paper is based in part on
observations made with the Isaac Newton and William Herschel
telescopes, operated on the island of La Palma by the Isaac Newton
Group in the Spanish Observatorio del Roque de los Muchachos of the
Instituto de Astrof\'{\i}sica de Canarias.

\end{acknowledgements}

{\scriptsize
\begin{center}
\begin{longtable}{lccccccccccrrr}
\caption[]{Physical parameters for NGC\,3293:  $V$, $B-V$ and spectral types are 
from Paper~I; Atmopheric parameters (\teff, \logg) and projected rotational
velocities (\vsini) were estimated using the methodologies discussed in Sect.
\ref{analysis}. The specific method used to detemine the atmospheric
parameters are designated as follows --- He: from neutral helium and hydrogen
equivalent widths;  He+: from ionized helium, \ion{Si}{iii}/\ion{Si}{iv}
ionization equilibrium and hydrogen line profiles; A: \teff adopted; I:
\teff interpolated. The radial velocities ($v_r$) are taken from Paper I,
whilst the distances from the cluster centre, $r\arcmin$, are in 
arcminutes. }
\\
\hline\hline\noalign{\smallskip}
ID & $V$  & $B-V$ & Sp.\ Type & \teff & \logg       & Method & $V$sin$i$ & $E(B-V)$ & \logL & Mass & $v_{\rm r}$ & $r\arcmin$\\
   & mag. &       &           & K     & cm~s$^{-2}$ &          & \kms    &          &        &      &   \kms         \\  
\noalign{\smallskip}\hline\noalign{\smallskip}
\endhead
\hline 
\multicolumn{7}{r}{\it{continued on next page}} \\
\endfoot
\hline 
\endlastfoot
3293-001 &\o6.52  & 0.00   & B0 Iab        & 27500 & 3.10  & Si&  80  &   0.22 & 5.55 &   40  &     0\p &  0.00    \\ 
3293-002 &\o6.73  & 0.07   & B0.7 Ib       & 23000 & 2.80  & Si& 100  &   0.27 & 5.35 &   29  &  $-$7\p &  2.11    \\ 
3293-003 &\o7.61  & 0.11   & B1 III        & 21100 & 2.90  & Si&  80  &   0.30 & 4.95 &   18  & $-$16\p &  0.82    \\ 
3293-004 &\o8.03  & 0.02   & B1 III        & 21500 & 3.00  & A & 105  &   0.21 & 4.69 &   14  & $-$21\p &  1.11    \\ 
3293-005 &\o8.12  & 0.08   & B1 III        & 21500 & 3.05  & A & 195  &   0.27 & 4.73 &   15  &  $-$8\p &  1.47    \\ 
3293-006 &\o8.21  & 0.07   & B1 III        & 21500 & 3.15  & A & 200  &   0.26 & 4.68 &   14  & $-$10\p &  1.59    \\ 
3293-007 &\o8.25  & 0.18   & B1 III        & 22800 & 3.10  & Si&  65  &   0.38 & 4.87 &   17  & $-$16\p &  4.74    \\ 
3293-008 &\o8.59  & 0.05   & B1 III        & 21500 & 3.25  & A & 140  &   0.24 & 4.50 &   13  &    57\p &  2.09    \\ 
3293-010 &\o8.77  & 0.00   & B1 III        & 21000 & 3.15  & Si&  70  &   0.18 & 4.34 &   11  & $-$16\p &  1.35    \\ 
3293-012 &\o8.95  & 0.06   & B1 III        & 21500 & 3.30  & A & 100  &   0.25 & 4.37 &   12  & $-$28\p &  2.34    \\ 
3293-013 &\o9.03&$-$0.04\pp& B1 III        & 21500 & 3.40  & A & 105  &   0.15 & 4.22 &   11  & $-$60\p &  2.53    \\ 
3293-014 &\o9.09  & 0.19   & B0.5 IIIn     & 23500 & 3.60 :& em & 290:&   0.40 & 4.59 &  14.8 &    50\p &  1.53    \\ 
3293-015 &\o9.11&$-$0.01\pp& B1 V          & 25000 & 3.80  & A & 260  &   0.21 & 4.40 &  13.5 & $-$16\p &  0.88    \\ 
3293-016 &\o9.21  & 0.03   & B2.5 V        & 19200 & 4.00  & He& 110  &   0.20 & 4.09 &   9.5 & $-$14\p &  0.79    \\ 
3293-017 &\o9.22  & 0.04   & B1 V          & 25000 & 3.90  & A & 145  &   0.26 & 4.42 &  13.7 &  $-$4\p &  1.45    \\ 
3293-018 &\o9.26&$-$0.04\pp& B1 V          & 22500 & 3.70  & Si&  26  &   0.16 & 4.18 &  11.1 &  $-$7\p &  1.57    \\ 
3293-019 &\o9.27&$-$0.04\pp& B1 V          & 25000 & 3.85  & A & 120  &   0.18 & 4.30 &  12.7 & $-$17\p &  0.92    \\ 
3293-020 &\o9.55  &  0.03  & B1.5 III      & 20500 & 3.15  & I &  60  &   0.21 & 4.04 &   9   & $-$16\p &  2.86    \\ 
3293-021 &\o9.85  &  0.03  & B1.5 III      & 20500 & 3.15  & I & 230  &   0.21 & 3.92 &   9   &  $-$8\p &  0.80    \\ 
3293-022 &\o9.97  &  0.12  & Be (B0.5-1.5n)& 23500 & 4.25  & A & 280  &   0.33 & 4.15 &  11.2 & $-$20\p &  2.94    \\ 
3293-023 & 10.01&$-$0.05\pp& B1.5 III      & 20500 & 3.40  & I & 160  &   0.13 & 3.76 &   8   & $-$10\p &  1.38    \\ 
3293-024 & 10.01&$-$0.01\pp& B1.5 III      & 20500 & 3.50  & I & 135  &   0.17 & 3.81 &   8   & $-$14\p &  3.42    \\ 
3293-025 & 10.01  &  0.00  & B2 III        & 21100 & 3.70  & He& 215  &   0.19 & 3.85 &   8.9 & $-$18:  &  1.20    \\ 
3293-026 & 10.16&$-$0.02\pp& B2 III        & 19900 & 3.65  & He& 30   &   0.15 & 3.70 &   7.9 & $-$21\p &  2.16    \\ 
3293-027 & 10.22  &  0.07  & Be (B0.5-1.5n)& 23500 & 3.75  & A & 315  &   0.27 & 3.98 &   10  & $-$13\p &  1.33    \\ 
3293-028 & 10.26  &  0.01  & B2 V          & 19400 & 3.65  & He& 215  &   0.18 & 3.66 &   7.7 & $-$10\p &  1.24    \\ 
3293-029 & 10.32&$-$0.01\pp& B0.5-B1.5 Vn  & 25000 & 4.20  & A & 370  &   0.21 & 3.92 &  10.5 &     8\p &  2.40    \\ 
3293-030 & 10.51&$-$0.04\pp& B2 V          & 19800 & 3.70  & He& 205  &   0.13 & 3.52 &   7.3 &  $-$7\p &  1.28    \\ 
3293-031 & 10.66  &  0.05  & B2 V          & 17400 & 3.45  & He& 230  &   0.21 & 3.42 &   7   &  $-$7\p &  2.22    \\ 
3293-032 & 10.69  &  0.01  & B0.5-B1.5 Vn  & 25000 & 4.30  & A & 365  &   0.23 & 3.80 &  10.0 &    0\p  &  0.79    \\ 
3293-033$\dagger$
 & 10.72  &  0.66  & B8 III        & 12300 & 3.30  & He& 120  &   0.78 & 3.75 &   ... &     8\p &  9.99    \\ 
3293-034 & 10.74  &  0.11  & B2 IIIh       & 26100 & 4.25  & He& 120  &   0.34 & 3.95 &  11.1 & $-$16\p &  1.44    \\ 
3293-035 & 10.81  &  0.09  & B2 V          & 20600 & 3.80  & He& 250  &   0.27 & 3.61 &   7.8 &  $-$9\p &  10.11   \\ 
3293-037 & 10.94  &  0.04  & B2 V          & 20800 & 3.95  & He&  70  &   0.22 & 3.51 &   7.5 & $-$12:  &  3.05    \\ 
3293-038 & 11.00&$-$0.04\pp& B2.5 V        & 20600 & 3.95  & He& 235  &   0.14 & 3.38 &   7.1 & $-$13:  &  6.32    \\ 
3293-040 & 11.21  &  0.05  & Be (B3n)      & 18100 & 3.65  & He& 335  &   0.21 & 3.24 &   6.1 & $-$12\p &  0.76    \\ 
3293-041 & 11.22  &  0.04  & B2.5 V        & 20600 & 4.00  & He& 205  &   0.22 & 3.39 &   7.1 & $-$12\p &  1.21    \\ 
3293-043 & 11.32  &  0.07  & B3 V          & 17900 & 4.05  & He& 14   &   0.22 & 3.21 &   5.9 &     0\p &  2.81    \\ 
3293-045 & 11.42  &  0.32  & Be (B1-2n)    & 21600 & 3.70  & He& 375  &   0.51 & 3.71 &   8.5 & $-$18\p &  5.19    \\ 
3293-047 & 11.55  &  0.13  & B2.5 V        & 19700 & 4.15  & He& 170  &   0.30 & 3.31 &   6.7 & $-$16\p &  3.10    \\ 
3293-048 & 11.56  &  0.16  & B2.5 V        & 20500 & 4.05  & He& 180  &   0.34 & 3.40 &   7.1 & $-$12\p &  3.18    \\ 
3293-049 & 11.64  &  0.07  & B2.5 V        & 18200 & 4.05  & He& 125  &   0.23 & 3.10 &   5.8 & $-$18\p &  11.66   \\ 
3293-050 & 11.69  &  0.05  & B3 Vn         & 17200 & 3.75  & He& 355  &   0.20 & 2.99 &   5.3 & $-$12\p &  1.80    \\ 
3293-053 & 11.83  &  0.03  & B3 V          & 16300 & 4.00  & He& 255  &   0.17 & 2.84 &   4.8 & $-$15\p &  1.78    \\ 
3293-056 & 11.91  &  0.00  & B3 V          & 15100 & 3.45  & He& 240  &   0.14 & 2.69 &   4.5 &  $-$2\p &  5.23    \\ 
3293-057 & 11.92  &  0.21  & B3 V          & 16200 & 3.95  & He& 250  &   0.35 & 3.02 &   5.1 & $-$11\p &  2.08    \\ 
3293-059 & 12.00  &  0.29  & B5 III-Vn     & 16500 & 3.90  & He& 355  &   0.43 & 3.11 &   5.4 &     0\p &  4.98    \\ 
3293-061 & 12.03  &  0.05  & B5 V          & 15700 & 4.00  & He& 230  &   0.18 & 2.73 &   4.6 & $-$14\p &  1.10    \\ 
3293-062 & 12.03  &  0.15  & B3 V          & 17900 & 4.20  & He& 24   &   0.30 & 3.02 &   5.6 & $-$62\p &  3.91    \\ 
3293-063 & 12.05  &  0.09  & B5 V          & 16000 & 4.05  & He& 150  &   0.22 & 2.80 &   4.7 & $-$10\p &  10.49   \\ 
3293-065 & 12.06  &  0.14  & B5 III-V      & 15100 & 3.70  & He& 150  &   0.26 & 2.79 &   4.5 & $-$18\p &  9.13    \\ 
3293-066 & 12.07  &  0.16  & B5 V          & 14300 & 3.90  & He& 110  &   0.28 & 2.74 &   4.3 & $-$49:  &  2.16    \\ 
3293-067 & 12.11  &  0.32  & B3 V          & 16300 & 4.00  & He& 185  &   0.46 & 3.08 &   5.3 & $-$18:  &  3.82    \\ 
3293-069 & 12.22  &  0.07  & B5 V          & 15300 & 4.00  & He& 205  &   0.20 & 2.65 &   4.3 & $-$16\p &  3.32    \\ 
3293-070$\dagger$ 
& 12.22  &  0.14  & B5 III-V      & 14300 & 3.70  & He& 265  &   0.26 & 2.65 &   4.1 &    12\p &  9.42    \\ 
3293-073 & 12.25  &  0.11  & B6-7 V        & 13700 & 3.90  & He& 160  &   0.22 & 2.55 &   3.9 & $-$16\p &  1.47    \\ 
3293-074 & 12.28  &  0.09  & B8 III        & 11700 & 3.60  & He&  70  &   0.17 & 2.33 &   3.3 & $-$16:  &  3.65    \\ 
3293-075 & 12.29  &  0.12  & B5 IIIn       & 15600 & 3.85  & He& 335  &   0.25 & 2.71 &   4.5 & $-$11\p &  5.07    \\ 
3293-077 & 12.32  &  0.20  & B6-7 V        & 13800 & 3.95  & He& 36   &   0.31 & 2.65 &   4.0 &  $-$1\p &  7.06    \\ 
3293-080 & 12.36  &  0.14  & B5 V          & 15300 & 4.05  & He& 255  &   0.27 & 2.68 &   4.4 & $-$20\p &  6.93    \\ 
3293-082 & 12.40  &  0.21  & B5 III        & 16500 & 4.00  & He& 300  &   0.35 & 2.85 &   4.9 &  $-$8:  &  8.24    \\ 
3293-084 & 12.48  &  0.11  & B5 V          & 14400 & 3.95  & He& 295  &   0.23 & 2.52 &   4.0 & $-$17\p &  1.78    \\ 
3293-085 & 12.49  &  0.13  & B5 V          & 14500 & 3.75  & He& 320  &   0.25 & 2.55 &   4.1 & $-$10\p &  7.26    \\ 
3293-086 & 12.52  &  0.16  & B5 V          & 14700 & 4.10  & He& 130  &   0.28 & 2.59 &   4.1 & $-$13:  &  6.68    \\ 
3293-087 & 12.60  &  0.04  & B5 V          & 14200 & 3.75  & He& 120  &   0.15 & 2.37 &   3.8 & $-$14\p &  5.92    \\ 
3293-089 & 12.64  &  0.15  & B8 III        & 12800 & 4.00  & He& 70   &   0.25 & 2.37 &   3.5 & $-$13\p &  11.75   \\ 
3293-090 & 12.67  &  0.31  & B6-7 V        & 13800 & 4.35  & He& 170  &   0.42 & 2.64 &   4.0 & $-$11\p &  2.58    \\ 
3293-093 & 12.71  &  0.14  & B6-7 V        & 14400 & 4.55  & He& 110  &   0.26 & 2.47 &   3.9 & $-$12\p &  1.86    \\ 
3293-094 & 12.71  &  0.20  & B5 V          & 14800 & 4.15  & He& 30   &   0.32 & 2.58 &   4.1 & $-$34:  &  3.90    \\ 
3293-095 & 12.72  &  0.13  & B6-7 V        & 14100 & 4.20  & He& 305  &   0.24 & 2.42 &   3.8 & $-$18\p &  3.10    \\ 
3293-096 & 12.74  &  0.23  & B6-7 III      & 13600 & 3.75  & He& 285  &   0.34 & 2.50 &   3.8 &    12:  &  6.14    \\ 
3293-097 & 12.75  &  0.13  & B6-7 III      & 13900 & 3.80  & He& 275  &   0.24 & 2.40 &   3.8 & $-$10\p &  3.94    \\ 
3293-098 & 12.75  &  0.14  & B8 III-V      & 13300 & 4.20  & He& 32   &   0.24 & 2.35 &   3.6 & $-$13\p &  0.53    \\ 
3293-099 & 12.78  &  0.18  & B5 Vn         & 13800 & 3.70  & He& 260  &   0.29 & 2.44 &   3.8 & $-$14\p &  8.51    \\ 
3293-100 & 12.81  &  0.14  & B6-7 III-Vn   & 13800 & 3.90  & He& 335  &   0.25 & 2.38 &   3.7 & $-$15:  &  2.85    \\ 
3293-104 & 12.88  &  0.11  & B6-7 V        & 13800 & 4.10  & He& 250  &   0.22 & 2.31 &   3.7 &  $-$8:  &  1.27    \\ 
3293-105 & 12.89  &  0.19  & B8 III-V      & 12400 & 4.00  & He&  80  &   0.28 & 2.28 &   3.4 & $-$11\p &  1.94    \\ 
3293-106 & 12.90  &  0.08  & B6-7 V        & 13100 & 3.75  & He& 235  &   0.18 & 2.20 &   3.4 & $-$22\p &  1.26    \\ 
3293-107 & 12.91  &  0.22  & B8 III-V      & 12900 & 4.05  & He&  60  &   0.32 & 2.35 &   3.5 & $-$11\p &  3.68    \\ 
3293-108 & 12.92  &  0.13  & B6-7 V        & 12900 & 3.80  & He& 200  &   0.23 & 2.24 &   3.4 &  $-$6\p &  8.18    \\ 
3293-109 & 12.92  &  0.15  & B5 V          & 13900 & 3.80  & He& 150  &   0.26 & 2.35 &   3.7 & $-$11\p &  3.86    \\ 
3293-111 & 12.96  &  0.16  & B6-7 III-V    & 13100 & 3.85  & He& 255  &   0.26 & 2.28 &   3.5 & $-$13\p &  2.30    \\ 
3293-113 & 13.04  &  0.14  & B6-7 V        & 13000 & 3.95  & He& 280  &   0.24 & 2.21 &   3.4 & $-$16\p &  1.71    \\ 
3293-114 & 13.07  &  0.25  & B6-7 III-V    & 13800 & 3.90  & He& 260  &   0.36 & 2.41 &   3.7 & $-$15\p &  6.53    \\ 
3293-115 & 13.11  &  0.14  & B8 III-V      & 12700 & 4.10  & He& 170  &   0.24 & 2.16 &   3.3 & $-$13\p &  1.73    \\ 
3293-116 & 13.12  &  0.11  & B6-7 V        & 13000 & 3.90  & He& 260  &   0.21 & 2.14 &   3.4 & $-$14\p &  1.81    \\ 
3293-117 & 13.15  &  0.14  & B8 III-V      & 13200 & 3.95  & He& 34   &   0.24 & 2.19 &   3.4 &  $-$8\p &  3.67    \\ 
3293-118 & 13.17  &  0.06  & B8 III-V      & 12100 & 3.95  & He& 205  &   0.15 & 1.98 &   3.1 & $-$10\p &  10.06   \\ 
3293-120 & 13.20  &  0.33  & B5 V          & 14600 & 4.10  & He& 245  &   0.45 & 2.52 &   4.0 & $-$12:  &  10.11   \\ 
3293-123 & 13.26  &  0.20  & B8 III        & 12400 & 3.85  & He& 275  &   0.29 & 2.14 &   3.2 & $-$27\p &  1.09    \\ 
3293-124 & 13.28  &  0.21  & B8 III        & 13200 & 3.95  & He& 235  &   0.31 & 2.22 &   3.5 &  $-$9\p &  8.29    \\ 
3293-125 & 13.29  &  0.10  & B8 III-V      & 12300 & 3.95  & He& 210  &   0.19 & 2.00 &   3.1 & $-$14:  &  0.72    \\ 
\label{tab:3293vrot}  
\end{longtable}
\end{center}
}

{\scriptsize
\begin{center}
\begin{longtable}{lccccccccccrrr}
\caption[]{Physical parameters for NGC\,4755. The columns and sources of
information are as for Table \ref{tab:3293vrot}.}
\\
\hline\hline\noalign{\smallskip}
ID & $V$  & $B-V$ & Sp.\ Type & \teff & \logg       & Method & $V$sin$i$ & $E(B-V)$ & \logL & Mass & $v_{\rm r}$ & $r\arcmin$\\
   & mag. &       &           & K     & cm~s$^{-2}$ &          & \kms    &          &        &      &   \kms         \\  
\noalign{\smallskip}\hline\noalign{\smallskip}
\endhead
\hline 
\multicolumn{7}{r}{\it{continued on next page}} \\
\endfoot
\hline 
\endlastfoot
4755-001  &\o5.77 &  0.32 & B9 Ia	 & 13500 & 2.00 & He &  38 & 0.44 &   5.04 &  20  & $-$16\p\p\p & 2.85   \\ 
4755-002  &\o5.98 &  0.22 & B3 Ia	 & 16300 & 2.25 & Si &  70 & 0.37 &   5.18 &  23  & $-$10\p & 1.53    \\ 
4755-003  &\o6.80 &  0.24 & B2 III	 & 24400 & 3.40 & He &  38 & 0.45 &   5.35 &  30  & $-$21\p & 3.01    \\ 
4755-004  &\o6.92 &  0.20 & B1.5 Ib	 & 20000 & 2.75 & Si &  75 & 0.38 &   5.02 &  20  & $-$19\p & 0.32    \\ 
4755-005  &\o7.96 &  0.20 & B2 III	 & 16900 & 2.80 & He & 110 & 0.36 &   4.41 &  12  & $-$53\p & 2.05    \\ 
4755-006  &\o8.37 &  0.14 & B1 III	 & 21500 & 3.15 & A  & 100 & 0.33 &   4.45 &  13  & $-$20\p & 2.68    \\ 
4755-007  &\o8.58 &  0.11 & B1 V	 & 25000 & 3.70 & A  & 170 & 0.33 &   4.52 & 14.5 & $-$12\p & 1.88    \\ 
4755-008  &\o8.6l &  0.14 & B0.5 V	 & 27000 & 3.65 & A  & 170 & 0.37 &   4.61 & 16.1 &  $-$2\p & 0.48    \\ 
4755-009  &\o9.01 &  0.11 & B1 V	 & 25000 & 3.90 & A  & 115 & 0.33 &   4.34 & 13.0 & $-$15\p & 4.00    \\ 
4755-010  &\o9.38 &  0.16 & B1 V	 & 25000 & 4.10 & A  & 180 & 0.38 &   4.26 & 12.4 &     0\p & 1.71    \\ 
4755-011  &\o9.58 &  0.17 & B1.5 V	 & 22500 & 4.10 & I  &  22 & 0.37 &   4.06 & 10.4 & $-$24\p & 1.12    \\ 
4755-012  &\o9.60 &  0.16 & B1.5 V	 & 22500 & 4.00 & I  & 140 & 0.36 &   4.04 & 10.3 & $-$55\p & 1.81    \\ 
4755-013  &\o9.68 &  0.15 & B1.5 V	 & 22500 & 3.90 & I  & 190 & 0.35 &   4.00 & 10.0 &  $-$3\p & 2.13    \\ 
4755-015  &\o9.74 &  0.16 & B1 V	 & 25300 & 4.00 & Si &  48 & 0.38 &   4.13 & 11.6 & $-$18\p & 2.76    \\ 
4755-016  &\o9.76 &  0.12 & B1.5 V	 & 22500 & 3.80 & I  & 195 & 0.32 &   3.93 &  9.7 &  $-$8\p & 1.00    \\ 
4755-017  &\o9.90 &  0.18 & B1.5 V	 & 22500 & 4.10 & I  &  75 & 0.38 &   3.95 &  9.8 & $-$15\p & 2.94    \\ 
4755-019  & 10.04 &  0.17 & B1.5 V	 & 22500 & 4.10 & I  &  80 & 0.37 &   3.88 &  9.4 & $-$20\p & 5.57    \\ 
4755-020  & 10.05 &  0.13 & B2 V	 & 19500 & 3.80 & He &  38 & 0.30 &   3.65 &  7.7 & $-$27\p & 6.68    \\ 
4755-021  & 10.29 &  0.20 & B1.5 Vn	 & 22500 & 3.80 & I  & 300 & 0.40 &   3.82 &  9.2 & $-$11\p & 3.26    \\ 
4755-022  & 10.89 &  0.12 & B2.5 Vn	 & 18800 & 3.70 & He & 300 & 0.28 &   3.26 &  6.3 & $-$18\p & 1.83    \\ 
4755-023  & 10.99 &  0.18 & B2 V	 & 19800 & 3.95 & He & 120 & 0.35 &   3.35 &  6.8 & $-$26\p & 1.85    \\ 
4755-025  & 11.02 &  0.14 & B2.5 V	 & 19600 & 3.95 & He & 235 & 0.31 &   3.28 &  6.6 & $-$20\p & 4.75    \\ 
4755-026  & 11.19 &  0.24 & B2.5 V	 & 18600 & 3.95 & He & 205 & 0.40 &   3.27 &  6.3 & $-$15\p & 0.97    \\ 
4755-027  & 11.21 &  0.17 & B2.5 Vn	 & 17700 & 3.80 & He & 210 & 0.32 &   3.11 &  5.7 & $-$18\p & 1.05    \\ 
4755-029  & 11.23 &  0.19 & B2 V	 & 20200 & 4.00 & He & 190 & 0.37 &   3.30 &  6.8 & $-$19\p & 3.02    \\ 
4755-030  & 11.31 &  0.21 & B2.5 Vn	 & 17500 & 3.70 & He & 330 & 0.36 &   3.11 &  5.6 & $-$35\p & 3.66    \\ 
4755-031  & 11.34 &  0.24 & B2.5 V	 & 18000 & 3.95 & He & 150 & 0.40 &   3.17 &  5.9 & $-$20\p & 7.13    \\ 
4755-032  & 11.35 &  0.19 & B2.5 V	 & 19000 & 4.00 & He & 195 & 0.36 &   3.17 &  6.1 & $-$13\p & 3.68    \\ 
4755-033  & 11.35 &  0.24 & B3 V	 & 17600 & 4.00 & He &  75 & 0.39 &   3.14 &  5.7 & $-$19\p & 1.76    \\ 
4755-034  & 11.39 &  0.23 & B3 V	 & 15200 & 3.85 & He & 265 & 0.36 &   2.92 &  4.8 & $-$19\p & 2.91    \\ 
4755-035  & 11.41 &  0.23 & B5 V	 & 15100 & 3.95 & He & 190 & 0.35 &   2.91 &  4.7 & $-$31\p & 0.11    \\ 
4755-036  & 11.45 &  0.16 & B3 Vn	 & 17500 & 3.75 & He & 330 & 0.31 &   2.99 &  5.4 & $-$29\p & 2.40    \\ 
4755-037  & 11.48 &  0.13 & B2.5 V	 & 18800 & 4.05 & He & 195 & 0.29 &   3.03 &  5.8 & $-$19\p & 3.74    \\ 
4755-038  & 11.54 &  0.16 & B3 Ve	 & 15400 & 3.60 & He & 115 & 0.29 &   2.79 &  4.6 & $-$19\p & 2.72    \\ 
4755-039  & 11.54 &  0.16 & B2.5 V	 & 18000 & 4.00 & He & 235 & 0.32 &   2.99 &  5.5 & $-$23\p & 1.24    \\ 
4755-040  & 11.54 &  0.32 & B2.5 V	 & 19200 & 4.10 & He &  65 & 0.49 &   3.27 &  6.4 & $-$20\p & 10.04   \\ 
4755-041  & 11.58 &  0.21 & B3 V	 & 17100 & 4.00 & He & 165 & 0.36 &   2.97 &  5.3 & $-$16\p & 0.96    \\ 
4755-042  & 11.58 &  0.29 & B3 Vn	 & 17300 & 3.80 & He & 345 & 0.44 &   3.08 &  5.5 & $-$14\p & 3.11    \\ 
4755-043  & 11.59 &  0.22 & B3 V	 & 16700 & 4.10 & He & 125 & 0.36 &   2.95 &  5.1 & $-$29\p & 1.48    \\ 
4755-044  & 11.59 &  0.28 & B5 V	 & 14300 & 3.60 & He & 265 & 0.40 &   2.83 &  4.4 & $-$12\p & 8.80    \\ 
4755-045  & 11.61 &  0.22 & B3 V	 & 17300 & 3.95 & He & 245 & 0.37 &   2.99 &  5.3 & $-$17\p & 1.25    \\ 
4755-046  & 11.62 &  0.20 & B3 Vn	 & 16300 & 3.65 & He & 325 & 0.34 &   2.88 &  4.9 & $-$29:  & 3.80    \\ 
4755-047  & 11.62 &  0.23 & B3 V	 & 17100 & 3.90 & He & 280 & 0.38 &   2.98 &  5.3 & $-$20\p & 1.97    \\ 
4755-048  & 11.63 &  0.19 & B3 V	 & 19000 & 4.15 & He &  55 & 0.36 &   3.06 &  5.9 & $-$20\p & 4.42    \\ 
4755-049  & 11.65 &  0.34 & B5 V	 & 14700 & 4.00 & He & 140 & 0.46 &   2.91 &  4.7 & $-$36:  & 8.46    \\ 
4755-051  & 11.75 &  0.23 & B3 Vn	 & 16200 & 3.75 & He & 320 & 0.37 &   2.86 &  4.9 & $-$28\p & 2.25    \\ 
4755-052  & 11.77 &  0.17 & B3 V	 & 16600 & 4.00 & He & 165 & 0.31 &   2.81 &  4.9 & $-$20\p & 4.71    \\ 
4755-053  & 11.78 &  0.18 & B3 Vn	 & 16000 & 3.80 & He & 305 & 0.31 &   2.77 &  4.7 & $-$22\p & 2.96    \\ 
4755-054  & 11.78 &  0.19 & B3 V	 & 15300 & 3.45 & He & 110 & 0.33 &   2.75 &  4.7 & $-$20\p & 3.98    \\ 
4755-056  & 11.86 &  0.29 & B3 Vn	 & 15600 & 3.75 & He & 300 & 0.42 &   2.84 &  4.7 & $-$22\p & 1.25    \\ 
4755-057  & 11.94 &  0.36 & B6-7 III-Ve  & 13200 & 3.65 & He & 235 & 0.46 &   2.69 &  4.0 & $-$44:  & 8.45    \\ 
4755-058  & 12.07 &  0.26 & B3 Vn	 & 15200 & 3.75 & He & 320 & 0.39 &   2.69 &  4.4 & $-$28\p & 0.98    \\ 
4755-060  & 12.12 &  0.19 & B5 V	 & 15400 & 4.05 & He & 145 & 0.32 &   2.60 &  4.3 & $-$19\p & 3.77    \\ 
4755-061  & 12.15 &  0.25 & B5 V	 & 14500 & 3.30 & He &   6 & 0.39 &   2.61 &  4.3 & $-$20\p & 3.73    \\ 
4755-062  & 12.18 &  0.28 & B6-7 III-V   & 14000 & 4.00 & He & 200 & 0.39 &   2.57 &  4.0 & $-$14\p & 3.13    \\ 
4755-063  & 12.18 &  0.33 & B8 V	 & 12400 & 4.20 & He & 235 & 0.42 &   2.49 &  3.6 & $-$32\p & 5.72    \\ 
4755-064  & 12.22 &  0.27 & B6-7 V	 & 13500 & 4.30 & He & 160 & 0.38 &   2.50 &  3.8 & $-$23\p & 1.05    \\ 
4755-065  & 12.25 &  0.18 & B3 V	 & 15200 & 3.70 & He & 105 & 0.31 &   2.52 &  4.2 & $-$22\p & 5.82    \\ 
4755-066  & 12.25 &  0.27 & B3 V	 & 15500 & 4.10 & He & 165 & 0.40 &   2.65 &  4.4 & $-$24\p & 11.21   \\ 
4755-067  & 12.28 &  0.24 & B5 V	 & 14000 & 4.05 & He & 290 & 0.35 &   2.48 &  3.9 & $-$23\p & 2.73    \\ 
4755-068  & 12.31 &  0.27 & B3 Vn	 & 14800 & 3.95 & He & 270 & 0.39 &   2.57 &  4.1 & $-$20\p & 0.22    \\ 
4755-069  & 12.31 &  0.28 & B5 V	 & 15000 & 4.10 & He & 185 & 0.40 &   2.60 &  4.2 & $-$17\p & 2.16    \\ 
4755-070  & 12.32 &  0.23 & B3 V	 & 15200 & 3.60 & He & 175 & 0.36 &   2.55 &  4.2 & $-$24\p & 3.75    \\ 
4755-072  & 12.43 &  0.24 & B5 V	 & 14200 & 4.10 & He & 100 & 0.35 &   2.44 &  3.9 & $-$19\p & 1.59    \\ 
4755-074  & 12.47 &  0.29 & B5 III-V	 & 13800 & 3.50 & He & 120 & 0.40 &   2.45 &  3.8 & $-$36:  & 2.58    \\ 
4755-075  & 12.52 &  0.24 & B5 III-V	 & 14100 & 3.55 & He & 115 & 0.35 &   2.39 &  3.8 & $-$22\p & 1.71    \\ 
4755-076  & 12.58 &  0.36 & B5 V	 & 13800 & 3.95 & He & 190 & 0.47 &   2.49 &  3.8 & $-$18\p & 0.43    \\ 
4755-077  & 12.59 &  0.32 & B6-7 III	 & 12800 & 3.25 & He & 170 & 0.44 &   2.38 &  3.8 & $-$27\p & 2.26    \\ 
4755-078  & 12.60 &  0.32 & B5 Vn	 & 14200 & 3.90 & He & 315 & 0.43 &   2.47 &  3.9 & $-$23\p & 9.62    \\ 
4755-079  & 12.62 &  0.28 & B5 V	 & 13700 & 3.85 & He & 250 & 0.39 &   2.37 &  3.7 & $-$20\p & 2.41    \\ 
4755-080  & 12.62 &  0.35 & B8 III-V	 & 12200 & 3.95 & He & 220 & 0.44 &   2.32 &  3.4 & $-$19\p & 3.86    \\ 
4755-081  & 12.63 &  0.28 & B6-7 IIIn	 & 13200 & 3.65 & He & 285 & 0.38 &   2.32 &  3.5 & $-$18\p & 1.67    \\ 
4755-082  & 12.65 &  0.30 & B8 III-V	 & 12000 & 3.80 & He & 245 & 0.39 &   2.22 &  3.2 & $-$25\p & 2.93    \\ 
4755-087  & 12.81 &  0.25 & B8 III	 & 13500 & 4.10 & He & 280 & 0.36 &   2.24 &  3.5 & $-$19\p & 1.18    \\ 
4755-088  & 12.83 &  0.25 & B8 IIIn	 & 12900 & 3.80 & He & 260 & 0.35 &   2.17 &  3.4 & $-$24\p & 6.34    \\ 
4755-089  & 12.84 &  0.36 & B8 III	 & 12600 & 3.35 & He & 160 & 0.48 &   2.31 &  3.6 & $-$19\p & 6.94    \\ 
4755-090  & 12.88 &  0.27 & B8 IIIn	 & 12600 & 3.65 & He & 290 & 0.36 &   2.15 &  3.3 & $-$16\p & 2.39    \\ 
4755-091  & 12.90 &  0.37 & B8 III	 & 14100 & 4.40 & He &  50 & 0.48 &   2.40 &  3.8 & $-$20\p & 4.11    \\ 
4755-093  & 12.91 &  0.28 & B8 III-V	 & 13200 & 4.35 & He & 190 & 0.38 &   2.21 &  3.4 & $-$29\p & 1.09    \\ 
4755-094  & 12.93 &  0.31 & B8 III	 & 13400 & 3.80 & He &  90 & 0.41 &   2.25 &  3.5 & $-$30\p & 4.52    \\ 
4755-095  & 12.96 &  0.20 & B8 III	 & 12700 & 4.00 & He & 225 & 0.30 &   2.04 &  3.2 & $-$22\p & 5.51    \\ 
4755-096  & 12.96 &  0.29 & B8 III	 & 13300 & 4.25 & He & 180 & 0.39 &   2.21 &  3.5 & $-$13\p & 1.48    \\ 
4755-098  & 13.00 &  0.25 & B8 IIIn	 & 12700 & 3.95 & He & 280 & 0.35 &   2.09 &  3.3 & $-$20\p & 3.93    \\ 
4755-099  & 13.01 &  0.41 & B8 III	 & 12600 & 3.75 & He & 250 & 0.50 &   2.27 &  3.4 & $-$20\p & 9.25    \\ 
4755-100  & 13.09 &  0.26 & B8 III	 & 13300 & 4.40 & He & 220 & 0.36 &   2.12 &  3.4 & $-$15\p & 1.09    \\ 
4755-106  & 13.16 &  0.32 & B8 III	 & 12200 & 3.80 & He & 290 & 0.41 &   2.06 &  3.1 & $-$18\p & 9.89     \\ 
4755-107  & 13.18 &  0.29 & B8 III-V	 & 12600 & 4.10 & He & 165 & 0.38 &   2.06 &  3.2 & $-$20\p & 2.43     \\ 
4755-108  & 13.22 &  0.31 & B8 III-V	 & 12400 & 4.05 & He & 165 & 0.40 &   2.05 &  3.2 & $-$22\p & 3.70     \\ 
\label{tab:4755vrot}
\end{longtable} 					   
\end{center}

{\scriptsize
\begin{center}
\begin{longtable}{lccccccccccrrr}
\caption[]{Physical parameters for NGC\,6611. The columns  and sources of
information are as for Table \ref{tab:3293vrot}.}
\label{tab:6611vrot} \\
\hline\hline\noalign{\smallskip}
ID & $V$  & $B-V$ & Sp.\ Type & \teff & \logg       & Method & $V$sin$i$ & $E(B-V)$ & \logL & Mass & $v_{\rm r}$ & $r\arcmin$\\
   & mag. &       &           & K     & cm~s$^{-2}$ &          & \kms    &          &        &      &   \kms         \\  
\noalign{\smallskip}\hline\noalign{\smallskip}
\endhead
\hline 
\multicolumn{7}{r}{\it{continued on next page}} \\
\endfoot
\hline 
\endlastfoot
6611-001$\dagger$ & \o8.18 &  0.34 & B0 III & 32000 & 3.60 & He+ & 142 &  0.60 &  5.39 & 40   &    86\p  &  13.17    \\
6611-002  &\o8.18 &  0.43 & O4 III((f$^+$)) & 41500 & 3.90 & He+ & 102 &  0.71 &  5.86 & 63   & 14\p     &  1.29     \\
6611-003  &\o8.73 &  0.45 & O6-7 V((f))     & 40000 & 3.90 & He+ &  87 &  0.73 &  5.63 & 50   &    17\p  &  1.08     \\
6611-004  &\o8.90 &  0.04 & O8.5 V	    & 37000 & 4.00 & He+ &  76 &  0.32 &  4.85 & 24.0 &    23\p  &  4.18     \\
6611-005  &\o9.13 &  0.44 & O8 III	    & 36000 & 3.80 & He+ &  90 &  0.71 &  5.32 & 25   &     5\p  &  29.49    \\
6611-006  &\o9.39 &  0.24 & O9.7 IIIp	    & 34000 & 4.10 & He+ &  43 &  0.51 &  4.84 & 21.9 &  $-$1\p  &  4.04     \\
6611-008  &\o9.46 &  0.82 & O7 II(f)	    & 36000 & 3.50 & He+ & 100 &  1.09 &  5.75 & 60   &    15\p  &  1.79     \\
6611-010  &\o9.75 &  0.49 & B1: e	    &	--  & --   & He  & 225 &       &       &      &     4\p  &  12.22    \\
6611-011  &\o9.85 &  0.58 & O9 V	    & 36000 & 4.20 & He+ &  66 &  0.85 &  5.24 & 29.5 &    17\p  &  1.54     \\
6611-012  &\o9.85 &  0.48 & B0.5 V	    & 27000 & 4.00 & Si  &  75 &  0.71 &  4.70 & 17.0 &    14\p  &  3.75     \\
6611-015  & 10.12 &  0.43 & O9.5 Vn	    & 32500 & 3.90 & He+ & 410 &  0.69 &  4.77 & 20.3 &    23:   &  0.72     \\
6611-017  & 10.37 &  0.57 & O9 V	    & 36000 & 3.95 & He+ &  95 &  0.84 &  5.02 & 25.6 &    19\p  &  2.54     \\
6611-018  & 10.56 &  0.35 & B8 III	    & 12900 & 3.90 & He  &  90 &  0.45 &  3.28 & ...  &    16\p  &  9.34     \\
6611-019$\dagger$  
& 10.68 &  0.36 & B1.5 V	    & 22500 & 3.85 & I   & 235 &  0.56 &  3.96 &  9.8 & $-$47\p  &  12.12    \\
6611-020  & 10.69 &  0.40 & B0.5 Vn	    & 27000 & 4.10 & A   & 220 &  0.63 &  4.24 & 13.0 &    20\p  &  6.17     \\
6611-021  & 10.80 &  0.47 & B1 V	    & 26300 & 4.25 & Si  &  32 &  0.70 &  4.27 & 12.9 & $-$18\p  &  0.30     \\
6611-022  & 10.98 &  0.82 & Herbig Be	    &	--  &  --  & He  & 375 &       &       &      &   $-$\p  &  0.46     \\
6611-023  & 10.99 &  0.41 & B3 V	    & 14600 & 3.35 & He  & 165 &  0.55 &  3.38 & ...  & $-$19:   &  7.37     \\
6611-025  & 11.20 &  0.59 & B1 V	    & 25000 & 4.10 & A   &  95 &  0.81 &  4.23 & 12.2 &     8\p  &  0.73     \\
6611-027  & 11.26 &  0.45 & B1 V	    & 25000 & 4.10 & A   & 255 &  0.67 &  4.00 & 10.9 &     0\p  &  2.85     \\
6611-029  & 11.29 &  1.05 & O8.5 V	    & 36000 & 3.85 & He+ & 135 &  1.32 &  5.37 & 35   &  $-$6\p  &  4.53     \\
6611-032  & 11.48 &  0.36 & B1.5 V	    & 22500 & 4.15 & I   &  70 &  0.56 &  3.64 &  8.5 &     4\p  &  7.46     \\
6611-033  & 11.47 &  0.56 & B1 V	    & 25600 & 4.00 & Si  &  30 &  0.78 &  4.11 & 11.7 &     8\p  &  1.07     \\
6611-034  & 11.47 &  0.55 & B8 III	    & 13600 & 4.05 & He  &  60 &  0.66 &  3.28 &  ... & $-$18\p  &  7.76     \\
6611-035  & 11.51 &  0.75 & B0.5 V	    & 27000 & 4.10 & A   & 120 &  0.98 &  4.44 & 14.5 &    16\p  &  1.62     \\
6611-038  & 11.63 &  0.47 & B5 IIIn	    & 13600 & 3.30 & He  & 250 &  0.60 &  3.13 &  ... & $-$15\p  &  9.78     \\
6611-040  & 11.73 &  0.38 & B5 V	    & 15800 & 4.20 & He  & 250 &  0.51 &  3.11 &  5.3 &    30\p  &  4.89     \\
6611-041  & 11.72 &  0.45 & B5 V	    & 14100 & 3.75 & He  &  90 &  0.56 &  3.08 &  4.9 &     6\p  &  10.64    \\
6611-042  & 11.78 &  0.50 & B1.5 V	    & 22500 & 3.90 & I   & 155 &  0.70 &  3.73 &  8.8 &    14\p  &  1.41     \\
6611-045  & 12.02 &  1.05 & O9 V	    & 35000 & 4.00 & He+ &  25 &  1.32 &  5.04 & 25.4 &     6\p  &  11.35    \\
6611-048  & 12.09 &  0.48 & B2.5 V	    & 17900 & 4.00 & He  & 230 &  0.63 &  3.28 &  6.1 &     8\p  &  9.34     \\
6611-052  & 12.16 &  0.60 & B2 V	    & 19200 & 3.85 & He  & 120 &  0.77 &  3.53 &  7.1 &     7\p  &  1.39     \\
6611-056  & 12.36 &  0.52 & B8 III	    & 13500 & 4.35 & He  &  70 &  0.63 &  2.87 &  4.4 & $-$21\p  &  6.84     \\
6611-062  & 12.57 &  0.54 & B3 V	    & 18400 & 4.05 & He  & 165 &  0.70 &  3.22 &  6.1 &     8\p  &  2.09     \\
6611-063  & 12.59 &  0.95 & B1.5 V	    & 22500 & 4.00 & I   &  95 &  1.15 &  4.08 & 10.5 &     7\p  &  6.65     \\
6611-064  & 12.74 &  0.56 & B3 V	    & 15900 & 4.00 & He  & 165 &  0.69 &  2.99 &  5.0 &    14\p  &  1.57     \\
6611-066  & 12.79 &  0.64 & B2 V	    & 20800 & 4.20 & He  &  80 &  0.82 &  3.44 &  7.3 & $-$50:   &  0.39     \\
6611-069  & 12.86 &  0.42 & B2.5 V	    & 17800 & 4.15 & He  & 190 &  0.57 &  2.88 &  5.3 &    17\p  &  6.62     \\
6611-071  & 12.85 &  0.70 & B2 Vn	    & 20700 & 3.95 & He  & 270 &  0.88 &  3.50 &  7.5 &     1\p  &  1.70     \\
6611-072  & 12.90 &  0.51 & B5 III	    & 13700 & 3.60 & He  & 230 &  0.62 &  2.66 &  4.0 &    23\p  &  2.73     \\
6611-078  & 13.07 &  0.54 & B3 V	    & 18600 & 4.10 & He  & 125 &  0.70 &  3.03 &  5.8 &     5\p  &  0.56     \\
6611-080  & 13.00 &  1.39 & O7 V((f))	    & 40000 & 4.00 & He+ &  95 &  1.67 &  5.33 & 38   &    16\p  &  3.50     \\
6611-082  & 13.17 &  0.66 & B1-3 V	    & 20900 & 4.90 & He  &  70 &  0.84 &  3.32 &  7.1 &     2\p  &  10.04    \\
6611-085  & 13.30 &  0.50 & B5 III	    & 13100 & 3.05 & He  & 265 &  0.62 &  2.46 &  3.9 &     5\p  &  2.43     \\
\end{longtable}
\end{center}

\appendix
\section{Finding charts for Galactic clusters}
With the benefit of hindsight on the part of the authors, we now also
include finding charts for the targets in the three Galactic clusters
reported in Paper~I.  The NGC\,6611 field is shown in
Figure~\ref{fchart_6611}.  The cores of NGC\,3293 and NGC\,4755 are
fairly dense, so we show the central 7$'$ of each cluster in
Figures~\ref{fchart_3293c} and \ref{fchart_4755c}, with the full
fields shown in Figures~\ref{fchart_3293} and \ref{fchart_4755}.

\begin{center} 
\begin{figure}
\caption{VLT-FLAMES targets (open circles) and FEROS targets (open squares) in the
NGC\,6611 field.}\label{fchart_6611} 
\end{figure} 
\end{center}

\begin{center} 
\begin{figure}
\caption{VLT-FLAMES targets (open circles) and FEROS targets (open squares) in the
inner 7$'$ of the NGC\,3293 field.}\label{fchart_3293c} 
\end{figure}
\end{center}

\begin{center} 
\begin{figure}
\caption{VLT-FLAMES targets (open circles) and FEROS targets (open squares) in the
NGC\,3293 field.  Most of the objects observed in the inner region are shown separately 
in Figure \ref{fchart_3293c}.}\label{fchart_3293} 
\end{figure}
\end{center}

\begin{center} 
\begin{figure}
\caption{VLT-FLAMES targets (open circles) and FEROS targets (open squares) in the
inner 7$'$ of the NGC\,4755 field.}\label{fchart_4755c} 
\end{figure}
\end{center}

\begin{center} 
\begin{figure}
\caption{VLT-FLAMES targets (open circles) and FEROS targets (open squares) in the
NGC\,4755 field.  Most of the objects observed in the inner region are shown separately 
in Figure \ref{fchart_4755c}.}\label{fchart_4755} 
\end{figure}
\end{center}


\begin{thebibliography}{}

\bibitem[2002]{2002ApJ...573..359A} Abt, H.~A., Levato, H., \& 
Grosso, M.\ 2002, \apj, 573, 359 
 
\bibitem[1997]{Aff97} 
  Afflerbach, A., Churchwell, E., \& Werner, M.W. 1997, \apj, 478, 190 

\bibitem[1994]{Bal94a}
  Balona, L.A. 1994, MNRAS, 268, 119

\bibitem[1994]{Bal94b}
  Balona, L.A., \& Koen, C., 1994, MNRAS, 267, 1071

\bibitem[2003]{Bau03}
  Baume G., Vasquez R.A., G. Carraro, \& A. Feinstein, 2003, A\&A, 402, 549

\bibitem[1999]{Bel99}
  Belikov, A. N., Kharchenko, N. V., Piskunov, A. E., \& Schilbach, E. 1999,
  A\&AS, 134, 525
  
\bibitem[1974]{Ber74}
   Bernacca, P.L., \& Perinotto, M. 1974, A\&A, 33, 443
  
\bibitem[2003]{Ble03}
   Blecha, A., North, P., Royer, F., \& Simond, G. 2003, BLDR Software -
   Reference Manual, 1st. edition

\bibitem[1997]{1997A&A...319..811B} Brown, A.~G.~A., 
\& Verschueren, W.\ 1997, \aap, 319, 811 

\bibitem[2003]{Bou03}
   Bouret, J.-C., Lanz, T., Hillier, D.J., et al. 2003, ApJ, 595, 1182


\bibitem[1950]{1950ApJ...111..142C} 
Chandrasekhar, S., M\"{u}nch, G.\ 1950, \apj, 111, 142 

\bibitem[1990]{Chi90}
   Chini, R., \& Wargau, W.F. 1990, A\&A, 227, 213
   
\bibitem[2002]{Cro02}
   Crowther, P.A., Hillier, D.J., Evans, C.J., et al. 2003, ApJ, 579, 774

   
\bibitem[2006]{Cro05}
   Crowther, P.A., Lennon, D.J., \& Walborn, N.R. 2006, A\&A, 446, 279
   
\bibitem[Daflon \& Cunha(2004)]{2004ApJ...617.1115D} 
Daflon, S., \& Cunha, K.\ 2004, \apj, 617, 1115 


\bibitem[1999]{Duf00}
   Dufton, P.L., Smartt, S.J., \& Hambly, N.C. 1999, A\&AS, 139, 231

\bibitem[1985]{1985ApJS...57...91E} 
Elias, J.~H., Frogel,  J.~A., \& Humphreys, R.~M.\ 1985, \apjs, 57, 91 

\bibitem[2004]{Eva04}
   Evans, C.J., Crowther P.A., Fullerton A. W., Hillier D. J., 2004, ApJ, 610, 1021

\bibitem[2005]{Eva05}
   Evans, C.J., Smartt, S.J., Lee, J.K. \& et al. 2005, A\&A, 437, 467

\bibitem[2006]{Eva06}
   Evans, C.J., et al. 2006, A\&A, submitted

\bibitem[2005]{Duf05}
   Dufton, P.L., Ryans, R.S.I., Trundle, C., Lennon, D.J., Hubeny, I.,
   Lanz, T. \& Allende Prieto, C.  2005, A\&A, 

\bibitem[1958]{Fea58}
 Feast, M.~W.\ 1958, \mnras, 118, 618 

\bibitem[1963]{Fea63}
 Feast, M.~W.\ 1963, \mnras, 126, 111

\bibitem[2005]{Fre05}
   Freyhammer, L.M., Hensberge, H., Sterken, C., et al. 2005, A\&A, 429, 631

\bibitem[1992]{Gie92}
   Gies, D.R., \& lambert, D.L. 1992, ApJ, 387, 673

\bibitem[2000]{Gir00}
   Girardi, L., Bressan, A., Bertelli, G., \& Chiosi, C. 2000, A\&AS, 141, 371

\bibitem[1992]{Gra92} 
   Gray D.F., 1992, The observation and analysis of stellar photospheres. 
   Cambridge Univ. Press, 2nd ed., Cambridge.

\bibitem[Gummersbach et al.(1998)]{1998A&A...338..881G} Gummersbach, C.~A., 
Kaufer, A., Schaefer, D.~R., Szeifert, T., \& Wolf, B.\ 1998, \aap, 338, 
881 

\bibitem[2000]{Heg00}
   Heger, A., \& Langer, N. 2000, ApJ, 544, 1016


\bibitem[2002]{Her02}
  Herrero, A., Puls, J., Najarro, F. 2002, A\&A, 396, 949  

\bibitem[1993]{Hil93}
  Hillenbrand, L.A., Massey, P., Strom, S.E., \& Merrill, K.M. 1993,
  AJ 106, 1906


\bibitem[2003]{Hil03}
   Hillier, D.J.,Lanz, T., \& Heap, S.R., et al. 2003, ApJ, 588, 1039

\bibitem[2003]{Hjo03}
   Hjorth, J., Sollerman, J., Moller, P., et al. Nature, 423, 847

\bibitem[2005]{Hua05a}
 Huang, W. Gies, D.R, 2005, 2005, ApJ, submitted, astro-ph/0510450

\bibitem[1997]{1997MNRAS.284..265H} 
   Howarth, I.~D., Siebert, K.~W., Hussain, G.~A.~J., \& Prinja, R.~K.\ 
   1997, \mnras, 284, 265 

\bibitem[2001]{How02}
   Howarth, I.D., \& Smith, K.C., 2001, MNRAS, 327, 353

\bibitem[1988]{Hub88}
   Hubeny, I., 1988, Computer Physics Comm., 52, 103

\bibitem[1995]{Hub95}
   Hubeny, I., \& Lanz, T. 1995, ApJ, 439, 875

\bibitem[1998]{Hub98}
   Hubeny, I., Heap, S.R., \& Lanz, T., 1998, in {\em ASP Conf. Series -
   Boulder-Munich: Properties of Hot, Luminous Stars}, ed. I.D.~Howarth, 
   131, 108
   
\bibitem[2006]{Hun06}
   Hunter, I., Dufton, P.L., Smartt, S.J. et al, 2006, A\&A, 
   in preparation 

\bibitem[1999]{Kau99} 
   Kaufer, A., Stahl, O., Tubbesing, S., \& et al. 1999, The Messanger, 95, 8
   
\bibitem[2006]{Kau06} 
   Kaufer, A., Stahl, O., Prinja, R.~K., \& Witherick, D.\ 2006, 
   A\&A, 447, 325    

\bibitem[2004]{Kel04}
   Keller, S.C. 2004, Publ. Astron. Soc. Australia, 21, 310
  
\bibitem[2000]{Kud00}
   Kudritzki, R.-P., \& Puls, J. 2000, ARA\&A, 38, 613 

\bibitem[2005]{Kur05} 
   Kurucz, R.~L.\ 2005, Memorie  della Societa Astronomica Italiana 
   Supplement, 8, 14 
 
\bibitem[1987]{Lyn87}
   Lynga, G. 1987, 10 European Regional Astronomy Meeting of the IAU, 
   Prague, Czechoslovakia, Proceedings. Volume 4 (A89-32590 13-90). 
   Ondrejov, Czechoslovakia, Czechoslovak Academy of Sciences 121

\bibitem[1991]{Mae91} 
   Maeder A., \& Meynet G.,  1991, A\&AS, 89, 451

\bibitem[2000]{Mae00} 
   Maeder A., \& Meynet G.,  2000, A\&A, 361, 159

\bibitem[2001]{Mae01} 
   Maeder A., \& Meynet G.,  2001, A\&A, 373, 555

\bibitem[2006]{2006A&A...445..931M} Martayan C., 
Fr{\'e}mat, Y, Hubert, A.~M., Floquet, M., Floquet M., Zorec, J., 
Neiner, C., 2006, \aap, astro-ph/0601240
  
\bibitem[1995]{1995ApJ...454..151M} 
   Massey, P., Johnson,  K.~E., \& De gioia-Eastwood, K.\ 1995, \apj, 454, 151 

\bibitem[1999]{1999A&A...349..553M} McErlean, N.~D., 
   Lennon, D.~J., \& Dufton, P.~L.\ 1999, \aap, 349, 553 

\bibitem[1989]{Mas89} Massey, P., Parker, J.~W., 
   \& Garmany, C.~D.\ 1989, \aj, 98, 1305 

\bibitem[1994]{Mey94} 
   Meynet, G., Maeder A., Schaller, G., Schaerer, D., \& Charbonnel, G.
   1994, A\&AS, 103, 97

\bibitem[2000]{2000A&A...361..101M} 
   Meynet, G., \&  Maeder, A.\ 2000, \aap, 361, 101 

\bibitem[2003]{2003A&A...404..975M} 
   Meynet, G., \&  Maeder, A.\ 2003, \aap, 404, 975

\bibitem[2006]{Mok06}
   Mokiem, M.R., de Koter, A., Evans, C.J., \& et al. 2006, A\&A submitted

\bibitem[1996]{1996ApJ...463..737P} Penny, L.~R.\ 1996, \apj, 463, 737 
	
\bibitem[2002]{Pas02}
   Pasquini, L., Avila, G., Blecha, A., \& et al. 2002, The Messenger, 110, 1 
\bibitem[2005]{Pul05}
   Puls, J., Urbaneja, M.A., Venero, R., \& et al. 2005, A\&A, accepted 

\bibitem[2004]{Rix04}
   Rix, S., Pettini, M., Leitherer, C., et al. 2004, ApJ, 615, 98

\bibitem[2000]{2000A&A...363..537R} Rolleston, W.~R.~J., 
Smartt, S.~J., Dufton, P.~L., \& Ryans, R.~S.~I.\ 2000, \aap, 363, 537 

\bibitem[2002]{Rya02}
   Ryans, R.S.I., Dufton, P.L., Rolleston, W.J.R., Lennon, D.J., Keenan, F.P.,
et al. 2002, MNRAS, 336, 577

\bibitem[2003]{Rya03}
   Ryans, R.S.I., Dufton, P.L., Mooney, C.J., Rolleston, W.J.R., Keenan, F.P.,
   et al. 2003, A\&A, 401, 1119

\bibitem[1995]{Sag95}
   Sagar, R., \& Cannon, R.D. 1995, A\&AS, 111, 75
   
\bibitem[1995]{San01}
   Sanner, J., Bruzendorf, J., Will, J.-M., \& Geffert, M. 2001, A\&A, 369, 511

\bibitem[1986]{1986FCPh...11....1S} Scalo, J.~M.\ 1986, Fundamentals 
of Cosmic Physics, 11, 1 

\bibitem[1992]{Sch92}
   Schaller G., Schaerer D., Meynet G., \& Maeder A. 1992, A\&AS 96, 269


\bibitem[1983]{Sho83}
   Shobbrook, R.R. 1983, MNRAS, 205, 1215

\bibitem[1992]{Sle02}
   Slesnick, C.L., Hillenbrand, L.A., \& Massey, P. 2002, ApJ, 576, 880
   
\bibitem[2004]{Sma04}
   Smartt, S.J., Maund, J.R., Hendry, M.A., et al. 2004, Science, 303, 499
  
\bibitem[2005]{Str05}
   Strom, S.E., Wolff, S.C., \& Dror, D.H.A., 2005, ApJ, 129, 809
   
\bibitem[1997]{Tow97} 
   Townsend, R.~H.~D.\ 1997, \mnras, 284, 839 

\bibitem[2004]{Tru04}
   Trundle, C., Lennon, D.J., Puls, J., \& Dufton, P.L.  2004, A\&A, 417, 217

\bibitem[2005]{Tru05}
   Trundle, C., \& Lennon, D.J. 2005, A\&A, in press

\bibitem[2004]{Vac96}
   Vacca, W.D., Garmany, C.D., \& Shull, M.J. 1996, ApJ, 460, 914

\bibitem[2004]{Val04}
   V\'alquez, G.A., Leitherer, C., Heckmann, T.M., et al., 2004, ApJ, 600, 162

\bibitem[2004]{Vin01}
   Vink, J.S., de Koter. A., Lamers, H.J.G.L.M. 2001, A\&A, 269,574

\bibitem[2004]{Wal04}
   Walborn, N.R., Morrell, N.I., Howarth, I.D., et al., 2004, ApJ, 608, 1028

\bibitem[1982]{Wof82}
   Wolff, S.C., Edwards, S., \& Preston, G.W. 1982, ApJ, 252, 322

\bibitem[2006]{Wol06}
   Wolff, S.C., Strom, S.E., Drod, D., Lanz, L, \& Venn, K., 
   2006, ApJ in press
   
\end{thebibliography}
\end{document}